\documentclass{article}

\RequirePackage{amsthm,amsmath,amsfonts,amssymb}
\RequirePackage[authoryear]{natbib}
\RequirePackage[colorlinks,citecolor=blue,urlcolor=blue]{hyperref}
\RequirePackage{graphicx}
\usepackage{comment}
\usepackage{multirow}
\usepackage[table,xcdraw]{xcolor}
\usepackage[left=2.5cm, right=2.5cm, top=2cm, bottom=2cm]{geometry}
\usepackage{authblk} 
\RequirePackage[authoryear]{natbib}
\title{Bayesian model-based clustering for populations of network data}

\author[1]{Anastasia Mantziou}
\author[2]{Sim\'{o}n Lunag\'{o}mez}
\author[3]{Robin Mitra}
\affil[1]{The Alan Turing Institute}
\affil[2]{Departamento de Estad\'{i}stica, ITAM}
\affil[3]{Department of Statistical Science,
University College London}

\date{ }

\begin{document}

\maketitle

\begin{abstract}

\textcolor{black}{There is increasing appetite for analysing populations of network data due to the fast-growing body of applications demanding such methods. While methods exist to provide readily interpretable summaries of heterogeneous network populations, these are often descriptive or ad hoc, lacking any formal justification. In contrast, principled analysis methods often provide results difficult to relate back to the applied problem of interest. Motivated by two complementary applied examples, we develop a Bayesian framework to appropriately model complex heterogeneous network populations, whilst also allowing analysts to gain insights from the data, and make inferences most relevant to their needs. The first application involves a study in Computer Science measuring human movements across a University. The second analyses data from Neuroscience investigating relationships between different regions of the brain.
While both applications entail analysis of a heterogeneous population of networks, network sizes vary considerably. We focus on the problem of clustering the elements of a network population, where each cluster is characterised by a network representative. We take advantage of the Bayesian machinery to simultaneously infer the cluster membership, the representatives, and the community structure of the representatives, thus allowing intuitive inferences to be made. The implementation of our method on the human movement study reveals interesting movement patterns of individuals in clusters, readily characterised by their network representative. For the brain networks application, our model reveals a cluster of individuals with different network properties of particular interest in Neuroscience. The performance of our method is additionally validated in extensive simulation studies. 
}
\end{abstract}
Keywords: Bayesian models, Clustering, Mixture models, Populations of network data, Object data analysis

\section{Introduction}

Conventional statistical methods for modelling and analysing data, such as standard regression models, assume each observation or outcome value is a scalar. While this is appropriate for many applications there is sometimes a need to handle more complex types of data. For example, each observation may constitute a set of interconnected points where the distribution of the connections between the points determines the observation's properties. Such data are typically referred to as networks in the literature, and modelling them is fundamentally important in some applications. For example, suppose we are interested in monitoring movements of subjects across a geographical area via displays at fixed locations. Each display represents a point, or node, of the network and when a subject moves from one display to another a connection, or edge, between the two nodes is assumed. The pattern of movement across the displays then characterises a network for that subject. \textcolor{black}{Effectively modelling such data can provide analysts with important insights into problems of interest}. For example, we may be interested in distinguishing different patterns of behaviour among the different subjects. Are some subjects more likely to visit certain displays than others? Do subjects have different or unusual patterns of movements to the majority of subjects? These are all important applied questions that present significant challenges to address due to the complexity of dealing with network data.

The availability of \textcolor{black}{populations of network} data has risen substantially in recent years, due to the advancement of technological means that record this type of data \citep{white1986structure, fields1989novel}. This has inspired many researchers to develop statistical models that most accurately describe the probabilistic mechanism that generates a network population. Specifically, there are three different frameworks considered in the literature for modelling \textcolor{black}{populations of network} data: 1) the latent space models, 2) the distance-based models, and 3) the measurement error models. 

For a single network observation, the fundamental idea behind the latent space class of models is that the occurrence of an edge between two nodes depends on the positions of the nodes in a latent space \citep{hoff2002latent,young2007random}. Recent studies \citep{gollini2016,levin2017central,durante2017,wang2019joint,nielsen2018multiple,arroyo2021inference} on modelling \textcolor{black}{populations of network} data have extended this idea to build models for populations of networks with aligned vertex sets, assuming that the nodes lie in a common, unobserved subspace. 

Another approach to modelling \textcolor{black}{populations of network} data is the utilisation of distance metrics that measure similarities among networks with respect to global or local characteristics of the networks \citep{donnat2018tracking}. Under this framework, researchers rely on the notion of an average network that represents a network population, with respect to a specified distance metric \citep{lunagomez,kolaczyk2020averages,ginestet2017hypothesis}. 

The third class of models, measurement error models, account for the erroneous nature of the networks. A fundamental source of noise found in network data originates from the various measurement tools used for the construction of networks, i.e. the processes used to measure an interaction (edge) between two objects (nodes). Researchers focusing on the statistical analysis of networks as single observations have developed methods to incorporate the uncertainty of falsely observing edges or non-edges in a network. Such studies involve predicting network topologies accounting for the falsely non-observed edges \citep{jiang2011network}, estimating the adjacency matrix from a set of noisy entries \citep{chatterjee2015matrix}, classifying nodes of networks with errorful edges \citep{priebe2015statistical}, developing a regression model for networks assuming that the observed network is a perturbed version of a true unobserved network \citep{le2020linear} and performing Bayesian inference on the network's structure utilising information from measurements \citep{young2020bayesian}. Another group of studies focuses on the propagation of the error to network summary statistics \citep{balachandran2017propagation,chang2020estimation}, and to estimators of average causal effects under network interference when the error arises from a measurement process used to construct the network \citep{li2021causal}. \cite{le2018estimating}, \cite{newman2018estimating} and \cite{peixoto2018reconstructing} develop this idea to model \textcolor{black}{populations of network} observations in order to infer the probabilistic mechanism that generates the network population. Specifically they assume the networks are noisy realisations of a true unobserved network. \textcolor{black}{Similarly, \cite{josephs2021network} consider the problem of network recovery from multiple noisy network realisations, for unlabeled networks.}

Despite the growing research interest on modelling \textcolor{black}{populations of network} data, only few studies developed to date consider the heterogeneity that can exist in a network population. Notably, \cite{mukher2017} were the first to consider the problem of clustering \textcolor{black}{populations of network} data. They assume two different scenarios, (a) the networks in the population share the same set of nodes, and (b) the networks in the population do not share the same set of nodes. Our paper focuses on Scenario (a). In Scenario (a), the authors obtain a mixture model of graphons and implement a spectral clustering algorithm to infer the membership allocation of each network observation. 

An application driven study on clustering \textcolor{black}{populations of network} data is introduced by \cite{diquigiovanni2019analysis}, who aim to cluster a population of networks where each network observation represents the playing style of a football team at a specific match. 
The clustering approach seen in this study involves the specification of an ad hoc measure of similarity between networks, and the implementation of an agglomerative method for clustering the networks according to their similarities. 

To the best of our knowledge, the third and last study that examines the problem of clustering network populations is that of \cite{signorelli2020model}. In this study, the authors deal with the problem of clustering using a mixture model whose components can be any statistical network model, under the restriction that it can be specified as a Generalised Linear Model (GLM). For estimating the parameters of their model, they implement an Expectation Maximisation (EM) algorithm for a predefined number of clusters. To determine the network model for their mixture, the authors propose the initial use of a mixture of saturated network models, to reveal information about the structure of the data at hand. The saturated network model assumes that each edge in each network in the population is generated with some unique, unconstrained probability.

Another group of studies that accounts for the heterogeneity in a set of network observations are the studies that perform the task of network classification. Some of these studies consider either specific network summary measures \citep{prasad2015brain}, or vectorise only the important entries (edges) of the adjacency matrix \citep{richiardi2011decoding,zhang2012pattern} to classify networks. Thus, they ignore the overall networks' structure. In contrast to these studies, \cite{relion2019network} perform prediction of the class membership of networks using a linear classifier with the adjacency matrices of the networks as predictors. Their approach accounts for the networks' structure by using a penalty to select important nodes and edges.

These contributions provide interesting approaches for identifying variations between network data, but there are some key limitations associated with these:
\begin{itemize}
\item In the studies of \cite{mukher2017}, \cite{diquigiovanni2019analysis} and \cite{relion2019network}, the methods proposed are non model-based. \cite{mukher2017} and \cite{diquigiovanni2019analysis} propose algorithms that detect underlying network clusters in the data and \cite{relion2019network} predict class membership of the networks. In all three studies the groups of networks identified cannot be interpreted using a parametric representation. Interpretability of the different groups of networks in a population is crucial in many applications in order to infer group specific properties and differences.
\item While \cite{signorelli2020model} provide a model-based approach for clustering \textcolor{black}{populations of network} data, the mixture components must conform to rigid modelling assumptions. This means that only specific characteristics of the networks can be inferred depending on what these model assumptions allow. It would be ideal to have a framework that is flexible enough to incorporate different modelling assumptions as deemed appropriate to application allowing the most scientifically relevant inferences to be made.
\item In addition, \cite{signorelli2020model} propose to initially obtain a mixture of saturated network models, thus resulting in an overly complex model with a large number of parameters to estimate. This can substantially increase the computational time needed for the EM algorithm to converge as well as increasing the potential for non-convergence due to having to explore a very high dimensional parameter space. 
\item The supervised approach of \cite{relion2019network} requires a training data set to predict the class.
The class labels of the networks in the training data set must be pre-specified, which can be restrictive for some network applications for which we do not have a priori information about the networks. 
\end{itemize}

To address these limitations, in this paper we propose a mixture model for identifying clusters of networks in a network population, with respect to similarities detected in the connectivity patterns of the networks' nodes.  We consider the case when the networks in the population share the same set of $n$ nodes, and each network could belong to one of, a predefined number of, $C$ clusters. Inspired by the approach of \cite{le2018estimating}, we adopt a measurement error formulation, assuming networks lying within each of the $C$ clusters are noisy realisations of a true underlying network representative. The attractive feature of this approach is that it decouples the statistical model for the network data from the underlying cluster specific network properties. We are thus able to provide a flexible model-based approach for detecting clusters of networks in a network population, as well as interpret these clusters with respect to our model parameterisation. Our framework is also flexible enough to incorporate, and thus exploit, any underlying assumptions about the structure of the networks within the clusters that are of scientific interest or otherwise supported by the data.

\cite{le2018estimating} develop a model for \textcolor{black}{populations of network} observations assuming that noisy network-valued observations arise from a true underlying adjacency matrix. The inferential framework built in their study consists of two steps. First, they use a Spectral Clustering algorithm to infer the community structure formed by the nodes of the true underlying network, and second they implement an EM algorithm to estimate the model parameters. An evident limitation of their inferential framework is that their algorithm does not simultaneously update the parameters of their model for the network data and the parameters characterising the underlying network structure, as this would require the development of new techniques. In addition, the assumption of a sole true underlying adjacency matrix is quite restrictive, especially for a large sample of networks where a degree of heterogeneity is expected.

We adopt a Bayesian modelling approach which provides some unique advantages over previous approaches. In particular by utilising Markov Chain Monte Carlo (MCMC) methods, we are able to simultaneously infer the cluster membership of the networks, together with model parameters characterising the distribution of the networks within each cluster as well as those that characterise the structure of the underlying cluster specific network representatives. To best of our knowledge, there is no coherent framework in the literature that permits this type of complete inference from the network data. 
\textcolor{black}{Our framework is flexible enough to answer a diverse range of applied questions with respect to the heterogeneity in a network population. These include being able to detect clusters of networks as well as inferring key different features between clusters through comparisons between the underlying representatives. In addition, interest may lie in identifying observations that do not follow the distribution of the majority of the network data, and the framework can also be formulated to detect outlying network observations.}

Our approach is motivated by two applied examples in very different fields, one involving monitoring movement of people across a University Campus and another measuring individuals’ connectivity patterns across different regions of the brain. In this second example we are particularly interested in determining whether any individuals possess unusual connectivity patterns, and if so how these differ to the rest of the sample. This applied question can be readily addressed using the proposed framework with the outlier formulation described above. Other principled methods, by contrast, would struggle to produce readily interpretable summaries.

The remainder of this article is organised as follows. In Section \ref{sec2}, we describe the applied examples that motivated the development of the methods. In Section \ref{sec3}, we provide background to our modelling framework. In Section \ref{sec4}, we develop the Bayesian formulation of a mixture of measurement error models for network data, along with the MCMC scheme to make inferences. \textcolor{black}{We further present the Sparse Finite Mixture (SFM) extension of our model that allows inferences for the number of clusters.} 
In Section \ref{sec5}, we present simulation studies to assess the performance of our method
for various network sizes and sample sizes. \textcolor{black}{In Section \ref{sec6}, we analyse the two different motivating \textcolor{black}{populations of network} data examples to provide important insights into applied questions of interest. This also serves to demonstrate the broad applicability of our methods.} Lastly, in Section \ref{sec7}, we give some concluding remarks.

\section{Two motivating examples}\label{sec2}
In this section we introduce two different applied examples that have triggered research questions which we aim to answer in our study. While both applications address very different areas, a common feature in both data sets is the heterogeneity of the network data in the sample.
\subsection{Data on movements of subjects across a University Campus}\label{sec21}
The first example comes from data collected on movement of people across Lancaster University Campus in the UK. The study was performed by members of the Computing and Communications department at the university \citep{shaw2018tacita}. 

\begin{figure}[h!]
\centering
\includegraphics[width=3.2in]{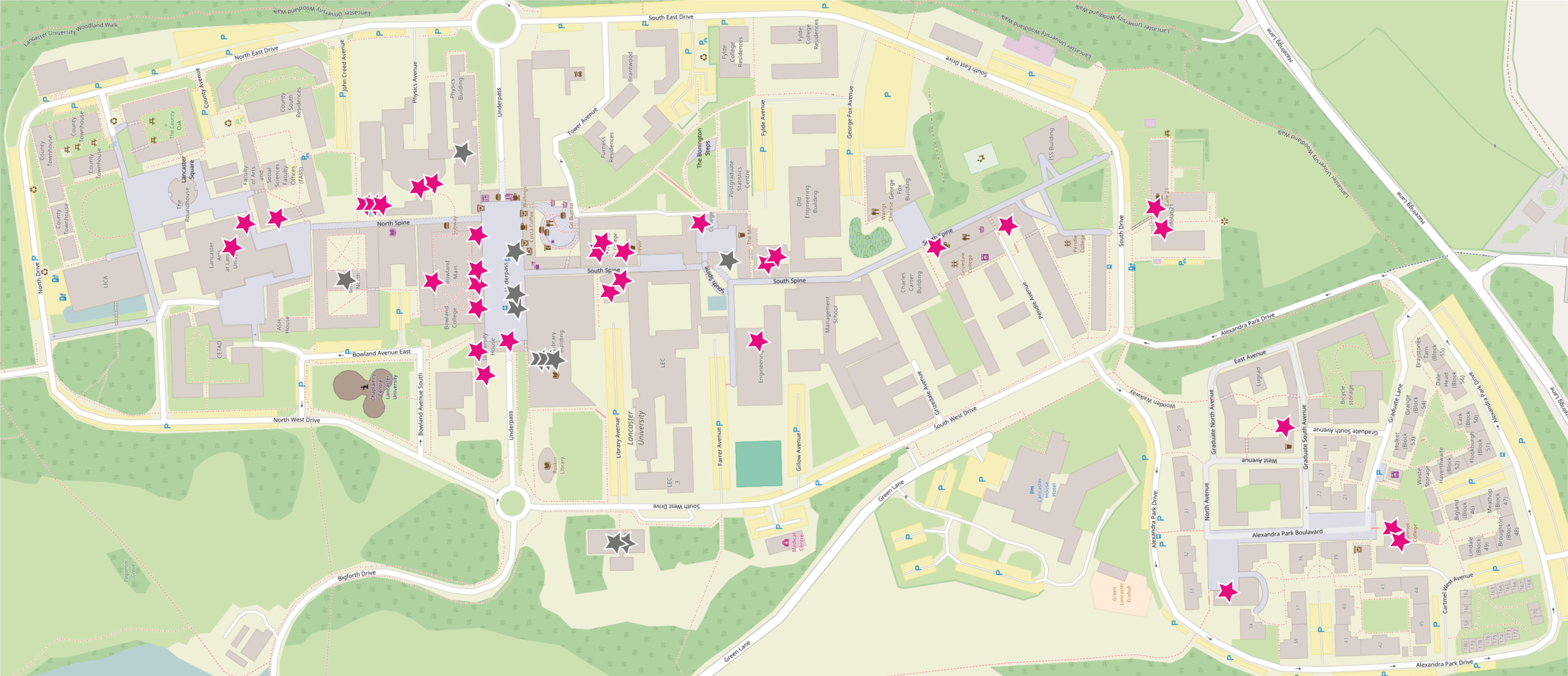}
\hspace{4mm}
\includegraphics[width=1.9in]{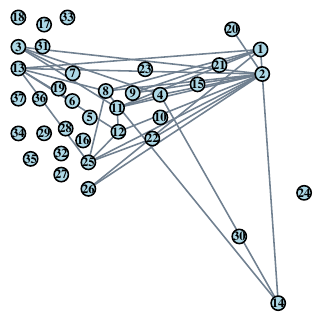}
\caption{Left: Lancaster University campus map with stars indicating displays' location. Right: Network visualisation of movements of one individual from the sample. The network layout corresponds to the physical location of the displays.}\label{tacita_app}
\end{figure}

A series of fixed displays are located across the campus (Figure \ref{tacita_app} left). Individuals taking part in the study installed a Tacita mobile application on their phone, and whenever they pass one of these displays this application registers their presence at that location. The application can also serve as a \textcolor{black}{means} of communication between a display and a viewer, in order for the viewers to be able to see content relevant to their interests. Specifically, the viewer can request what to see on the screen of the display, but also the display can detect when a user is in its proximity in order to show content aligned with their interests. Thus, the application records the consecutive displays visited by the users, along with the time visited, and the type of content shown.

Consequently, the Tacita data set serves as an example of \textcolor{black}{a population of network} observations, as the movements of each individual can be represented by a network. Thus, for each individual we can obtain a network where nodes represent displays, and edges represent the movements of the user among the displays. In this example each network corresponds to the aggregated movements of an individual across the campus during different times of the day, resulting in 120 network observations \textcolor{black}{sharing the same set of 37 nodes}. \textcolor{black}{In Figure \ref{tacita_app} (right), we illustrate the movements among the 37 displays (nodes) on campus, for one of the users of the Tacita application. The nodes' positions in Figure \ref{tacita_app} correspond to the physical location of the displays. We note here that there is not direct correspondence between the display locations indicated on the campus map and the nodes positions in Figure \ref{tacita_app} for two reasons: first, after data manipulation and consultation with our collaborators, some displays were not considered in our analysis, and second, the data collected involved newly activated displays not depicted on the campus map in Figure \ref{tacita_app}.} The following questions arise:

\begin{itemize}
\item \textcolor{black}{Can we infer a meaningful number of clusters based on the observed population of individuals?}
\item Can we detect different patterns among the users' movements?
\item Can we cluster the users according to their movements?
\item How informative can the clustering be for the users in our data?
\end{itemize}

\subsection{Data measuring connectivity patterns across different regions of the brain}\label{sec22}
\begin{figure}[h!]
\centering
\includegraphics[height=1.9in]{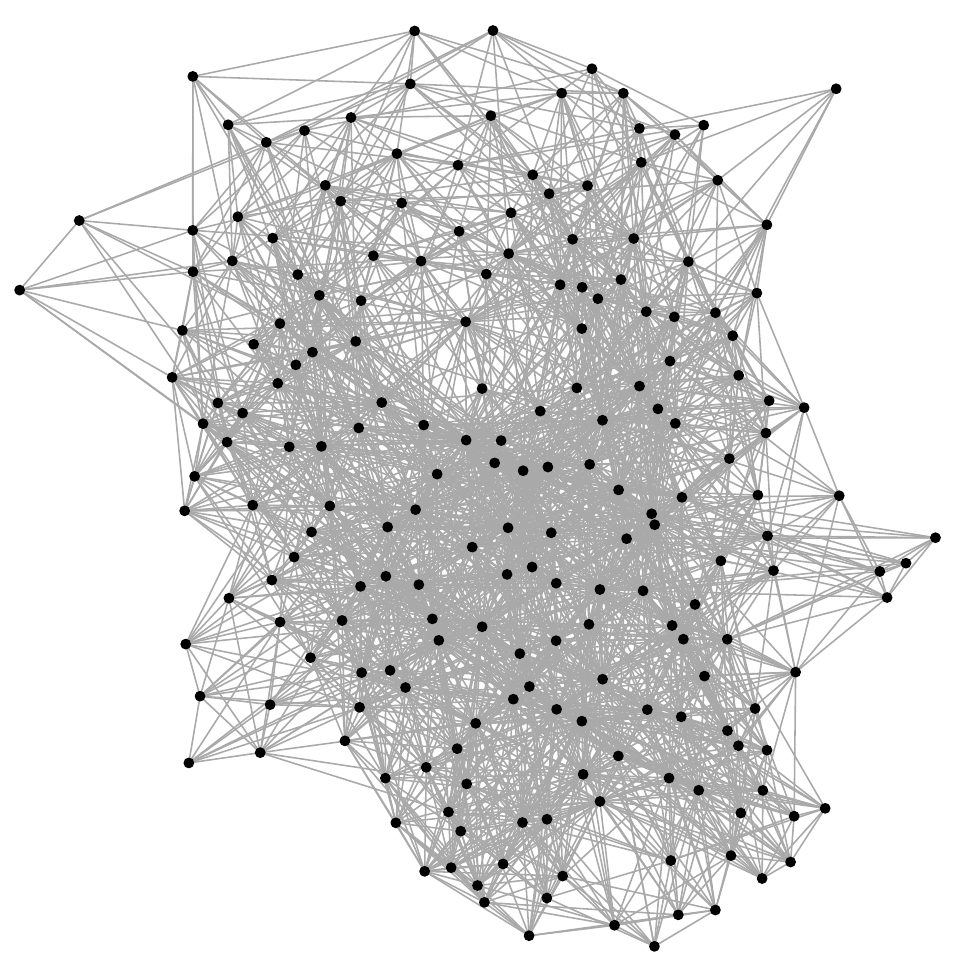}
\caption{Network representation of brain scan taken for one of the individuals in the sample.}\label{brain_app}
\end{figure}

The second example is a \textcolor{black}{population of networks} data set arising from the field of Neuroscience. In this data example, connectivity patterns across different regions of the brain were measured for 30 healthy individuals at resting state. For each individual a series of 10 measurements were taken using diffusion magnetic resonance imaging (dMRI). The measurements are represented as networks with the nodes corresponding to fixed regions of the brain, and edges denoting the connections recorded among those regions. Specifically, the network data consist of 200 nodes (regions of the brain) according to the CC200 atlas \citep{craddock2012whole}, \textcolor{black}{and the resulting data set consists of 300 network observations. In Figure \ref{brain_app}, we illustrate the network representation of one brain scan for one of the individuals in the data set.} 

This data set has been also discussed in the study of \cite{zuo2014open}, \cite{arroyo2021inference} and \cite{lunagomez}, with the latter two studies analysing the data from a modelling perspective. Specifically, \cite{arroyo2021inference} investigate the ability of their method to identify differences among individuals with respect to communities formed from the networks' nodes, while \cite{lunagomez} assume unimodality of the probabilistic mechanism that generates the network population and infer a representative network for the population of individuals, according to a pre-specified distance metric. None of these approaches seek to determine and interpret clusters of networks. \textcolor{black}{In particular, a relevant research question here is determining whether any individuals meaningfully differ from the rest of the sample in terms of their network characteristics, and if so, in what way.
Motivated by this objective, we can formulate our model to capture, and subsequently interpret, outlying network data, through an appropriate cluster specification, as presented in Section \ref{sec44}.}

\textcolor{black}{More generally,} our goal in this application is to explore possible heterogeneity amongst the networks. Questions include: 

\begin{itemize}
    \item Can we identify clusters of individuals with respect to similarities found in their connectivity patterns? \textcolor{black}{In particular, can we identify individuals with different network characteristics to the majority of the population?}
    \item \textcolor{black}{Can we interpret the clusters identified with respect to some network feature so they are relevant to Neuroscience applications?}
    \item \textcolor{black}{Are brain scans of the same individual assigned to the same cluster?}
\end{itemize}

These applied research questions have motivated the proposed mixture of network measurement error models, described in detail in Section \ref{sec6}.

\section{Background}\label{sec3}
A network can be represented as a graph $\mathcal{G}=(V,E)$, where $V= \{1,\ldots,n\}$ represents the set of $n$ nodes and $E$ represents the set of observed edges in $\mathcal{G}$, with $E\in\mathcal{E}_{n}$ and $\mathcal{E}_n=\{(i,j)|i,j\in V\}$. A common mathematical network representation is an $n\times n$ adjacency matrix $A_{\mathcal{G}}$, with the $A_{\mathcal{G}}(i,j)$ entry of the matrix denoting the state of the $(i,j)$ edge. The $(i,j)^{th}$ element of the adjacency matrix for a graph with binary edges is,
\begin{displaymath}
A_{\mathcal{G}}(i,j)=
\begin{cases}
1,\text{ if an edge occurs between nodes i and j,}\\
0,\text{ otherwise}.
\end{cases}
\end{displaymath}
The adjacency matrix of an undirected graph with no self-loops is symmetric with $A_{\mathcal{G}}(i,j)=A_{\mathcal{G}}(j,i)$ and  $A_{\mathcal{G}}(i,i)=0$, for $i,j \in \{1,\ldots,n\}$. By $\mathcal{G}_{1},\ldots,\mathcal{G}_{N}$ we represent a population of $N$ graphs, with corresponding adjacency matrices $A_{\mathcal{G}_{1}},\ldots,A_{\mathcal{G}_{N}}$. In this study, we assume that the networks in the population $\mathcal{G}_{1},\ldots,\mathcal{G}_{N}$ are undirected with no self-loops, and share the same set of $n$ nodes. We note here that our methods can be equally applied to populations of directed graphs.

When modelling a population of network observations it is often natural to assume that the networks have been subjected to some noise or measurement error during their construction. For example, in the application investigating the movement of subjects across campus, it is possible that the Tacita mobile application fails to register a subject at a display occasionally, while also sometimes incorrectly registering a subject at a display, particularly when displays are located fairly close together. This results in network data that might have edges missing as well as edges recorded that should not be present. 

Under this measurement error hypothesis, the researcher assumes that the observed network data correspond to noisy realisations of a true underlying network, which leads to recording some erroneous edges, due to the existence of an underlying measurement error process. \cite{le2018estimating} were the first to introduce this approach for modelling populations of networks, which has inspired the proposed modelling approach.

\cite{le2018estimating} assume that the information contained in the network population can be summarised by a representative network $\mathcal{G}^{*}$, and a measurement error process that does not allow us to accurately observe the representative network. Specifically, the authors assume a false positive probability $P_{ij}$ of observing an edge between nodes $i,j$ in the $k^{th}$ network observation $\mathcal{G}_{k}$, given that there is no edge between the same two nodes in the representative network $\mathcal{G}^{*}$; and respectively, a false negative probability $Q_{ij}$ of not observing an edge for the nodes $i,j$ in the $k^{th}$ network observation $\mathcal{G}_{k}$, while there is an edge for the same two nodes in the representative network $\mathcal{G}^{*}$. Thus, the entries of the matrices $P,\text{ }Q$, are the false positive/negative probabilities of seeing/not seeing an edge respectively between two nodes in the data. 

The mathematical formulation of the above set-up is the following. Let $A_{\mathcal{G}_{1}},\cdots,A_{\mathcal{G}_{N}}$ denote the adjacency matrices for the network population, and $A_{\mathcal{G}^{*}}$ the adjacency matrix of the representative network. The false positive and false negative probabilities $P_{ij},Q_{ij}$ can be described as follows,
\begin{gather*}
\text {if } A_{\mathcal{G}^{*}}(i,j)=1, \text{ then }A_{\mathcal{G}_{k}}(i,j)=
\begin{cases}
1, \quad \text{with prob } 1-Q_{ij} \\
0, \quad \text{with prob } Q_{ij}
\end{cases};
\end{gather*}
\vspace{-3mm}
\begin{gather}
\text {if } A_{\mathcal{G}^{*}}(i,j)=0, \text{ then }A_{\mathcal{G}_{k}}(i,j)=
\begin{cases}
1, \quad \text{with prob } P_{ij} \\
0, \quad \text{with prob } 1-P_{ij}
\end{cases}.
\end{gather}

\noindent From (1) it follows that the probability of the occurrence or non-occurrence of an edge between nodes $i,j$ in the $k^{th}$ network observation is,
\begin{displaymath}
P(A_{\mathcal{G}_{k}}(i,j)|A_{\mathcal{G}^{*}}(i,j)=1,P_{ij},Q_{ij})=(1-Q_{ij})^{A_{\mathcal{G}_{k}}(i,j)}\cdot Q_{ij}^{1-A_{\mathcal{G}_{k}}(i,j)},\text{ if }A_{\mathcal{G}^{*}}(i,j)=1;
\end{displaymath}
\begin{displaymath}
P(A_{\mathcal{G}_{k}}(i,j)|A_{\mathcal{G}^{*}}(i,j)=0,P_{ij},Q_{ij})=P_{ij}^{A_{\mathcal{G}_{k}}(i,j)}\cdot (1-P_{ij})^{1-A_{\mathcal{G}_{k}}(i,j)}, \text{ if }A_{\mathcal{G}^{*}}(i,j)=0.
\end{displaymath}

\cite{le2018estimating} treat the adjacency matrix of the representative network $A_{\mathcal{G}^{*}}$ as a latent variable, while the false positive and false negative probabilities $P_{ij}$ and $Q_{ij}$ are model parameters. Thus, the likelihood of the representative network $A_{\mathcal{G}^{*}}$ given the network data $A_{\mathcal{G}_{1}},\cdots,A_{\mathcal{G}_{N}}$, as seen in \cite{le2018estimating}, is
\begin{gather*}
\mathcal{L}(A_{\mathcal{G}^{*}};A_{\mathcal{G}_{1}},\ldots,A_{\mathcal{G}_{N}})=\prod_{k=1}^{N} \prod_{(i,j):i<j} [(1-Q_{ij})^{A_{\mathcal{G}_{k}}(i,j)}\cdot Q_{ij}^{1-A_{\mathcal{G}_{k}}(i,j)}\cdot W_{ij}]^{A_{\mathcal{G}^{*}}(i,j)}\cdot \\ [P_{ij}^{A_{\mathcal{G}_{k}}(i,j)}\cdot (1-P_{ij})^{1-A_{\mathcal{G}_{k}}(i,j)} \cdot (1-W_{ij})]^{1-A_{\mathcal{G}^{*}}(i,j)},
\end{gather*}
where $W_{ij}=\mathbb{E}A_{\mathcal{G}^{*}}(i,j)$ represents the probability of observing an edge between nodes $i,j$ in $A_{\mathcal{G}^{*}}$. 

\cite{le2018estimating} further assume that the nodes of the underlying true network form communities that can be described by a Stochastic Block Model (SBM). Under the SBM assumption, each node of the true network belongs to an unobserved block $k \in \{1,\cdots,K\}$, and the probability of observing an edge between two nodes depends on their block membership denoted by $\{b_{i}\}_{i=1}^{n}$, with $b_{i} \in \{1,\cdots,K\}$. The probability of observing an edge between nodes $(i,j) \text{ with } b_{i}=k, \text{ } b_{j}=l$ is represented by $\theta_{kl}$. \textcolor{black}{In addition, the corresponding block structure is assumed to be shared among the matrices $P$, $Q$ and $W$.}

The inference of the model parameters and the latent variable is conducted in two stages. First, a Spectral Clustering algorithm is applied to reveal the underlying block membership of the representative's nodes, and second, an EM algorithm is implemented to estimate the model parameters. While this formulation has appealing features it would be ideal to have a coherent modelling framework that can jointly infer block membership of the representative's nodes together with the parameters characterising the distribution of the network data. In addition, using an EM algorithm to estimate model parameters means that measures of uncertainty such as standard errors
rely on asymptotic approximations that may not be valid in many applications, particularly when involving small samples sizes.

In the next section we propose a mixture of measurement error models inspired by \cite{le2018estimating} for clustering heterogeneous network data. We adopt a Bayesian framework that allows us to jointly infer the parameters of the measurement error model as well as those characterising the underlying network representatives corresponding to each cluster. In addition, the Bayesian formulation is flexible enough to accommodate diverse modelling assumptions for the network representatives.

\section{A Mixture of Measurement Error models}\label{sec4}

In this section, we detail the formulation and implementation of the mixture of measurement error models. We first describe the Bayesian formulation of the measurement error model when there is only one cluster. We then extend this to multiple clusters. Following this we describe how posterior samples can be obtained using MCMC. Finally we describe a special case of this formulation that can correspond to detecting outlying networks in the data.

\subsection{Model formulation}\label{sec41}

To begin with, we assume underlying the network data there is a latent representative network with adjacency matrix denoted by $A_{\mathcal{G}^{*}}$. In addition, we assume that the probability of observing a false positive or false negative edge between two nodes in the network data is independent of the pair of nodes considered. Thus, the false positive probability, $p$, and false negative probability, $q$, can be viewed as scalars. \textcolor{black}{In this way, we limit model complexity in terms of the number of parameters to be inferred. 
The specification of component specific $n\times n$ matrices of false positive probabilities and false negative probabilities would lead to a drastic increase in the number of model parameters. A compromise
assumes matrices $P,Q$ share an SBM structure defined by the true network $A_{\mathcal{G}^{*}}$, as in \cite{le2018estimating}. However this assumption would only be relevant where an SBM structure is already known to be appropriate,
which might not be appropriate for some applications.
We discuss this further in Section \ref{sec6}.}

Under this specification, the probability mass function of the edge state between nodes $(i,j)$ in network observation $k$, given $A_{\mathcal{G}^{*}}(i,j)$, $p$, $q$, is
\begin{gather*}
P(A_{\mathcal{G}_{k}}(i,j)|A_{\mathcal{G}^{*}}(i,j),p,q)=[(1-q)^{A_{\mathcal{G}_{k}}(i,j)}\cdot q^{1-A_{\mathcal{G}_{k}}(i,j)}]^{A_{\mathcal{G}^{*}}(i,j)} \cdot \\ [p^{A_{\mathcal{G}_{k}}(i,j)}\cdot (1-p)^{1-A_{\mathcal{G}_{k}}(i,j)}]^{1-A_{\mathcal{G}^{*}}(i,j)}.
\end{gather*}
\noindent Hence, the conditional probability of observing a network population $A_{\mathcal{G}_{1}},\cdots,A_{\mathcal{G}_{N}}$ given $A_{\mathcal{G}^{*}}$,  $p$, $q$ is,
\begin{gather*}
P(A_{\mathcal{G}_{1}},\ldots,A_{\mathcal{G}_{N}}|p,q,A_{\mathcal{G}^{*}})=\prod_{k=1}^{N} \prod_{(i,j):i<j}	P(A_{\mathcal{G}_{k}}(i,j)|A_{\mathcal{G}^{*}}(i,j),p,q) = \\ \prod_{k=1}^{N} \prod_{(i,j):i<j}((1-q)^{A_{\mathcal{G}_{k}}(i,j)}\cdot q^{1-A_{\mathcal{G}_{k}}(i,j)})^{A_{\mathcal{G}^{*}}(i,j)} \cdot (p^{A_{\mathcal{G}_{k}}(i,j)}\cdot (1-p)^{1-A_{\mathcal{G}_{k}}(i,j)})^{1-A_{\mathcal{G}^{*}}(i,j)}.
\end{gather*}

An advantage of the measurement error formulation is that the model specification for the representative network $A_{\mathcal{G}^{*}}$ can vary depending on the type of information the analyst wants to capture for the data at hand. As previously discussed, \cite{le2018estimating} assume a SBM for the network representative. For illustration we also assume an SBM structure for the representative $A_{\mathcal{G}^{*}}$, but note that this can be easily modified, e.g. reduced to a simpler model such as the Erd\"{o}s-R\'{e}nyi if supported by the data. The SBM for the representative can be represented hierarchically in the following way,
\begin{gather*}
A_{\mathcal{G}^{*}}(i,j)|\boldsymbol{\theta},\boldsymbol{b} \sim \text{Bernoulli}(\theta_{b_{i}b_{j}});\\
\theta_{kl} \sim \text{Beta}(\epsilon_{kl},\zeta_{kl}); \\
\boldsymbol{b|w} \sim \text{Multinomial}(\boldsymbol{w});
\end{gather*}
where $w_{k}$ represents the probability of a node to belong to block $k\in \{1,\cdots,K\}$. For the probability vector $\boldsymbol{w}=\{w_{1},\cdots,w_{K} \}$ we assume a symmetric Dirichlet prior distribution with hyperparameter $\boldsymbol{\chi}$. Common choices for the hyperparameter vector $\boldsymbol{\chi}$ are setting all elements to 0.5 or 1.

Thus the hierarchical structure of the model is, 

\begin{gather*}
\prod_{k=1}^{N} \prod_{(i,j):i<j}P(A_{\mathcal{G}_{k}}(i,j)|A_{\mathcal{G}^{*}}(i,j),p,q)P(A_{\mathcal{G}^{*}(i,j)}|\boldsymbol{\theta} ,\boldsymbol{b})
\end{gather*}
where
\begin{displaymath}
P(A_{\mathcal{G}^{*}}(i,j)|\boldsymbol{\theta} ,\boldsymbol{b})=\theta_{b_{i}b_{j}}^{A_{\mathcal{G}^{*}}(i,j)} (1-\theta_{b_{i}b_{j}})^{1-A_{\mathcal{G}^{*}}(i,j)}.
\end{displaymath}

\noindent We further specify a Beta prior distribution for both the false positive $p$ and false negative $q$ probabilities,
\begin{displaymath}
p \sim \text{Beta}(\alpha_0,\beta_0) \text{, } q \sim \text{Beta}(\gamma_0,\delta_0),
\end{displaymath}
which facilitates posterior computations. A common choice sets the Beta prior hyperparameters to 0.5, corresponding to the Jeffreys prior.

\subsection{Mixture of measurement error models}\label{sec42}
We further extend the measurement error model to a mixture of measurement error models, with a predefined number of mixture components, $C$, in order to 
provide a model-based approach for identifying clusters of networks in a network population $A_{\mathcal{G}_{1}},\cdots,A_{\mathcal{G}_{N}}$. 
We assume each cluster $c$ of networks is described by a unique network representative $A_{\mathcal{G}^{*}_{c}}$, a false positive probability $p_{c}$, and a false negative probability $q_{c}$, where $c\in \{1,\ldots,C\}$. In this section, we present the Bayesian framework for this mixture of measurement error models. Each cluster-specific representative network is characterised by an SBM, and the block structure of each representative is allowed to vary.  

Let $\boldsymbol{z}=(z_{1},\cdots,z_{N})\in\{1,\cdots,C\}$ be the latent variables representing the cluster membership of the network data $A_{\mathcal{G}_{1}},\cdots,A_{\mathcal{G}_{N}}$. Then the conditional probability of the data given $\boldsymbol{z}$ takes the form
\begin{gather*}
P(A_{\mathcal{G}_{1}},\ldots, A_{\mathcal{G}_{N}}|\{ p_{c},q_{c},A_{\mathcal{G}^{*}_{c}} \}_{c=1}^{C},z_{1},\cdots,z_{N})= \\   \prod_{k=1}^{N} \Big ( \prod_{(i,j):i<j} \Big ( (1-q_{z_{k}})^{A_{\mathcal{G}_{k}}(i,j)}q_{z_{k}}^{1-A_{\mathcal{G}_{k}}(i,j)} \Big)^{A_{\mathcal{G}^{*}_{z_{k}}}(i,j)} \cdot \Big  (p_{z_{k}}^{A_{\mathcal{G}_{k}}(i,j)}(1-p_{z_{k}})^{1-A_{\mathcal{G}_{k}}(i,j)} \Big)^{1-A_{\mathcal{G}^{*}_{z_{k}}}(i,j)} \Big ).
\end{gather*}

We assume that the cluster labels $z_{1},\cdots,z_{N}$ follow a Multinomial distribution with parameter $\boldsymbol{\tau}=(\tau_{1},\cdots,\tau_{C})$, where $\tau_{c}$ represents the probability that a network observation belongs to cluster $c$, and $\sum_{c=1}^{C}\tau_{c}=1$. We assume a symmetric Dirichlet prior distribution for the vector of probabilities $\boldsymbol{\tau}$ which has the advantage of being conditionally conjugate with the distribution for $\boldsymbol{z}$. As commented previously
common choices set the Dirichlet hyperparameters all to 0.5 or to 1.

\subsection{MCMC scheme for mixture model}\label{sec43}

With the modelling framework described above we are able to draw samples from the joint posterior distribution of the parameters using MCMC. The joint posterior distribution is known up to a normalising constant, specifically
\begin{align}
\begin{gathered}
P(\boldsymbol{A_{\mathcal{G}^{*}}},\boldsymbol{p},\boldsymbol{q},\boldsymbol{z},\boldsymbol{\tau},\boldsymbol{w},\boldsymbol{b},\boldsymbol{\theta}|A_{\mathcal{G}_{1}},\ldots,A_{\mathcal{G}_{N}}) \\ \propto
P(A_{\mathcal{G}_{1}},\ldots, A_{\mathcal{G}_{N}}|\boldsymbol{A_{\mathcal{G}^{*}}},\boldsymbol{p},\boldsymbol{q},\boldsymbol{z})\cdot P(\boldsymbol{A_{\mathcal{G}^{*}}}|\boldsymbol{w},\boldsymbol{b},\boldsymbol{\theta})\cdot P(\boldsymbol{p}|\boldsymbol{\alpha_{0}},\boldsymbol{\beta_{0}})\\ \cdot P(\boldsymbol{q}|\boldsymbol{\gamma_{0}},\boldsymbol{\delta_{0}})\cdot P(\boldsymbol{z}|\boldsymbol{\tau}) \cdot P(\boldsymbol{\tau}|\boldsymbol{\psi}) \cdot P(\boldsymbol{\theta}|\boldsymbol{\epsilon},\boldsymbol{\zeta}) \cdot P(\boldsymbol{b}|\boldsymbol{w}) \cdot P(\boldsymbol{w}|\boldsymbol{\chi}),
\end{gathered}
\end{align}

\noindent where $P(A_{\mathcal{G}_{1}},\ldots, A_{\mathcal{G}_{N}}|\boldsymbol{A_{\mathcal{G}^{*}}},\boldsymbol{p},\boldsymbol{q},\boldsymbol{z})$ is the conditional probability of the network data, $P(\boldsymbol{A_{\mathcal{G}^{*}}}|\boldsymbol{w},\boldsymbol{b},\boldsymbol{\theta})$ is the conditional probability of the latent variable $\boldsymbol{A_{\mathcal{G}^{*}}}$ and the rest of the components of the right hand side of the expression are the prior distributions for the model parameters, as defined in Sections \ref{sec41} and \ref{sec42}.

To obtain posterior samples from the joint posterior in (2) we note that many of the full conditional distributions of the unknown quantities (parameters/latent data) are available in closed form, and when these are not available these can be approximated using Metropolis-Hastings. As a result we obtain posterior inferences through a component wise MCMC sampler, also known as a Metropolis-Hastings-within-Gibbs sampler. This closely follows a Gibbs sampler, where all parameters and latent data are updated from their full conditional distributions except for the full conditionals of $\{ A_{\mathcal{G}^{*}_{c}},p_{c},q_{c}\}_{c=1}^{C}$, which are approximated using Metropolis-Hastings proposal distributions. 

In the Metropolis-Hastings step, we use a mixture of kernels for updating the parameters of the measurement error model $\{ A_{\mathcal{G}^{*}_{c}},p_{c},q_{c}\}_{c=1}^{C}$, in analogy to the MCMC scheme seen in \cite{lunagomez}. Specifically in every iteration of the MCMC we update the adjacency matrix of the network representative of cluster $c$, $A_{\mathcal{G}^{*}_{c}}$, using either of the following two proposals with some fixed probability:
\begin{itemize}
\item[(I)] We perturb the edges of the current network representative $A_{\mathcal{G}^{*}_{c}}^{(curr)}$ of cluster $c$ in the following way:
\begin{displaymath}
A_{\mathcal{G}^{*}_{c}}^{(prop)}(i,j)=
\begin{cases}
1-A_{\mathcal{G}^{*}_{c}}^{(curr)}(i,j), \text{ with probability } \omega \\
A_{\mathcal{G}^{*}_{c}}^{(curr)}(i,j), \text{ with probability } 1-\omega
\end{cases}.
\end{displaymath}
\item[(II)] We propose a new network representative $A_{\mathcal{G}^{*}_{c}}^{(prop)}$ for cluster $c$ drawing each edge of the proposed representative $A_{\mathcal{G}^{*}_{c}}^{(prop)}(i,j)$ independently from a Bernoulli distribution with parameter $\frac{1}{N}\sum_{k=1}^{N}A_{\mathcal{G}_{k}}(i,j)$. 
\end{itemize}

Thus we accept the proposed network representative $A_{\mathcal{G}^{*}_{c}}^{(prop)}$ with probability

\begin{align}
\min \Bigg\{ 1, \frac{P(A_{\mathcal{G}_{1}},\cdots,A_{\mathcal{G}_{N}}|A_{\mathcal{G}^{*}_{c}}^{(prop)},p_{c}^{(curr)},q_{c}^{(curr)},\boldsymbol{z}^{(curr)}) P(A_{\mathcal{G}^{*}_{c}}^{(prop)}|\boldsymbol{b_{c}}^{(curr)},\boldsymbol{\theta_{c}}^{(curr)})}{P(A_{\mathcal{G}_{1}},\cdots,A_{\mathcal{G}_{N}}|A_{\mathcal{G}^{*}_{c}}^{(curr)},p_{c}^{(curr)},q_{c}^{(curr)},\boldsymbol{z}^{(curr)}) P(A_{\mathcal{G}^{*}_{c}}^{(curr)}|\boldsymbol{b_{c}}^{(curr)},\boldsymbol{\theta_{c}}^{(curr)})} \nonumber \\
\cdot \frac{Q(A_{\mathcal{G}^{*}_{c}}^{(curr)}|A_{\mathcal{G}^{*}_{c}}^{(prop)})}{Q(A_{\mathcal{G}^{*}_{c}}^{(prop)}|A_{\mathcal{G}^{*}_{c}}^{(curr)})}
\Bigg\},
\end{align}

\noindent where $P(A_{\mathcal{G}^{*}_{c}}^{(\cdot)}|\boldsymbol{b_{c}},\boldsymbol{\theta_{c}})$ is the SBM assumed for the representative defined in Section \ref{sec41} and $Q(A_{\mathcal{G}^{*}_{c}}^{(\cdot)}|A_{\mathcal{G}^{*}_{c}}^{(\cdot)})$ corresponds to the proposal distribution. The proposal distribution under case (I) proposal is symmetric, and so it cancels out from the Metropolis ratio in expression (3). 

To update the false positive probability $p_{c}$ of cluster $c$, we use a mixture of random walk proposals indexed by $l$ following \cite{lunagomez}. 
\begin{itemize}
\item Draw $v \sim \text{Unif}(-u_{l},u_{l})$, for $0 < u_l < 0.5$.
\item Calculate the candidate proposal value $y=p_{c}^{(curr)}+v$.
\item Propose a new value for $p_{c}$ (constrained to lie in the interval (0,0.5) for identifiability reasons) as follows,
\begin{displaymath}
p_{c}^{(prop)}=
\begin{cases}
y, \quad \text{if } 0<y<0.5; \\
-y, \quad \text{if } y<0; \\
1-y, \quad \text{if } y>0.5. \\
\end{cases}
\end{displaymath}
\end{itemize}
The mixture is over $\{u_1,\ldots,u_L\}$. Thus, we perturb the current state of the false positive probability $p_{c}^{(curr)}$ using various sizes of $u_{l}$, each imposing a less or more drastic change on $p_{c}^{(curr)}$. We accept the proposed value $p_{c}^{(prop)}$ with probability
\begin{equation}
\min \Bigg\{ 1,\frac{P(A_{\mathcal{G}_{1}},\cdots,A_{\mathcal{G}_{N}}|A_{\mathcal{G}^{*}_{c}}^{(curr)},p_{c}^{(prop)},q_{c}^{(curr)},\boldsymbol{z}^{(curr)})P(p_{c}^{(prop)}|\alpha_{0,c},\beta_{0,c})}{P(A_{\mathcal{G}_{1}},\cdots,A_{\mathcal{G}_{N}}|A_{\mathcal{G}^{*}_{c}}^{(curr)},p_{c}^{(curr)},q_{c}^{(curr)},\boldsymbol{z}^{(curr)}) P(p_{c}^{(curr)}|\alpha_{0,c},\beta_{0,c})} \Bigg\},
\end{equation}

\noindent where $P(p_{c}^{(\cdot)}|\alpha_{0,c},\beta_{0,c})$ is a Beta($\alpha_{0,c},\beta_{0,c}$) prior as in Section \ref{sec41}. The proposal distribution for $p_{c}$ is symmetric, thus it does not appear in the Metropolis ratio in expression (4). In exactly the same manner, we update the false negative probability $q_{c}$, for $c \in \{1,\ldots,C\}$.

The rest of the parameters are updated via Gibbs samplers, by drawing values from their full conditional posteriors.
The full conditional posterior for $\boldsymbol{\tau}$ is given by
\begin{align}
P(\boldsymbol{\tau}|\boldsymbol{A_{\mathcal{G}^{*}}},\boldsymbol{p},\boldsymbol{q},\boldsymbol{z},\boldsymbol{w},\boldsymbol{b},\boldsymbol{\theta},A_{\mathcal{G}_{1}},\ldots,A_{\mathcal{G}_{N}})= \text{Dirichlet}(\psi + \eta_{1}, \ldots ,\psi + \eta_{C}).
\end{align}
where $\eta_{c}=\sum_{j=1}^{N}1_{c}(z_{j})$, $c=1,\ldots,C$, denotes the number of networks that belong to cluster $c$.

We draw the latent cluster-membership $z_{k}$ for each network observation $k$ from a Multinomial distribution with unnormalised probabilities specified in the following way:

\begin{align}
P(z_{k}=c|\boldsymbol{\tau},\boldsymbol{A_{\mathcal{G}^{*}}},\boldsymbol{p},\boldsymbol{q},A_{\mathcal{G}_{k}}) \propto P(A_{\mathcal{G}_{k}}|z_{k}=c, \boldsymbol{A_{\mathcal{G}^{*}}},\boldsymbol{p},\boldsymbol{q}) \cdot P(z_{k}=c|\boldsymbol{\tau}) \nonumber \\
= \tau_{c} \cdot \prod_{(i,j):i<j} \Big ( (1-q_{c})^{A_{\mathcal{G}_{k}}(i,j)}q_{c}^{1-A_{\mathcal{G}_{k}}(i,j)} \Big)^{A_{\mathcal{G}^{*}_{c}}(i,j)} \cdot \Big  (p_{c}^{A_{\mathcal{G}_{k}}(i,j)}(1-p_{c})^{1-A_{\mathcal{G}_{k}}(i,j)} \Big)^{1-A_{\mathcal{G}^{*}_{c}}(i,j)} 
\end{align}

\noindent where $P(A_{\mathcal{G}_{k}}|z_{k}=c, \boldsymbol{A_{\mathcal{G}^{*}}},\boldsymbol{p},\boldsymbol{q})$ is the probability we observe network $k$ given cluster membership $z_{k}=c$, described by a measurement error model. The normalised probabilities are obtained via Bayes Theorem.

The full conditional posterior for $\boldsymbol{w_{c}}$ is
\begin{displaymath}
P(\boldsymbol{w_{c}}|A_{\mathcal{G}^{*}_{c}},p_{c},q_{c},\boldsymbol{z},\boldsymbol{\tau},\boldsymbol{b_{c}},\boldsymbol{\theta_{c}},A_{\mathcal{G}_{1}},\ldots,A_{\mathcal{G}_{N}})=\text{Dirichlet}(\chi + h_{1}, \ldots ,\chi + h_{K}).
\end{displaymath}
\noindent where $h_{k}$ denotes the number of the nodes that belong to block k. 

The full conditional posterior for the vector of the block-specific probabilities of an edge occurrence for the network representative of cluster $c$, $\boldsymbol{\theta_{c}}$, is
\begin{align}
P(\boldsymbol{\theta_{c}}|A_{\mathcal{G}^{*}_{c}},p_{c},q_{c},\boldsymbol{z},\boldsymbol{\tau},\boldsymbol{b_{c}},\boldsymbol{w_{c}},A_{\mathcal{G}_{1}},\ldots,A_{\mathcal{G}_{N}})=\text{Beta}(A_{\mathcal{G}^{*}_{c}}[kl]+\epsilon_{kl},\zeta_{kl}+n_{c,kl}-A_{\mathcal{G}^{*}_{c}}[kl]).
\end{align}
where $A_{\mathcal{G}^{*}_{c}}[kl]=\sum_{(i,j):b_{c,i}=k,b_{c,j}=l} A_{\mathcal{G}^{*}_{c}}(i,j)$ represents the sum of the entries for the pairs of nodes of the network representative for cluster $c$ that have block membership $k,l$ respectively, and $n_{c,kl}=\sum_{(i,j):i \neq j} \mathbb{I}(b_{c,i}=k,b_{c,j}=l)$ represents the number of the pair of nodes of the representative for cluster $c$ that have membership $k,l$ respectively. 

Similarly to the formulation obtained for updating the latent cluster-membership $\boldsymbol{z}$ of the network data, we obtain updates of the latent block-membership $\boldsymbol{b_{c}}$ for the nodes of the network representative of cluster $c$ from a Multinomial distribution with unnormalised probabilities specified as follows:

\begin{align}
P(b_{c,i}=k|A_{\mathcal{G}^{*}_{c}},p_{c},q_{c},\boldsymbol{z},\boldsymbol{\tau},\boldsymbol{\theta_{c}},\boldsymbol{w_{c}},A_{\mathcal{G}_{1}},\ldots,A_{\mathcal{G}_{N}}) \propto  P(A_{\mathcal{G}^{*}_{c}}|\boldsymbol{w_{c}},\boldsymbol{\theta_{c}},b_{c,i}=k) \cdot  P(b_{c,i}=k|\boldsymbol{w_{c}}) \nonumber \\
=w_{c,k} \cdot \prod_{j=1}^{n}\theta_{kb_{c,j}}^{A_{\mathcal{G}^{*}_{c}}(i,j)}(1-\theta_{kb_{c,j}})^{1-A_{\mathcal{G}^{*}_{c}}(i,j)}.
\end{align}
where $P(A_{\mathcal{G}^{*}_{c}}|\boldsymbol{w_{c}},\boldsymbol{\theta_{c}},b_{c,i}=k)$ is the probability of observing the representative of cluster $c$, $A_{\mathcal{G}^{*}_{c}}$, described by an SBM, given its $i^{th}$ node belongs to block $k$. Normalised probabilities are obtained by Bayes Theorem. 

For the detailed derivation of the full conditional posterior distributions refer to the Supplementary material Section 1 \citep{mantziousupp}. In addition, the MCMC algorithm for clustering is sketched in the Supplementary material Algorithm 1 \citep{mantziousupp}.

\subsection{Outlier network detection}\label{sec44}
\textcolor{black}{Motivated by the application on brain networks, in this section we present a modification of the mixture model presented in Section \ref{sec42} to further explore the heterogeneity in a network population. Specifically, we modify our mixture model to detect a cluster of outlier networks that are different to the majority of the networks in the population. Under this formulation, we are able to address additional applied research questions of interest.
In particular, we would like to be able to identify individuals with different brain connectivity patterns compared to the rest of the population.}

In contrast to the mixture model formulated for multiple cluster representatives, the outlier cluster detection model assumes a single network representative for the whole population of networks. Under this setup, we assume that there are ultimately two clusters of networks formed within the population of networks, one cluster being the majority cluster, while the other cluster determining the outlier networks in the population. Thus, while the false positive and false negative probabilities remain component specific for each of the two clusters, the network representative is no longer a component specific latent variable. 

Similar to the mixture model formulated in Section \ref{sec42}, we now specify the number of clusters to $C=2$, and $\boldsymbol{z}=(z_{1},\cdots,z_{N})\in \{1,2\}$ denotes the latent cluster membership of the network data $A_{\mathcal{G}_{1}},\ldots,A_{\mathcal{G}_{N}}$. Under the assumption of a single network representative, $A_{\mathcal{G}^{*}}$, the conditional probability of the data given the latent variables, $\boldsymbol{z}$, takes the form,
\begin{gather*}
P(A_{\mathcal{G}_{1}},\cdots, A_{\mathcal{G}_{N}}|\{ p_{c},q_{c} \}_{c=1}^{C},A_{\mathcal{G}^{*}},z_{1},\cdots,z_{N})= \nonumber\\ \prod_{k=1}^{N} \Big ( \prod_{(i,j):i<j} \Big ( (1-q_{z_{k}})^{A_{\mathcal{G}_{k}}(i,j)}q_{z_{k}}^{1-A_{\mathcal{G}_{k}}(i,j)} \Big)^{A_{\mathcal{G}^{*}}(i,j)} \cdot \Big  (p_{z_{k}}^{A_{\mathcal{G}_{k}}(i,j)}(1-p_{z_{k}})^{1-A_{\mathcal{G}_{k}}(i,j)} \Big)^{1-A_{\mathcal{G}^{*}}(i,j)} \Big ).
\end{gather*}
 
Again, the model specification for the representative network can vary depending on the type of information we want to capture for the data at hand. A common choice is to consider an SBM structure again for the representative. Due to having now a single representative only, the SBM model parameters are no more component specific to the cluster.

To sample from the joint posterior of this model, we develop a Metropolis-Hastings-within-Gibbs MCMC scheme, as presented in Section \ref{sec43}. The full conditional posterior distributions are obtained as seen in Section \ref{sec43} with the only difference that the parameters/latent variables characterising the representative, $A_{\mathcal{G}^{*}}$, namely, the block-specific edge probabilities, $\boldsymbol{\theta}$, the probability of a node to belong to a block, $\boldsymbol{w}$, and the block membership of the nodes, $\boldsymbol{b}$, are no longer component specific, i.e. not indexed by cluster $c$.

\subsection{\textcolor{black}{Sparse Finite Mixture extension}\label{sec45}}
\textcolor{black}{
In practice, the number of clusters is not known a priori, and so we need to be able to determine an appropriate number of clusters to specify in our model. This is a problem that has been extensively discussed in the literature, and a variety of approaches have been proposed.
For finite mixture models, a common approach for estimating $C$ is through information criteria such as BIC \citep{keribin2000consistent}. Alternative approaches can incorporate uncertainty in the number of clusters such as reversible jump MCMC methods \citep{richardson1997bayesian} or Dirichlet Process (DP) mixture models \citep{neal2000markov}, with the former being particularly challenging in network models, due to the MCMC moving between very different dimensions.}

\textcolor{black}{We adopt the approach in \cite{malsiner2016model}, who develop a method that conveniently extends the finite mixture model to make inferences with an unknown number of clusters, known as the Sparse Finite Mixture (SFM) model. A sparse symmetric Dirichlet prior distribution is specified for the weights of the mixture components, and the number of mixture components are assumed to be more than the number of clusters in the data. \cite{fruhwirth2019here} discuss how the SFM model can be easily extended to examples with non-Gaussian data, and make comparisons to DP mixtures with respect to their clustering performance. The convenient implementation of the SFM model for finite mixture models, together with its wide applicability for different types of data, led us to consider this as an extension to our finite mixture model that allows an unknown number of clusters $C$.}

\textcolor{black}{Extending our finite mixture model to the SFM model requires the specification of a symmetric Dirichlet prior distribution Dir$(e_0,\ldots,e_0)$ of order $C_{max}$, with $C_{max}$ being an upper bound on the number of clusters such that $C<C_{max}$ where $C$ is the number of clusters in the data, resulting in an overfitted mixture model. The size of $e_0$ should result in many of the weights $\boldsymbol{\tau}=(\tau_1,\ldots,\tau_{C_{max}})$ being close to 0, imposing sparsity on the number of clusters.  We specify a Gamma$(a_e,b_e)$ hyperprior on the hyperparameter, $e_0$, of the Dirichlet prior, and sample from the posterior of $e_0$ using a Metropolis-Hastings step as proposed in \cite{fruhwirth2019here}. Specifically, in our model we specify a Random Walk proposal for the MH step for $e_0$. The specification of $a_e,b_e$ plays a key role in the clustering performance of the model and should result in values of $e_0$ close to 0.}

\textcolor{black}{In Section \ref{sec53}, we explore the performance of the SFM extension of our model on simulated network populations, as well as implement it in the real data applications in Section \ref{sec6}.}

\section{Simulations}\label{sec5}
In this Section, we perform simulation studies to assess the performance of our algorithm in inferring the model parameters/latent variables and clustering network data. First, we explore the performance of our algorithm for moderate network sizes and various noise levels and SBM models, and second, we investigate the algorithm performance for various network and sample sizes.

\subsection{Moderate network sizes}\label{sec51}
In this simulation study we investigate the performance of our model in inferring model parameters for network populations with a moderate number of nodes in different scenarios. Specifically, we consider the case of networks with $n=21$ nodes, and a population of $N=180$ networks. We assume $C=3$ clusters of networks in the population, and consider 
SBMs for each representative network with $B=2$ blocks. We vary the model parameter values in order to explore performance.

\begin{figure}[htb!]
\centering
\includegraphics[height=1.6in,width=5.5in]{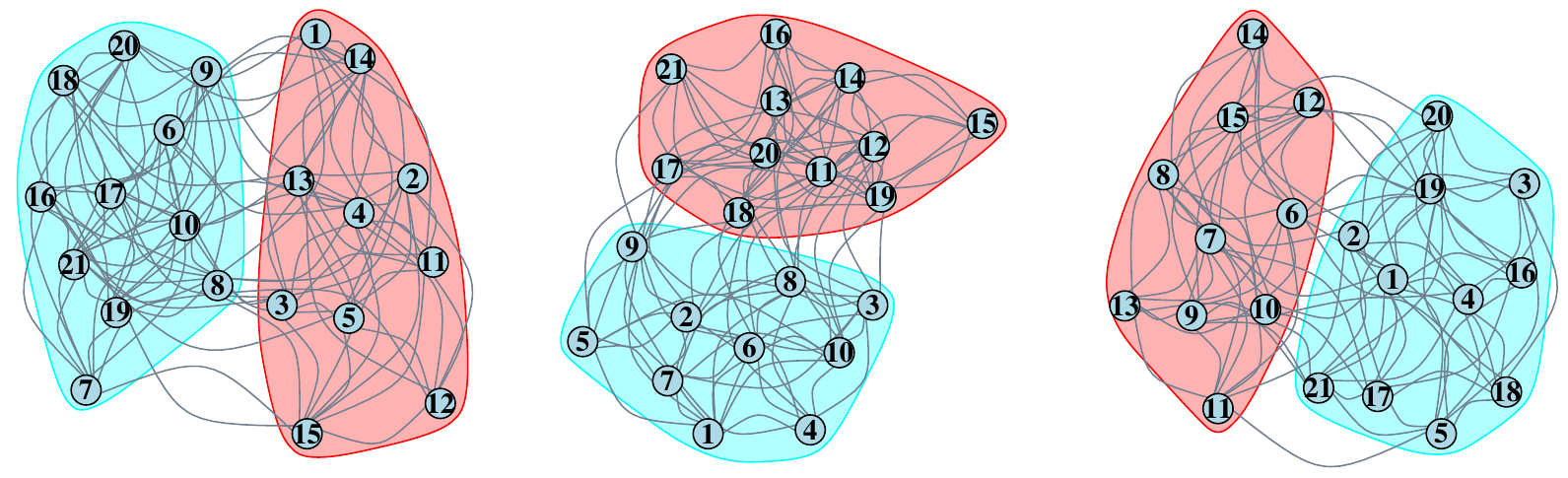}
\hfill
\includegraphics[height=1.6in,width=5.5in]{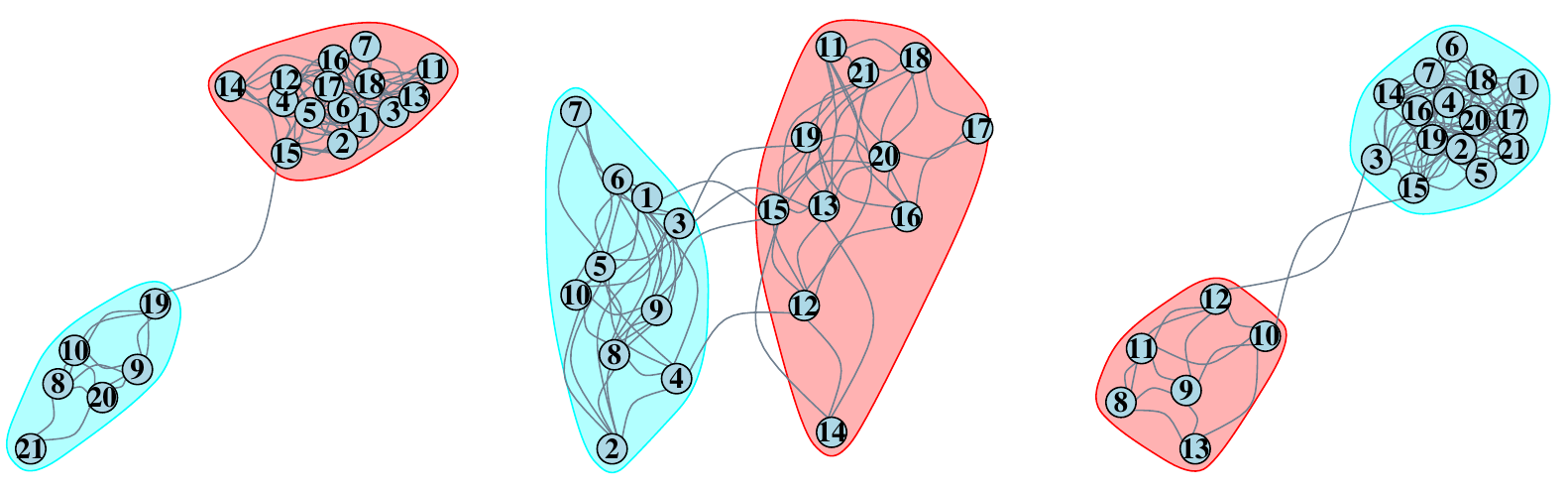}
\caption{Top: Network representatives for clusters $c=1,$ $2$ and $3$ respectively (from left to right), under SBM structure 1. Bottom: Network representatives for clusters $c=1,$ $2$ and $3$ respectively (from left to right), under SBM structure 2.}\label{sbm_1_sbm_2_rep}
\end{figure}

To simulate the network population, we first simulate the representative network of each cluster. We generate representatives of the clusters under two different SBM structures with \textcolor{black}{parameters as shown in Table \ref{sim_regime} (left).} In Figure \ref{sbm_1_sbm_2_rep} we visualise the 21-node representatives for each of the $C=3$ clusters under each of the SBM structures 1 and 2 assumed.

\begin{table}[ht!]
\centering
\begin{tabular}{clccccc}
\multicolumn{1}{r}{} & \multicolumn{1}{r}{} & \multicolumn{3}{c}{$\boldsymbol{\theta}$} & \multicolumn{2}{c}{$\boldsymbol{w}$} \\ \hline \hline
\multicolumn{1}{l}{\textbf{SBM}} & \textbf{c} & \multicolumn{1}{l}{$\boldsymbol{\theta_{11}}$} & \multicolumn{1}{l}{$\boldsymbol{\theta_{12}}$} & \multicolumn{1}{l}{$\boldsymbol{\theta_{22}}$} & \multicolumn{1}{l}{$\boldsymbol{w_1}$} & \multicolumn{1}{l}{$\boldsymbol{w_2}$} \\ \hline \hline
\multirow{3}{*}{1} & 1 & 0.8 & 0.2 & 0.8 & 0.5 & 0.5 \\
 & 2 & 0.8 & 0.2 & 0.8 & 0.5 & 0.5 \\
 & 3 & 0.8 & 0.2 & 0.8 & 0.5 & 0.5 \\ \hline
\multirow{3}{*}{2} & 1 & 0.7 & 0.05 & 0.8 & 0.7 & 0.3 \\
 & 2 & 0.7 & 0.05 & 0.8 & 0.5 & 0.5 \\
 & 3 & 0.7 & 0.05 & 0.8 & 0.3 & 0.7 \\ \hline
\end{tabular}
\hspace{2cm}
\begin{tabular}{cc}
\multicolumn{2}{c}{$\boldsymbol{\text{SBM}_i}$} \\ \hline \hline
\rowcolor[HTML]{FFFFFF} 
$\boldsymbol{p_c}$& $\boldsymbol{q_c}$ \\ \hline \hline
\rowcolor[HTML]{FFFFFF} 
\cellcolor[HTML]{FFFFFF} & 0.2 \\
\rowcolor[HTML]{FFFFFF} 
\multirow{-2}{*}{\cellcolor[HTML]{FFFFFF}0.1} & 0.3 \\ \hline
 & 0.1 \\
\multirow{-2}{*}{0.2} & 0.3 \\ \hline
 & 0.1 \\
\multirow{-2}{*}{0.3} & 0.2 \\ \hline
\end{tabular}
\caption{Simulation regimes for 21-node networks and $C=3$ clusters. Left Table: SBM structures 1 and 2 for simulating a network representative for each cluster $c$. Right Table: sizes of false positive $p_c$ and false negative $q_c$ probabilities used to simulate network data under each SBM structure.}\label{sim_regime}
\end{table}

Next, we generate a population of 180 networks by perturbing the edges of each representative through a measurement error process. Specifically, we generate edges for 60 networks in each of the $C=3$ clusters, depending on the existence or non-existence of an edge in the representative network of the corresponding cluster $c$, given a false positive $p_{c}$ and false negative $q_{c}$ probability. The simulation regimes considered for $p_{c},q_{c}$ are presented in Table \ref{sim_regime} (right).

\begin{figure}[ht!]
\centering
\begin{minipage}[b]{0.48\linewidth}
\centering
\includegraphics[height=1.8in,width=2in]{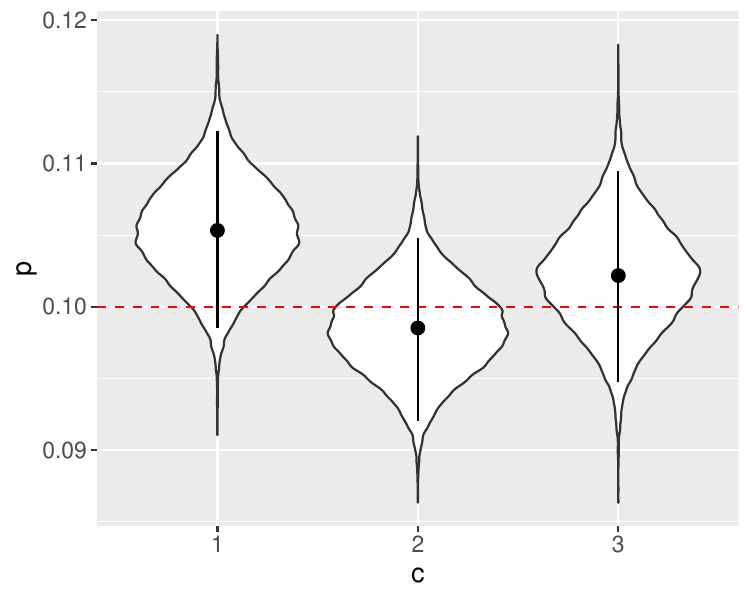}
\label{fix_q_plot}
\end{minipage}
\quad
\begin{minipage}[b]{0.48\linewidth}
\centering
\includegraphics[height=1.8in,width=2in]{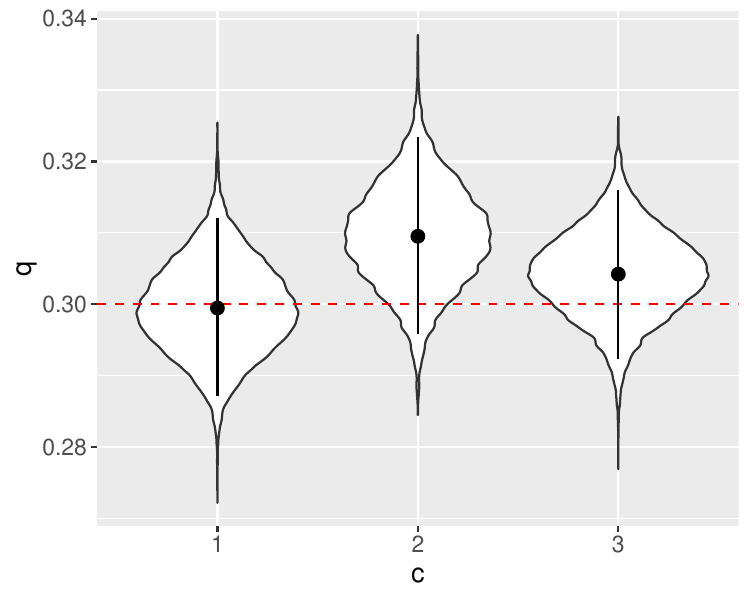}
\label{fix_p_plot}
\end{minipage}
\caption{Posterior distribution of false positive probabilities $p_{c}$ (left) and false negative $q_{c}$ (right) for $c\in\{1,2,3\}$, for simulation regime with $p_c=0.1$ and $q_c=0.3$. Red dotted lines indicate the true value of $p_c$ and $q_c$.}\label{viol_reg_4}
\end{figure}

\begin{figure}[ht!]
\centering
\begin{minipage}[b]{0.48\linewidth}
\centering
\includegraphics[height=1.8in,width=2in]{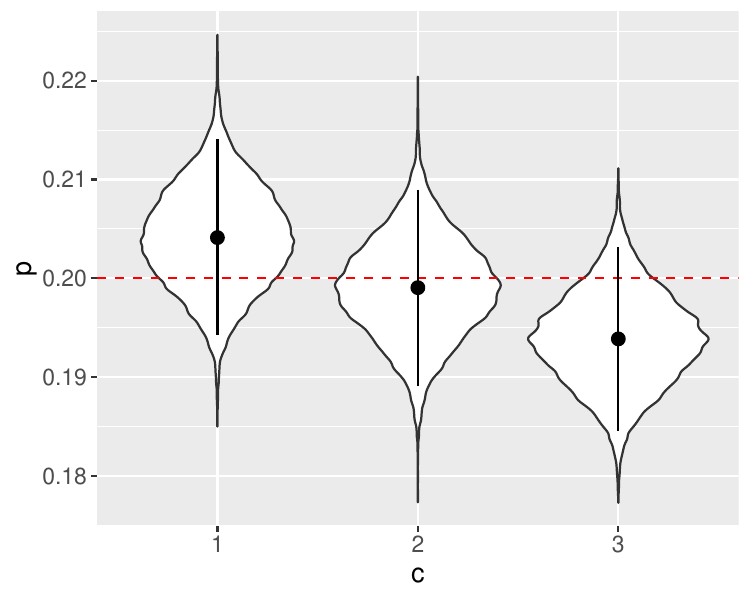}
\label{fix_q_plot}
\end{minipage}
\quad
\begin{minipage}[b]{0.48\linewidth}
\centering
\includegraphics[height=1.8in,width=2in]{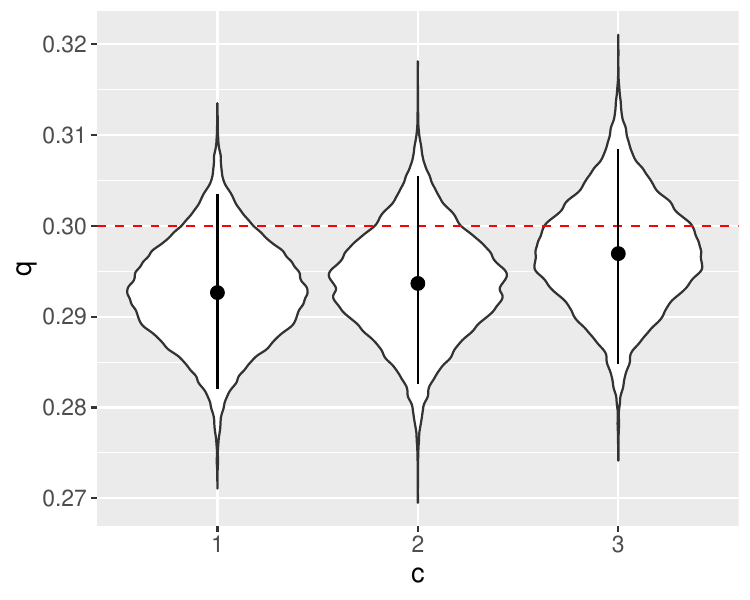}
\label{fix_p_plot}
\end{minipage}
\caption{Posterior distribution of false positive probabilities $p_{c}$ (left) and false negative $q_{c}$ (right) for $c\in\{1,2,3\}$, for simulation regime with $p_c=0.2$ and $q_c=0.3$. Red dotted lines indicate the true value of $p_c$ and $q_c$.}\label{viol_reg_7}
\end{figure}

For each simulation regime, we run our MCMC for 500,000 iterations with a burn-in of 150,000. \textcolor{black}{In Figure \ref{viol_reg_4}, we visualise the posterior distribution of $p_c$ and $q_c$ under the simulation regime with $p_c=0.1$ and $q_c=0.3$ for all $c$, and SBM structure 2 for the representative networks. In addition, in Figure \ref{viol_reg_7}, we visualise the posterior distribution of $p_c$ and $q_c$ under the simulation regime with $p_c=0.2$ and $q_c=0.3$ for all $c$, and SBM structure 1 for the representative networks. The bar in the violin plot indicates the 95\% credible interval and the point indicates the posterior mean. In both Figures, we observe that the posterior means are very close to the true values of the parameters, while the 95\% credible intervals enclose the true values of the parameter for all cases. This finding also holds for the rest of the simulation regimes. In Supplementary Material, Section 2.1 \citep{mantziousupp}, we summarise the results obtained for the rest of the simulation regimes in Tables 1-15.}

In order to investigate the performance of our algorithm in identifying the true representatives, we obtain the Hamming distance between the posterior representative samples and the true representatives, after a burn-in of 150,000 and a lag of 50, leaving 7,000 posterior samples. \textcolor{black}{The Hamming distance measures how dissimilar two graphs are with respect to their edges \citep{donnat2018tracking}. The maximum Hamming distance between two undirected, 21-node networks is equal to ${21 \choose 2} = 210$, meaning the two graphs have no edges in common}. We calculate the proportion of times that the distance is less or equal to 1, 5 and 10 respectively, similar to the summaries obtained in \cite{lunagomez}. For each simulation regime, we observe that all the posterior representative samples drawn for each cluster have a Hamming distance from the true representative less than or equal to 1, 5, and 10, 100\% of the time, as presented in the Supplementary material, Section 2.1, Tables 16-17 \citep{mantziousupp}. This result suggests that the true representatives are almost fully identified from our algorithm.

In addition, we assess the effectiveness of our algorithm in identifying the cluster membership of the networks using the clustering entropy and purity indices. \textcolor{black}{Both clustering entropy and clustering purity are indices for evaluating clustering performance when the true cluster labels are known 
\citep{kim2007sparse, schutze2008introduction}. We note that a clustering entropy value of 0 and clustering purity value of 1 indicate a perfect cluster allocation of the networks.} We obtain 7,000 posterior draws (after a burn-in of 150,000 and lag of 50) for the cluster membership $z$, calculate the clustering entropy and clustering purity with respect to the true membership of the networks, and calculate their mean for each simulation regime. The simulation results indicate the true cluster membership of the networks is fully recovered by our MCMC algorithm, with mean entropy 0 and mean purity 1 for each simulation regime. These results are in the Supplementary material, Section 2.1, Table 18 \citep{mantziousupp}. 

\textcolor{black}{We further compare our approach to the nonparameteric Bayesian approach for modelling network populations by \cite{durante2017}, and to the maximum likelihood approach for clustering network populations by \cite{signorelli2020model}. 
\cite{durante2017} were originally interested in flexibly modelling a population of networks with diverse characteristics, rather than
clustering network-valued data, although clustering is a natural extension of their approach. 
Both \cite{durante2017} and \cite{signorelli2020model} capture the heterogeneity in a network population through a model-based framework, as is also the case in our framework. However, neither approach permits easy interpretation of the clusters.
The fundamental difference in our approach 
is the ability to infer a network representing each cluster, thus providing a useful summary of the networks in each cluster which can be advantageous when making inferences for diverse real-data applications.}

\textcolor{black}{We implement both approaches on the diverse network populations simulated as described earlier in this section, and assess the performance of the methods in clustering the network observations with respect to the clustering entropy and purity indices.
\cite{durante2017} method perfectly recovers the underlying $C=3$ clusters in the simulated network populations, with mean clustering entropy 0 and mean clustering purity 1 for each simulation regime.}

\textcolor{black}{To implement \cite{signorelli2020model} we need to first specify a statistical network model for the components of their mixture model. We choose to use an SBM as it is the model that we assumed for the network representatives of the clusters for simulating the network populations. However, there are two key limitations in the approach of \cite{signorelli2020model}. First, it is assumed that all clusters share the same block structure, and second, the block structure should be pre-specified as it is not inferred in their scheme, in contrast to our model which infers the block structure of the representatives and allows it to vary between the network representatives. Thus, to obtain a single block structure for all three mixture components, we use majority vote to determine the block membership of each node using the block structures specified for the representatives in our simulations, for each SBM simulation scenario 1 and 2. The clustering entropy and clustering purity calculated from the results obtained after applying the mixture model of \cite{signorelli2020model}, for each simulation regime, are presented in Table \ref{sign_res}. We see our model outperforms \cite{signorelli2020model} on our simulated data, which can be attributed to their model's restrictive assumption of a single SBM structure for all mixture components. As a result we do not consider the approach of \cite{signorelli2020model} any further.}

\begin{table}[ht!]
\centering
\begin{tabular}{clllll}
\textbf{}&& \multicolumn{2}{c}{\textbf{\textbf{$\text{SBM}_1$}}}&\multicolumn{2}{c}{\textbf{$\text{SBM}_2$}}\\ \hline \hline
$\boldsymbol{p_c}$         & \multicolumn{1}{c}{$\boldsymbol{q_c}$} & \multicolumn{1}{c}{\textbf{Entropy}} & \multicolumn{1}{c}{\textbf{Purity}} & \multicolumn{1}{c}{\textbf{Entropy}} & \multicolumn{1}{c}{\textbf{Purity}} \\ \hline \hline
\multirow{2}{*}{0.1} & 0.2& 0.69& 0.7& 0.71& 0.69\\
& 0.3& 0.79& 0.57& 0.61& 0.78\\ \hline
\multirow{2}{*}{0.2} & 0.1& 0.65& 0.73& 0.55& 0.81\\
& 0.3& 0.79& 0.57& 0.41& 0.86\\ \hline
\multirow{2}{*}{0.3} & 0.1& 0.76& 0.63& 0.56& 0.81\\
& 0.2& 0.75& 0.65& 0.59& 0.74 \\ \hline            
\end{tabular}
\caption{Clustering entropy and purity for results obtained after implementing \cite{signorelli2020model} model on the simulated data presented in this section.}\label{sign_res}
\end{table} 

\textcolor{black}{We now explore the performance of our model on data simulated under a different parameterisation to a mixture of measurement error models. Specifically, we consider the simulated population of networks in \cite{durante2017}. The data comprises 100 networks, with each network generated using one of four possible parameterisations of the edge probabilities, resulting in four groups of networks with different properties. Specifically, the
four groups are characterised by a community structure, small-worldness, an Erd\"{o}s-R\'{e}nyi structure and scale-free properties respectively. We run our MCMC for 500,000 iterations and assess the accuracy of our algorithm in inferring the group membership of each network in the population using the clustering entropy and clustering purity indices described earlier. Specifically, we consider 7,000 posterior draws (after a burn-in of 150,000 iterations and lag of 50) for the cluster membership $z$ of the networks, and obtain mean clustering entropy of 0 and mean clustering purity of 1. This indicates our algorithm perfectly recovers the group membership of the networks in the population, despite the different generative process used to simulate the data.}

\begin{figure}[ht!]
\centering
\begin{minipage}[b]{0.48\linewidth}
\centering
\includegraphics[height=2.2in]{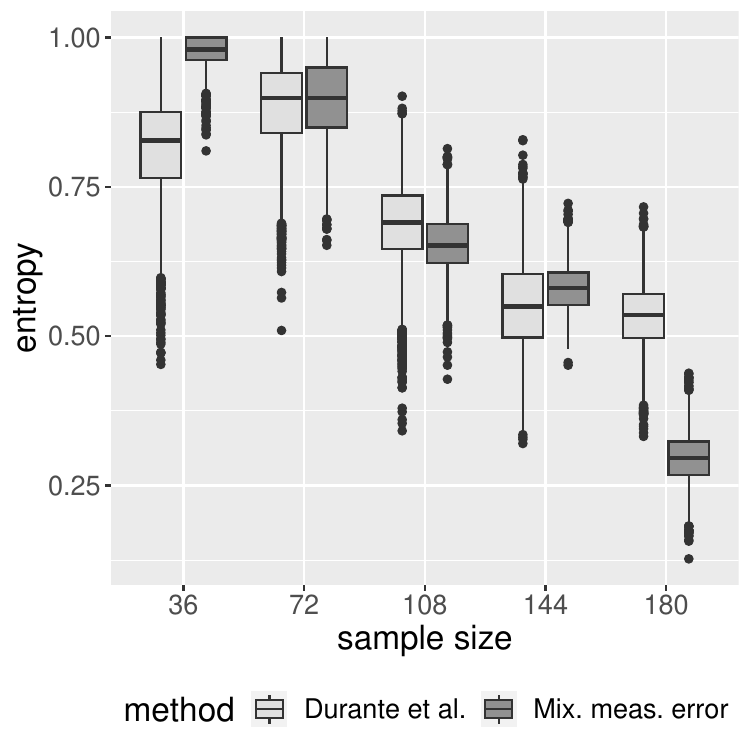}
\caption{Distribution of the clustering entropy across posterior draws, for our method and \cite{durante2017}, for varying sample sizes.}
\label{enr_comp}
\end{minipage}
\quad
\begin{minipage}[b]{0.47\linewidth}
\centering
\includegraphics[height=2.2in]{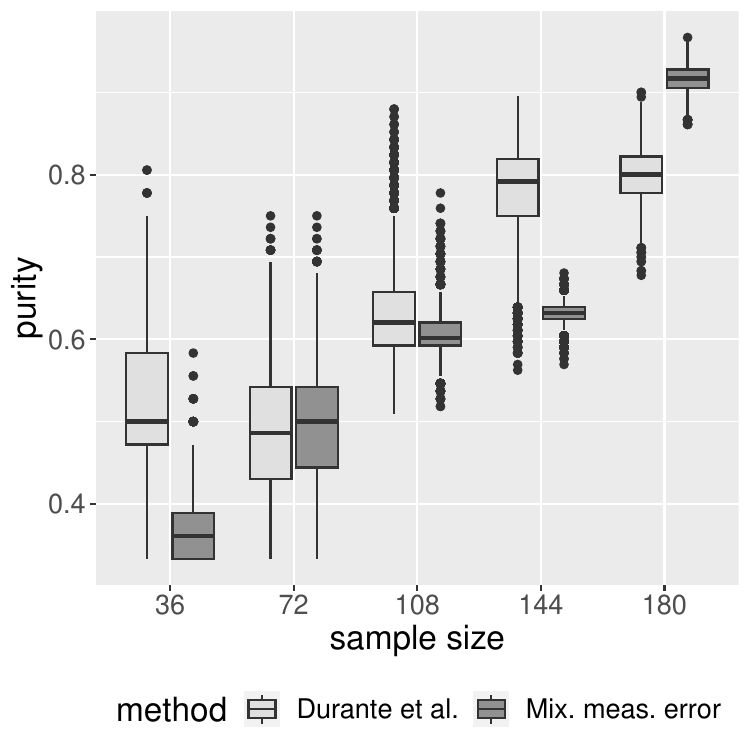}
\caption{Distribution of the clustering purity across posterior draws, for our method and \cite{durante2017}, for varying sample sizes.}
\label{pur_comp}
\end{minipage}
\end{figure}

\textcolor{black}{The simulation results so far show perfect performance of our model in recovering the true cluster labels for the network observations. To investigate clustering performance under more challenging scenarios, we consider simulating network populations with high false positive and negative probabilities ($p_c=q_c=0.4$) for a range of smaller population sizes, ranging from 36 to 180.
In each population, $C=3$ and network representatives have SBM structure 1 (Figure \ref{sbm_1_sbm_2_rep} top). We compare results to those obtained using the model in \cite{durante2017} on the same data. 
For our model, the MCMC is run for 500,000 iterations with 150,000 burn-in and lag 87 resulting in 4,023 MCMC draws, while for the \cite{durante2017} approach the MCMC is run for 5,000 iterations with 1,000 burn-in, leaving 4,000 posterior draws. Figures \ref{enr_comp} and \ref{pur_comp} illustrate the distribution of the clustering entropy and clustering purity for the posterior draws under each model, as well as for different sample sizes.}

\textcolor{black}{We observe that the clustering performance of the two methods have similar relationships with the sample size. As expected, both approaches perform least well in recovering the true cluster configurations for high noise levels and the smallest sample sizes, but steadily improve as sample sizes increase. We observe a slightly better performance with \cite{durante2017} over our approach for the smallest sample size of 36 networks, with the converse 
for the biggest sample size of 180 networks.
It is worth noting again that although both approaches have similar clustering performance overall, a key advantage of our method is the interpretability of the clusters with respect to a network representative, which is particularly relevant for our motivating data applications.}

\textcolor{black}{
We additionally investigate the performance of our model
in the network population size of 180 for various different noise levels. Specifically, we generate network populations by perturbing the edges of the representatives of SBM structure 1, illustrated in
Figure \ref{sbm_1_sbm_2_rep}, with varying sizes of the false positive and negative probabilities 
given in Tables \ref{fix_q_sim} and \ref{fix_p_sim}.} 

\begin{table}[h!]
\centering
\parbox{.48\linewidth}{
\centering
\begin{tabular}{lc}
\multicolumn{1}{c}{$\boldsymbol{p_c}$ } & $\boldsymbol{q_c}$ \\ \hline \hline
 0.01 &\multirow{10}{*}{0.1} \\
0.05 &\\
 0.1&\\
 0.15&\\
0.2&\\
0.25&\\
0.3&\\
0.35&\\
0.4&\\
0.45 &\\ \hline
\end{tabular}
\caption{Simulation regimes for varying sizes of false positive probabilities $p_c$ and fixed false negative probabilities $q_c$, for $c \in \{1,2,3\}$ and 21-node networks.}\label{fix_q_sim}
}
\hfill
\centering
\parbox{.48\linewidth}{
\centering
\begin{tabular}{cl}
$\boldsymbol{p_c}$& \multicolumn{1}{c}{$\boldsymbol{q_c}$ } \\ \hline \hline
\multirow{10}{*}{0.1} & 0.01\\
& 0.05\\
& 0.1\\
& 0.15\\
& 0.2\\
& 0.25\\
& 0.3\\
& 0.35\\
& 0.4\\
& 0.45 \\ \hline
\end{tabular}
\caption{Simulation regimes for varying sizes of false negative probabilities $q_c$ and fixed false positive probabilities $p_c$, for $c \in \{1,2,3\}$ and 21-node networks.}\label{fix_p_sim}
}
\end{table}

\begin{figure}[ht!]
\centering
\begin{minipage}[b]{0.48\linewidth}
\centering
\includegraphics[height=2.2in]{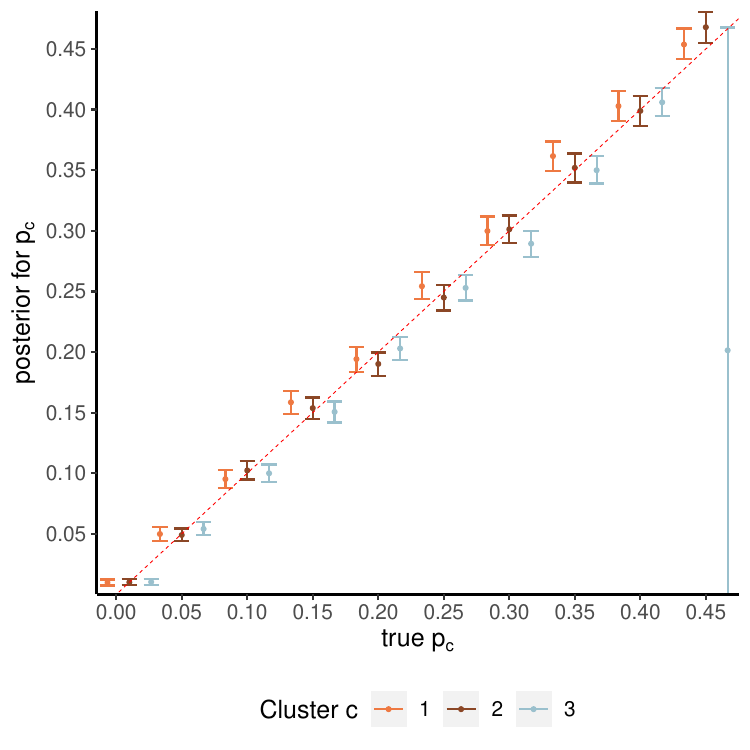}
\caption{Posterior means and 95\% credible intervals for false positive probabilities $p_{c}$ for $c\in\{1,2,3\}$ (y axis), plotted against the true values of $p_{c}$ (x axis). \textcolor{black}{Red dashed line is the y=x line.}}
\label{fix_q_plot}
\end{minipage}
\quad
\begin{minipage}[b]{0.47\linewidth}
\centering
\includegraphics[height=2.2in]{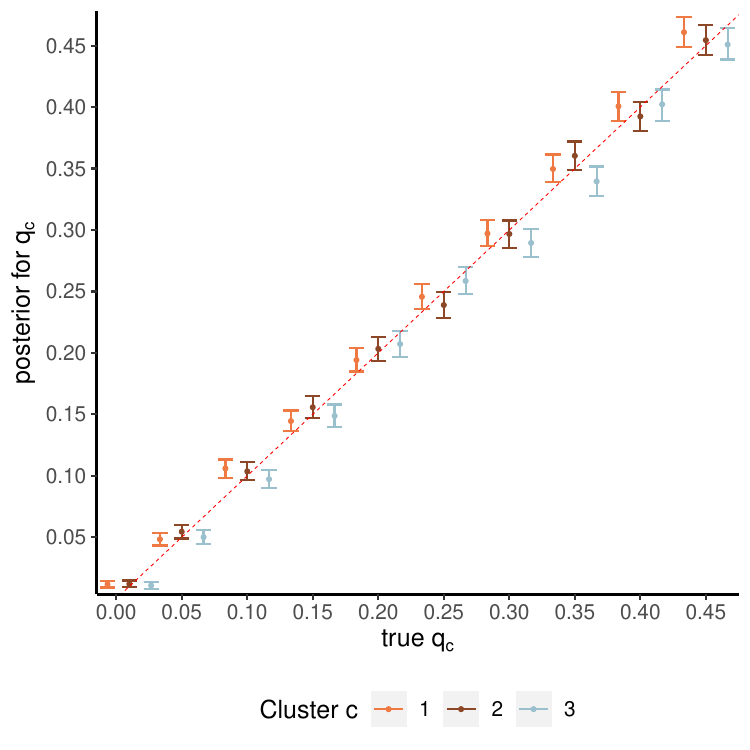}
\caption{Posterior means and 95\% credible intervals for false negative probabilities $q_{c}$ for $c\in\{1,2,3\}$ (y axis), plotted against the true values of $q_{c}$ (x axis). \textcolor{black}{Red dashed line is the y=x line.}}
\label{fix_p_plot}
\end{minipage}
\end{figure}

\textcolor{black}{For each regime presented in Tables \ref{fix_q_sim} and \ref{fix_p_sim}, we generate a network population and run the MCMC for 500,000 iterations with a burn-in of 150,000 iterations. 
In Figures \ref{fix_q_plot} and \ref{fix_p_plot}, for each cluster $c\in\{1,2,3\}$, we plot the posterior means and 95\% credible intervals (via errors bars) for the different false positive (Table \ref{fix_q_sim}) and false negative probabilities (Table \ref{fix_p_sim}) respectively against their true values. 
We see the posterior means lie mostly close to the $y=x$ line \textcolor{black}{(red dashed line)}, indicating our model performs well in inferring the true false positive and false negative probabilities, even for high noise levels. 
However, in Figure \ref{fix_q_plot}, for the highest noise value of $p_{c}=0.45$ for $c\in\{1,2,3\}$, we observe that the posterior mean of the false positive probability of cluster 3, $p_{3}$, is equal to 0.2, which is substantially different to its true value, while its 95\% credible interval covers a wide range of values, indicating that our MCMC chain struggles to make inferences here.}
These results suggest that our model performs well in most cases, even for high noise levels, but we must be cautious when making inferences for network populations with great variability in their structure.

\textcolor{black}{For the simulations reported in this section,
the computational time required to run our MCMC procedure for 500,000 iterations varied from 
approximately 50 minutes (for a population size of 36 networks) through to approximately 80 minutes (for a population size of 180 networks). 
We note here that in both our simulations and data analysis, we consider a large number of iterations to ensure convergence of our MCMC, even though for some scenarios less iterations would suffice.}

\subsection{Varying sizes of networks and network populations}\label{sec52}

We now explore how well our model infers the parameters with respect to various network sizes and sample sizes. We keep $C=3$ clusters and $B=2$ blocks. We consider four different network sizes of 25, 50, 75  and 100 nodes, and simulate populations of 45, 90, 135, 180, 225, 270 and 315 networks, for each network size respectively.

\begin{figure}[ht!]
\centering
\begin{minipage}[b]{0.48\linewidth}
\centering
\includegraphics[height=1.7in]{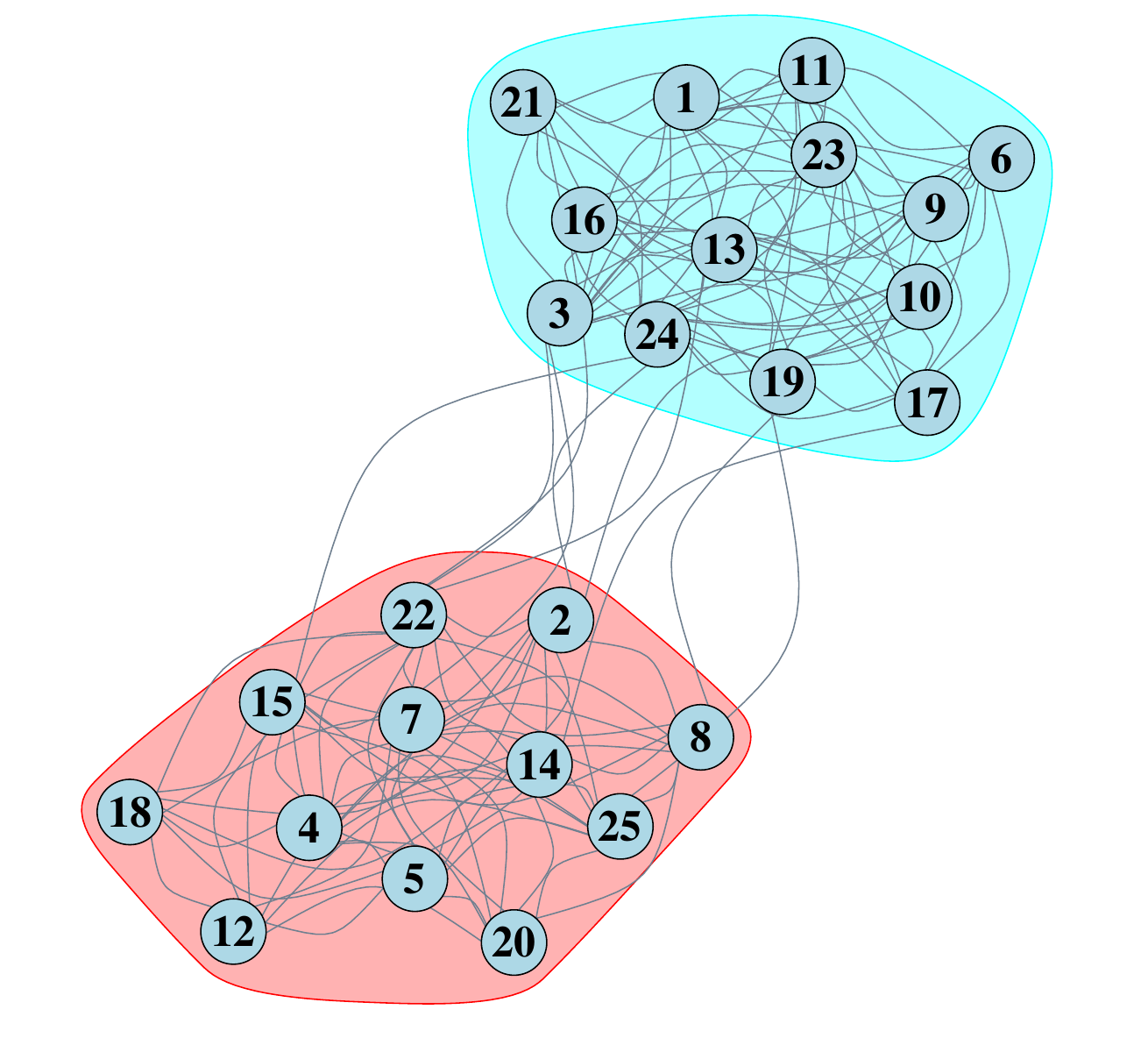}
\caption{25-node representative of cluster labelled 1.}\label{rep_25}
\label{fig:minipage1}
\end{minipage}
\quad
\begin{minipage}[b]{0.48\linewidth}
\centering
\includegraphics[height=1.7in]{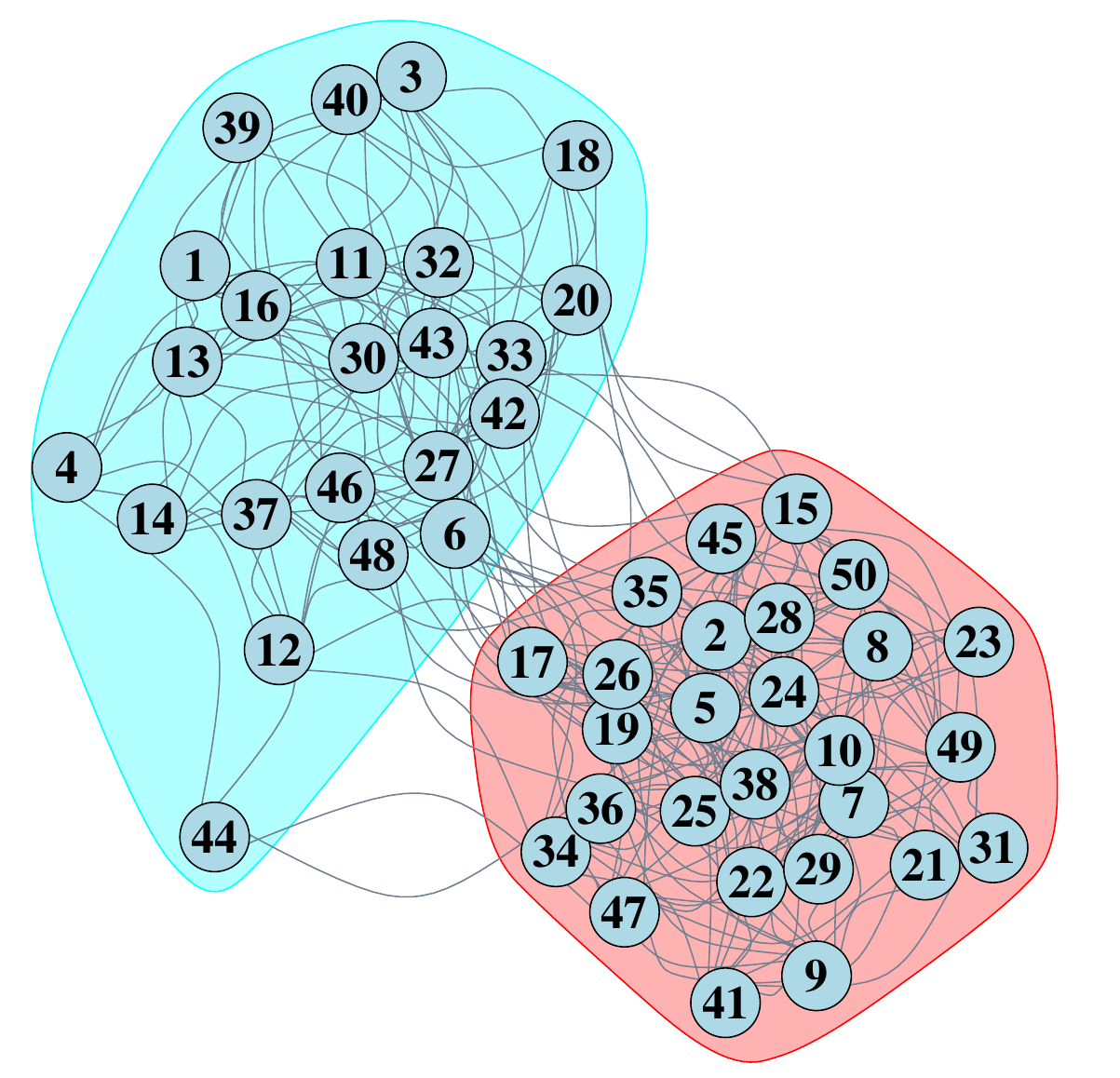}
\caption{50-node representative of cluster labelled 1.}\label{rep_50}
\label{fig:minipage2}
\end{minipage}
\end{figure}
\begin{figure}[ht!]
\begin{minipage}[b]{0.48\linewidth}
\centering
\includegraphics[height=1.7in]{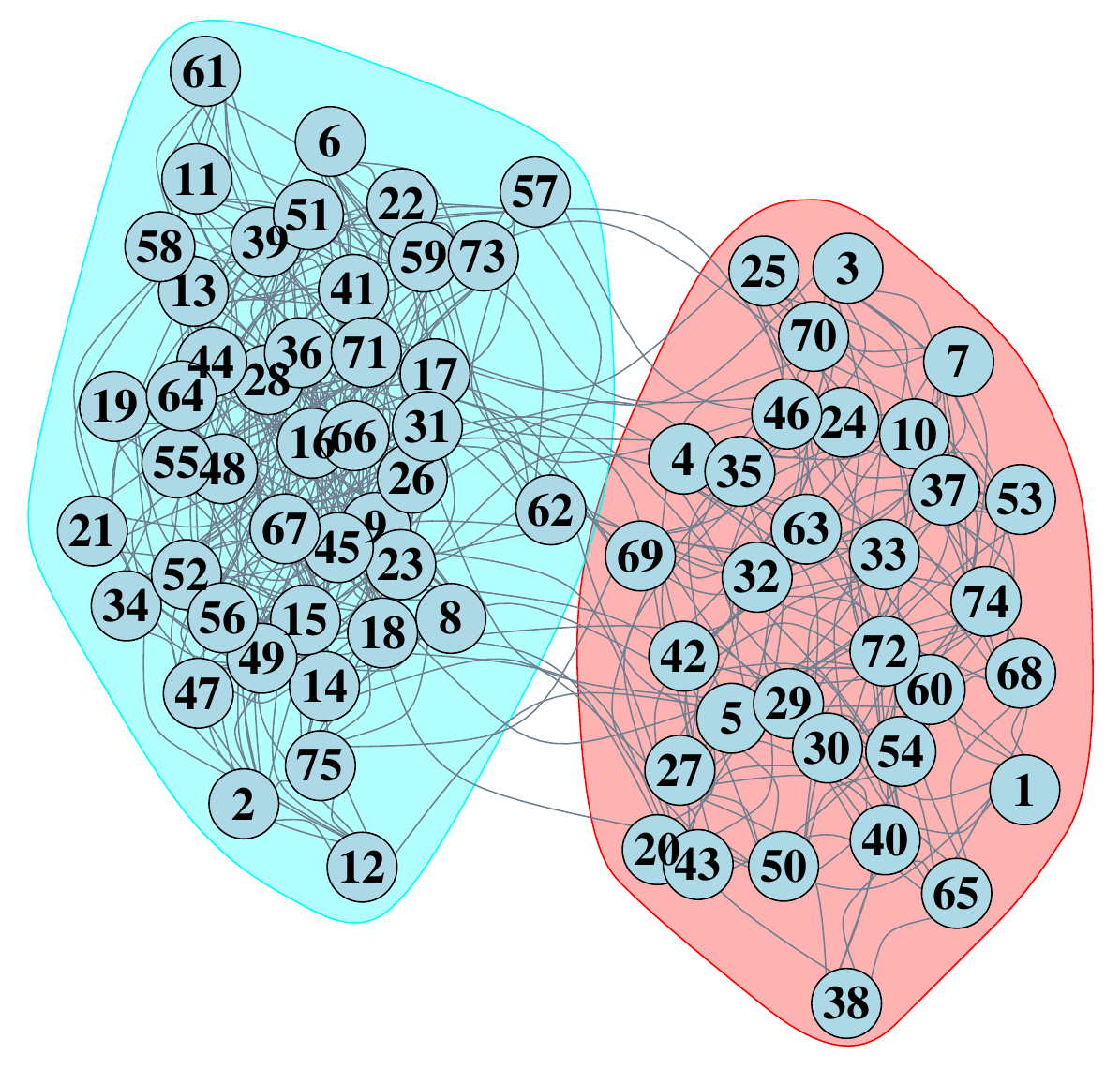}
\caption{75-node representative of cluster labelled 1.}\label{rep_75}
\label{fig:minipage1}
\end{minipage}
\quad
\begin{minipage}[b]{0.48\linewidth}
\centering
\includegraphics[height=1.7in]{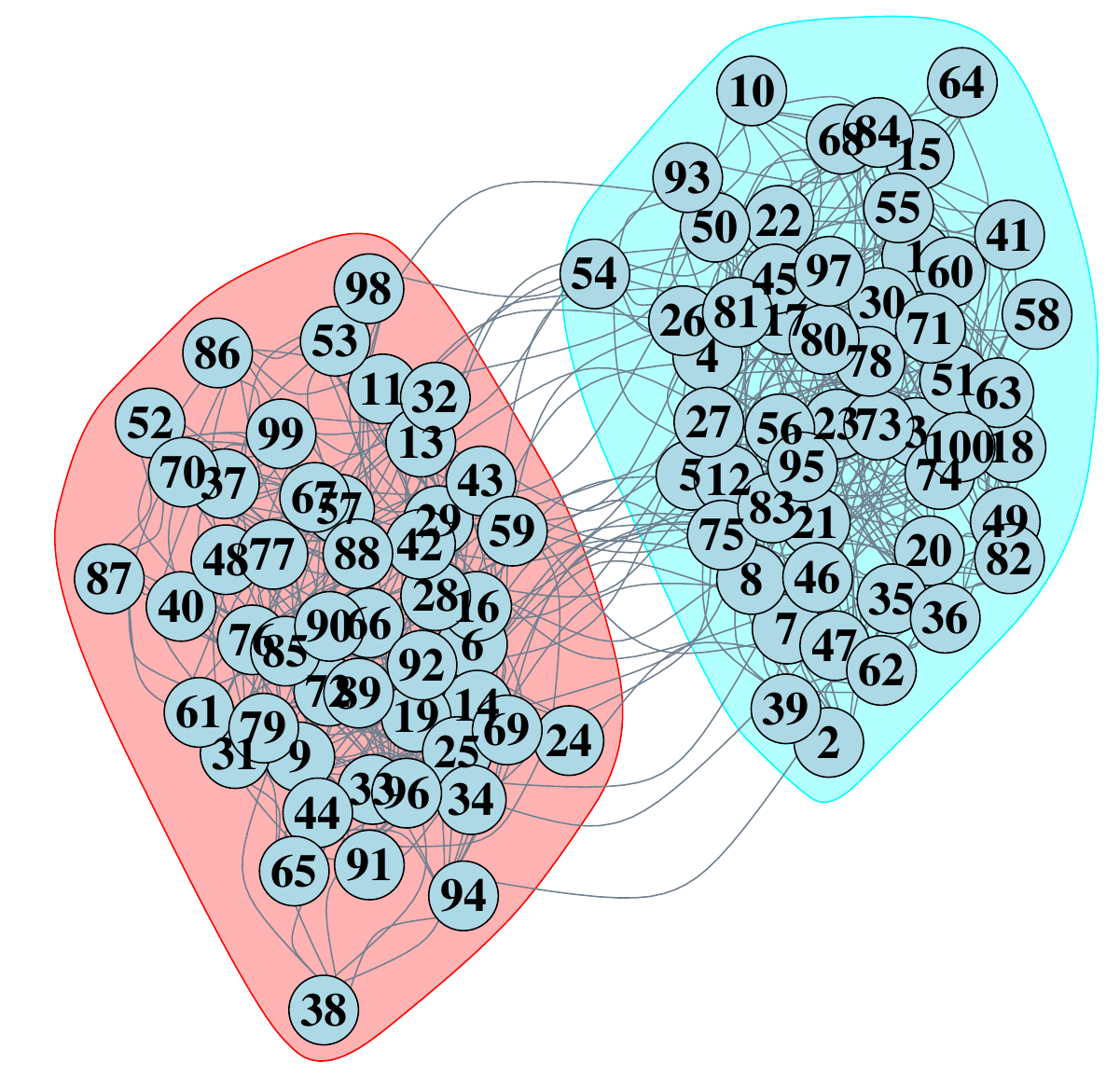}
\caption{100-node representative of cluster labelled 1.}\label{rep_100}
\label{fig:minipage2}
\end{minipage}
\end{figure}

To simulate the network populations, we first generate the network representatives of each cluster assuming each follows an SBM. We specify the parameters of the SBMs so that the expected degree of the resulting network representatives is preserved for all network sizes. The representatives obtained for cluster $c=1$ for the different network sizes are visualised in Figures \ref{rep_25}, \ref{rep_50}, \ref{rep_75} and \ref{rep_100}. In the Supplementary material, Section 2.2, Figures 1-4 \citep{mantziousupp}, we illustrate the rest of the representatives for $c=2,3$, for the network sizes considered. The resulting populations are generated by perturbing the edges of each network representative, for each network size considered, with a false positive $p_{c}$ and false negative $q_{c}$ probability fixed at 0.08, for $c\in\{1,2,3\}$.

\begin{figure}[h!]
\centering
\includegraphics[height=2.5in,width=0.49\textwidth]{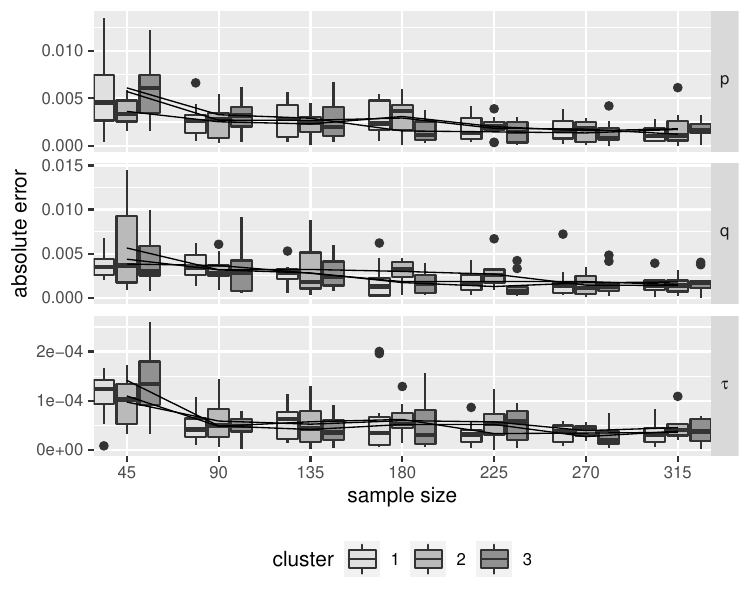}
\hfill 
\includegraphics[height=2.5in,width=0.49\textwidth]{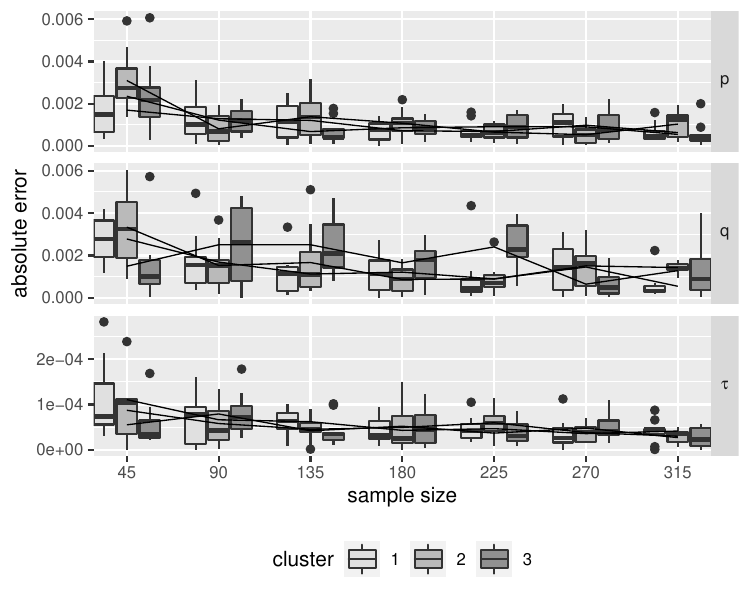}
\caption{Left: Absolute error (y axis) for model parameters $\boldsymbol{p}$, $\boldsymbol{q}$ and $\boldsymbol{\tau}$, for 25-node networks and varying population sizes (x axis). Right: Absolute error (y axis) for model parameters $\boldsymbol{p}$, $\boldsymbol{q}$ and $\boldsymbol{\tau}$, for 50-node networks and varying population sizes (x axis).}\label{var_sim}
\vfill
\includegraphics[height=2.2in,width=0.49\textwidth]{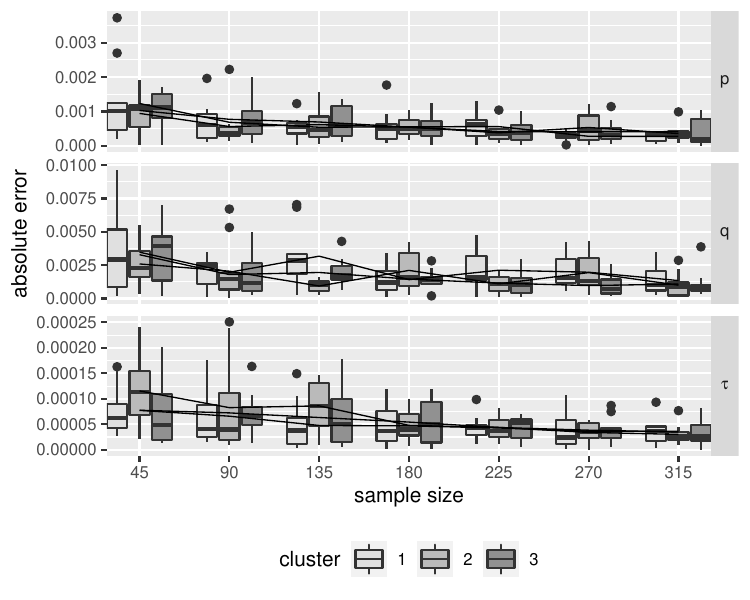}
\hfill 
\includegraphics[height=2.2in,width=0.49\textwidth]{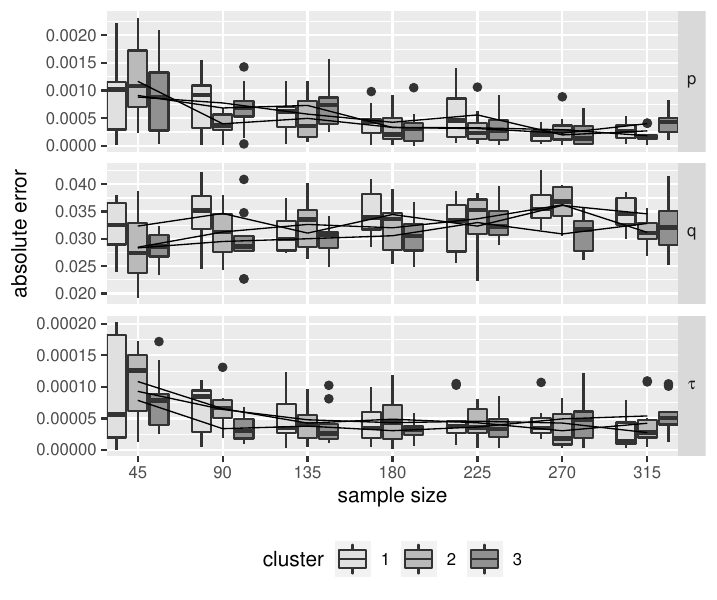}
\caption{Left: Absolute error (y axis) for model parameters $\boldsymbol{p}$, $\boldsymbol{q}$ and $\boldsymbol{\tau}$, for 75-node networks and varying population sizes (x axis). Right: Absolute error (y axis) for model parameters $\boldsymbol{p}$, $\boldsymbol{q}$ and $\boldsymbol{\tau}$, for 100-node networks and varying population sizes (x axis).}\label{var_sim2}
\end{figure}

\textcolor{black}{For each simulation regime, we consider 10 replications of our MCMC, each on a different randomly generated data set, and run our algorithm for 500,000 iterations with a burn-in of 150,000. \textcolor{black}{For the largest simulation scenario involving 100-node networks and 315 network observations, 500,000 iterations of our MCMC required a run time of approximately 24 hours. Thus, 10 replications was deemed reasonable, taking into account the increased computational burden associated with running these simulations.} We demonstrate the performance of our model by obtaining the distribution of the absolute error of the model parameters for each simulation regime, as seen in Figures \ref{var_sim} and \ref{var_sim2}. Specifically, the plots demonstrate how the distribution of the absolute error (y axis) scales for various sample sizes (x axis). The absolute error is the absolute value of the difference between the posterior means obtained after burn-in, and the true value of the parameter. The different grey shades of the boxplots correspond to the different clusters considered, and the lines connect the mean absolute error across replications to illustrate the trend.}

The plots indicate that the sample size affects the performance of our model for the network sizes considered. We observe that as the number of nodes increase, there is an increase in the required number of networks to preserve the same level of accuracy in estimation.
Nevertheless, even for large networks and small sample sizes, it is encouraging to see the posterior means obtained are not far away from their true values. We further illustrate the performance of our model in identifying the true representatives of each cluster for the various sample and network sizes, in the Supplementary Material, Section 2.2, Figure 5 \citep{mantziousupp}. 

\textcolor{black}{For the simulation regime with the largest size of networks and network population, $n=100$ nodes and $N=315$ networks respectively, we additionally explore the performance of our model for different hyperparameter settings for the Beta and Dirichlet priors of our model. Specifically, we consider two common non-informative priors, the Jeffrey's prior, Beta(0.5,0.5) and Dirichlet(0.5), and the Uniform prior, Beta(1,1) and Dirichlet(1). Additionally, we specify an informative prior such that the proportional reduction in variance from the Uniform prior to the informative prior is equal to the proportional reduction in variance from the Jeffrey's prior to the Uniform prior. This results in Beta(1.75,1.75) and Dir(1.75) priors. Under each hyperparameter setting, we run our MCMC for 500,000 iterations and obtain similar results, not presented herein due to length restrictions, indicating that our model is not sensitive to the hyperparameter specification.
}

\textcolor{black}{Lastly, for completeness, we explore the clustering performance of the model of \cite{durante2017} for varying sizes of networks. Specifically, we implement the model by \cite{durante2017} on the simulation regimes with $N=180$ and $n=25,50,75,100$. The model accurately clusters the networks with mean clustering entropy 0 and mean clustering purity 1 for each regime considered. As previously noted, despite the good clustering performance of this model, meaningful interpretations of the clusters can be challenging. 
}

\textcolor{black}{
\subsection{Number of clusters with SFM}\label{sec53}
In the simulations performed in Sections \ref{sec51} and \ref{sec52}, the number of clusters were known and specified according to the number of mixture components used to simulate the network populations, which is typically not the case in real data applications. The SFM extension of our model introduced in Section \ref{sec45} allows us to treat the number of clusters as unknown within our framework, and subsequently infer an appropriate number of clusters to specify.
}

\textcolor{black}{To explore the performance of the SFM extension of our model in identifying the true number of clusters and the true cluster labels of the networks in a population, we implement it on a range of simulation regimes described in Section \ref{sec51}. Specifically, we consider the simulated network data under SBM structure 1 and noise levels $p_c=0.1$ and $q_c=0.2$ as well as the case with noise levels $p_c=0.2$ and $q_c=0.3$, for $c\in\{1,2,3\}$. Similarly, we consider the simulated network populations under the same noise levels, for the representatives generated with SBM structure 2. The true number of clusters in all cases are $C=3$. We tune our MCMC specifying $C_{max}=10$ clusters and hyperparameters $a_e=1$ and $b_e=400$ for the Gamma prior on $e_0$ to impose strong shrinkage of $e_0$ to 0, and run our MCMC algorithm for 500,000 iterations. }

\textcolor{black}{We notice 
we quickly converge to the true number of clusters, $C=3$, not illustrated herein due to length restrictions. To assess the model performance in identifying the true cluster labels of the networks observations, we calculate the clustering entropy and clustering purity indices for $\boldsymbol{z}$ after a burn-in of 150,000 iterations and a lag of 50, leaving 7,000 posterior draws. We observe that for all four simulation regimes considered in this study, the model perfectly recovers the cluster membership of the networks with mean clustering entropy 0 and mean clustering purity 1.} 

\textcolor{black}{We note that the SFM model is highly sensitive to the tuning of $a_e,b_e$ in the Gamma hyperprior and $C_{max}$, as also discussed in \cite{fruhwirth2019here}. Different levels of shrinkage can lead to inferences with different numbers of clusters $C$ and cluster configurations. In the next section, involving the motivating real data applications, we consider two ways of determining the number of clusters $C$, one of which is the SFM model. 
}

\section{\textcolor{black}{Motivating data examples}}\label{sec6}
In this Section, we present the application of our mixture model on the two real-world \textcolor{black}{populations of network} data sets presented in Section \ref{sec2}. 
\subsection{Movement patterns across campus}\label{sec61}
As introduced in Section \ref{sec21}, Tacita is a mobile phone application that records the displays visited by users, along with the time visited and the type of content shown on the display. One way to represent the data collected from the Tacita application is through a network, where nodes correspond to displays, and edges correspond to movements of users among the displays. Consequently, we obtain a \textcolor{black}{population of network} data set where each network observation corresponds to a user's movements across displays. The final data sample consists of 120 undirected and unweighted network observations that share the same set of 37 nodes corresponding to the displays across campus. Names of the displays are presented in the Supplementary Material, Section 3.1, Table 19 \citep{mantziousupp}.
As our mixture model requires the pre-specification of the number of clusters $C$, \textcolor{black}{one way to determine an appropriate number of clusters in the data is through exploratory data analysis (EDA).}
We considered various network distance metrics, and for each metric we obtain a distance matrix that contains the pairwise distances of the networks in the population. 
We obtain the Multi Dimensional Scaling (MDS) plot for each distance matrix obtained. The MDS algorithm maps objects in a 2-d space, respecting their pairwise distances. The MDS plots obtained from the EDA are presented in the Supplementary Material, Section 3.1, Figures 6 and 7 \citep{mantziousupp}. 

\textcolor{black}{As anticipated different network distances reveal different type of similarities between the networks. In this regard, the MDS visualisations provide only a point of reference to determine the number of clusters. To begin with, the specification of $C=3$ seems a reasonable starting point for our analysis as per the EDA results. Later in this section, we compare this with the SFM extension of our model presented in Section \ref{sec53}.}
To meaningfully initialise the networks' cluster membership, we combine the results from four different distance metrics; the Hamming, the Jaccard, the $l_{2}$, and the wavelets metrics. For a descriptive review on the distance metrics refer to \cite{donnat2018tracking}. Specifically, we use a k-means algorithm using the R package \texttt{kmed} \cite{kmed} to determine four different cluster memberships, corresponding to the four different metrics considered, and determine the final cluster membership initialisation using majority vote, i.e. by determining which networks are consistently allocated to one of the three clusters among the four memberships obtained. We initialise the representative of each cluster by generating its edges using independent Bernoulli draws. The probability with which we draw an edge between two specific nodes of the representative corresponds to the proportion of times that we see that edge in the network data of the corresponding cluster. We then initialise the nodes' block membership of the representatives using SBM estimates from the R package \texttt{blockmodels} \cite{blockmodels} that suggest the presence of two underlying blocks. We note here that a simpler network model, namely the Erd\"{o}s-R\'{e}nyi model, could also be applied to describe the representative networks. We run our MCMC for 500,000 iterations with a burn-in of 100,000. 

\begin{figure}[h!]
\centering
\includegraphics[height=1.5in,width=0.27\textwidth]{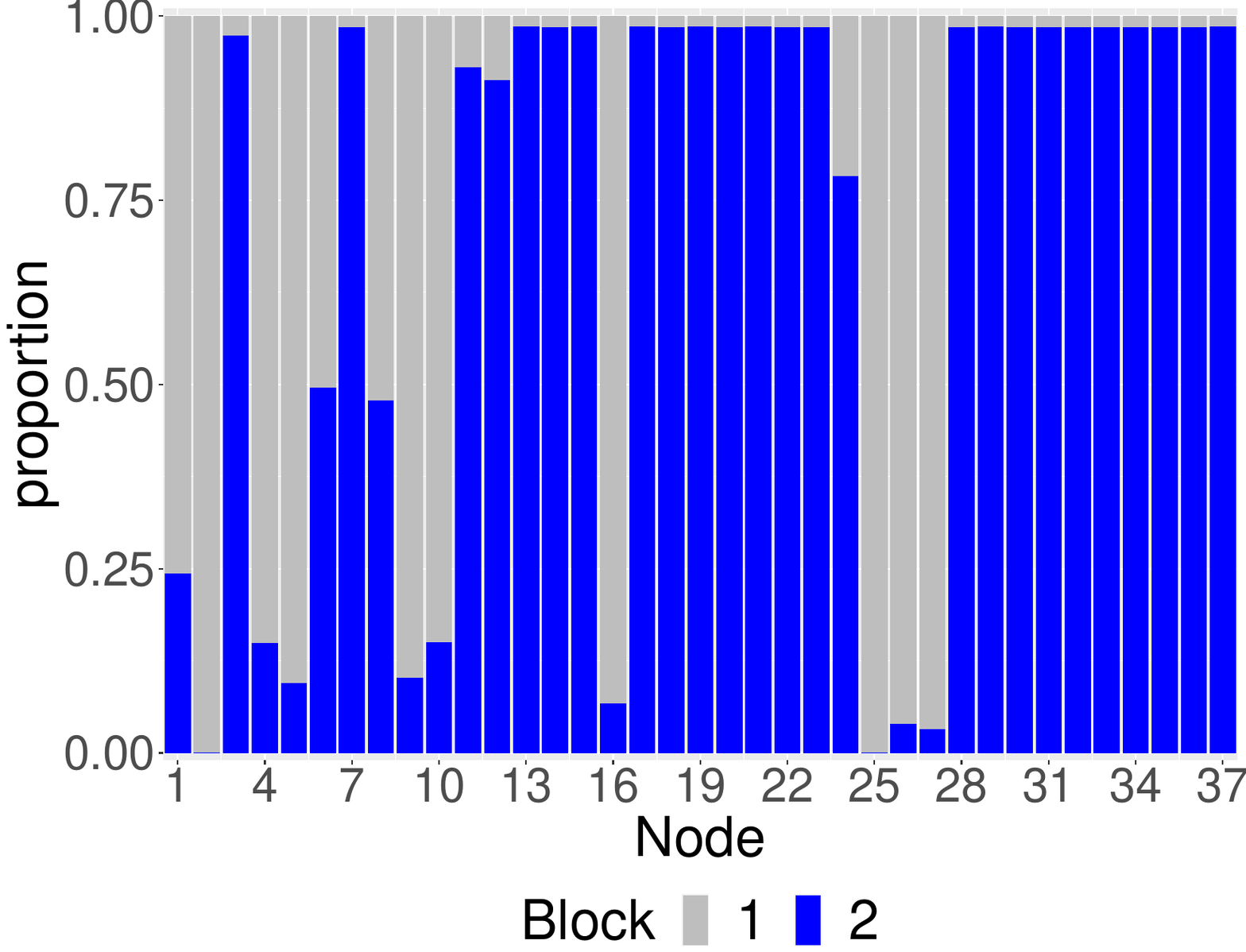}
\hfill 
\includegraphics[height=1.5in,width=0.27\textwidth]{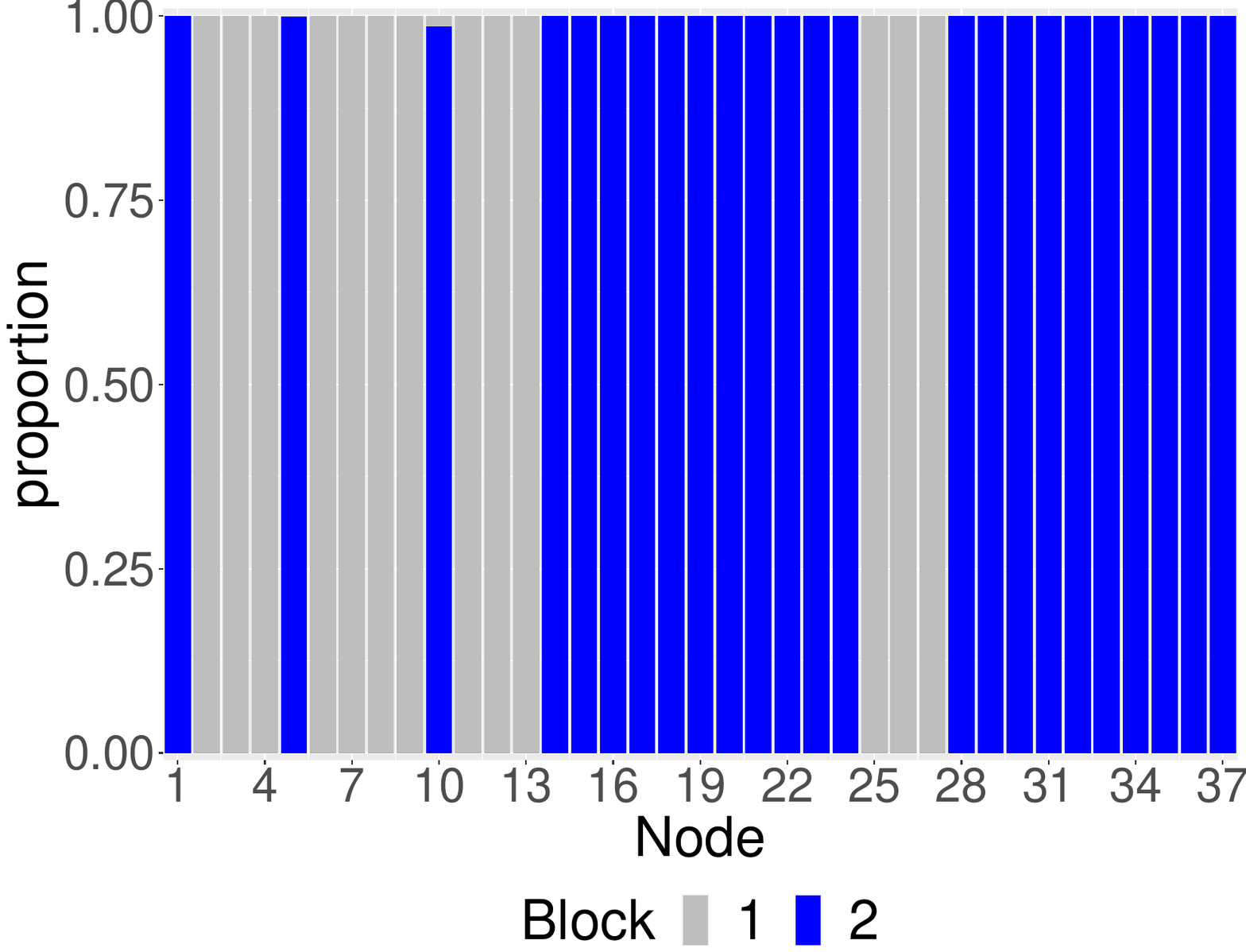}
\hfill
\includegraphics[height=1.5in,width=0.44\textwidth]{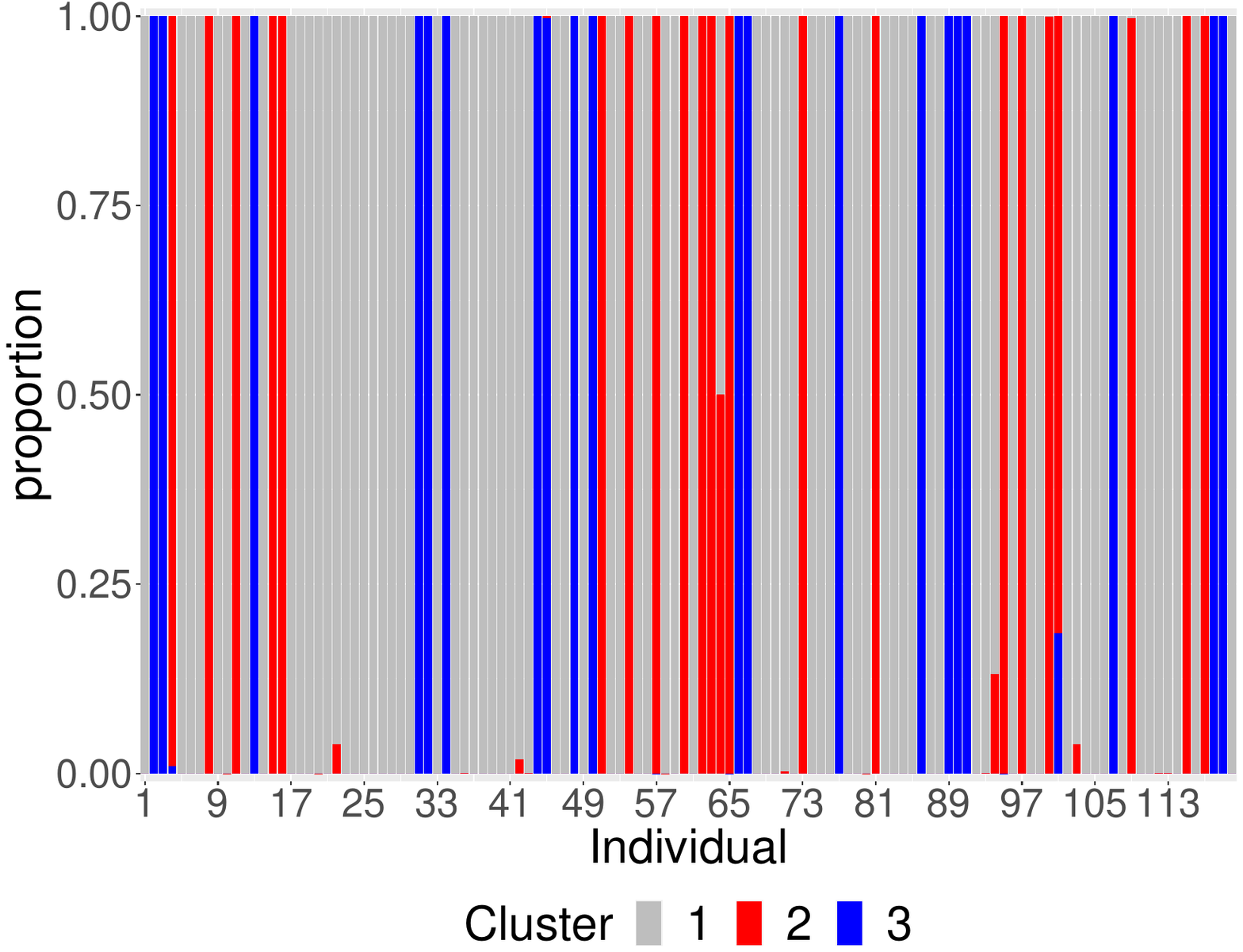}
\caption{Left: Proportion of times (y axis) that each node (x axis) of the representative network of cluster labelled 2 is allocated in Blocks 1 or 2, after a burn-in of 100,000 iterations. Middle: Proportion of times (y axis) that each node (x axis) of the representative network of cluster labelled 3 is allocated in Blocks 1 or 2, after a burn-in of 100,000 iterations. Right: Proportion of times (y axis) an individual's network (x axis) is allocated to each of the 3 clusters, after a burn-in of 100,000 iterations.}\label{memberships}
\end{figure}

In Figure \ref{memberships}, the left and middle plots present the proportion of times that the nodes of the representative networks of clusters 2 and 3 respectively, belong in each of the two blocks specified. The results for the representative of cluster $c=1$ are presented in the Supplementary material, Section 3.1, Figure 10 \citep{mantziousupp}. We note here, that a block structure is not identified for the representative of cluster $c=1$. From Figure \ref{memberships}, we observe a similar block membership is revealed for the representatives of clusters 2 and 3. However, we notice differences in the block allocation of some nodes between the two representatives, namely nodes labelled 1, 3, 5, 7, 10, 11, 12 and 13. In addition, for the representative of cluster $c=3$, the nodes are more clearly allocated to each of the two blocks, as seen from the proportions. In Figure \ref{memberships}, the right plot corresponds to the proportion of times that an individual is allocated to the clusters, showing that most individuals are clearly allocated to one of the three clusters.

\begin{figure}
\centering
\includegraphics[width=0.32\textwidth]{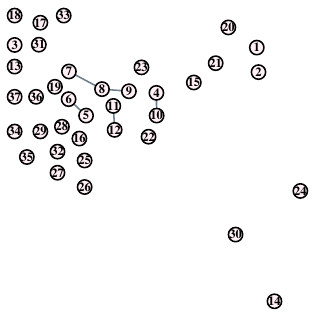}
\hfill 
\includegraphics[width=0.32\textwidth]{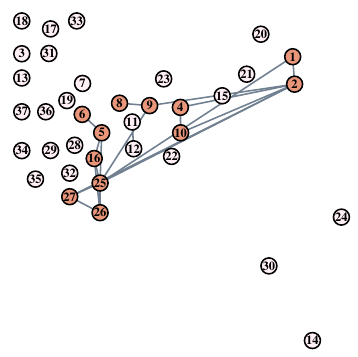}
\hfill 
\includegraphics[width=0.32\textwidth]{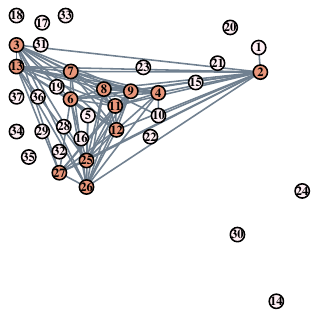}
\vspace{3mm}
\caption{Left: Posterior mode for the representative of cluster 1, with posterior mass 100\%, for the last 100,000 iterations. Middle: Posterior mode for the representative of cluster 2, with posterior mass 37\%, for the last 100,000 iterations. Right: Posterior mode for the representative of cluster 3, with posterior mass 15\%, for the last 100,000 iterations.}\label{tacita_rep}
\end{figure}

In addition, in Figure \ref{tacita_rep} we obtain the network representatives of each cluster, with node labels corresponding to the numbering seen in Table 19 given in the Supplementary material, Section 3.1 \citep{mantziousupp}, and layout similar to the true location of the displays on campus. The two different colours of the nodes denote the block membership inferred for each representative. In the Supplementary Material, Section 3.1, Figures 8 and 9 \citep{mantziousupp}, we present trace plots of the false negative and false positive probabilities for each of the three clusters.

For cluster $c=1$, we observe that the representative concentrating the whole posterior mass is sparse having only few edges and no SBM structure. Specifically, we note that the edges of the representative correspond to movements of individuals among displays that are very close to each other (e.g. edge between nodes 4 and 10 corresponding to displays both located in the same building). In addition, most of the networks in the population are allocated to this cluster with a very small false positive probability with posterior mean 0.003 and very large false negative probability with posterior mean 0.49. The small false positive probability indicates that the edges observed in the network data are correctly recorded, while the high false negative probability indicates a high possibility of edges in the network data that we do not observe. This is a reasonable finding as we would anticipate that the Tacita application might have missed movements of users among displays due to WiFi connection issues. This remark is also justified by the movements observed in the representative network that correspond to movements among displays within the same building, suggesting that when a user is in a building, the WiFi connection is preserved, and so the application can record the movements of the user. 

For cluster $c=2$, we observe that the posterior mode of the representative network concentrates a relatively smaller posterior mass of 37\%, while being slightly denser compared to the representative of cluster 1. Moreover, an SBM structure is revealed, with the mostly connected nodes belonging in the same block. This representative reveals a specific movement pattern of the users allocated in cluster 2, corresponding to the displays located at the
central part of the campus (nodes 5, 6, 16, 25, 26, 27), as well as displays located at Infolab (nodes 1 and 2), and Furness College (4 and 10). However, there is a smaller proportion of networks allocated to cluster 2 compared to cluster 1. However, the small false positive (posterior mean 0.02) and high false negative probability (posterior mean 0.49)
again indicates that there might be movements of users not recorded by the application.

Lastly, for cluster $c=3$ the posterior mode of the network representative concentrates a smaller posterior mass compared to the other two representatives equal to 15\%. We notice similarities both in the block structure and the connectivity patterns of the representatives of clusters 2 and 3. However, the representative of cluster 3 is notably denser, and some new movements of individuals at displays labelled 3 and 13 (Faraday College and County College) are discovered. We also note that for cluster 3, the posterior means obtained for the false positive and false negative probabilities are similar to cluster 2. In addition, clusters 2 and 3 have a similar proportion of individuals. Overall, a common movement pattern is discovered for individuals in clusters 2 and 3, indicating movements among displays located in the central part of the campus.

\begin{figure}[ht!]
\centering
\includegraphics[height=2.1in]{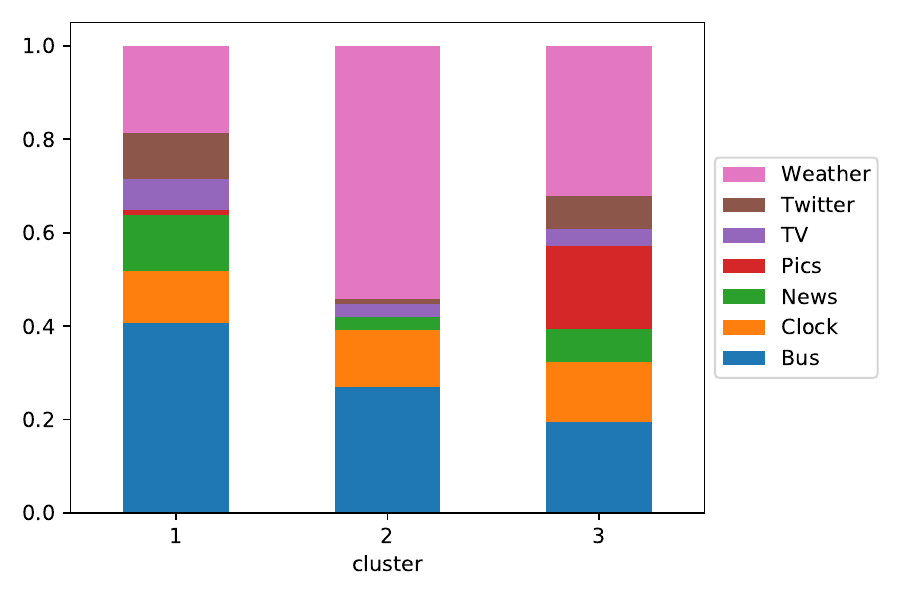}
\vspace{3mm}
\caption{Proportion of times (y axis) each type of content shown (colors) per cluster of individuals (x axis).}\label{content}
\end{figure}

\textcolor{black}{We additionally investigate whether users in each cluster interact differently with the displays. Specifically, Figure \ref{content} presents the proportion of times each type of content was shown per cluster of individuals. We have excluded the content "Welcome screen" as it is the default screen introducing the users to the application. We see that individuals assigned to cluster 3 are more active in terms of the range of content shown on the displays visited. This is reasonable considering that the representative of this cluster (Figure \ref{tacita_rep} right) is the densest, potentially indicating these users are most engaged with the application. For individuals in cluster 1, we see the most common type of content shown is the bus timetable. This can be explained by the movement patterns of the individuals in this cluster 1, as represented by the network in Figure \ref{tacita_rep} left, which are primarily between displays near to where the bus station is located. Lastly, the type of content mostly shown for individuals of cluster 2 is the weather, which seems to be relevant given the representative movement patterns for these individuals (Figure \ref{tacita_rep} middle), encompassing longer distances across campus, away from the central covered part.
}

\begin{figure}[ht!]
\centering
\begin{minipage}[t]{0.48\linewidth}
\centering
\includegraphics[height=2.3in]{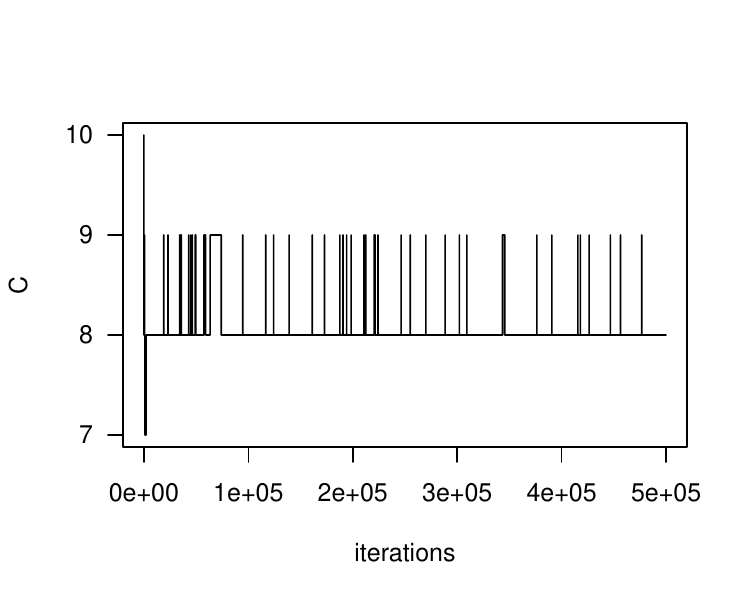}
\caption{Traceplot for number of clusters detected in each iterations of the MCMC for SFM extension of our model.}
\label{cplus}
\end{minipage}
\quad
\begin{minipage}[t]{0.48\linewidth}
\centering
\includegraphics[height=1.85in]{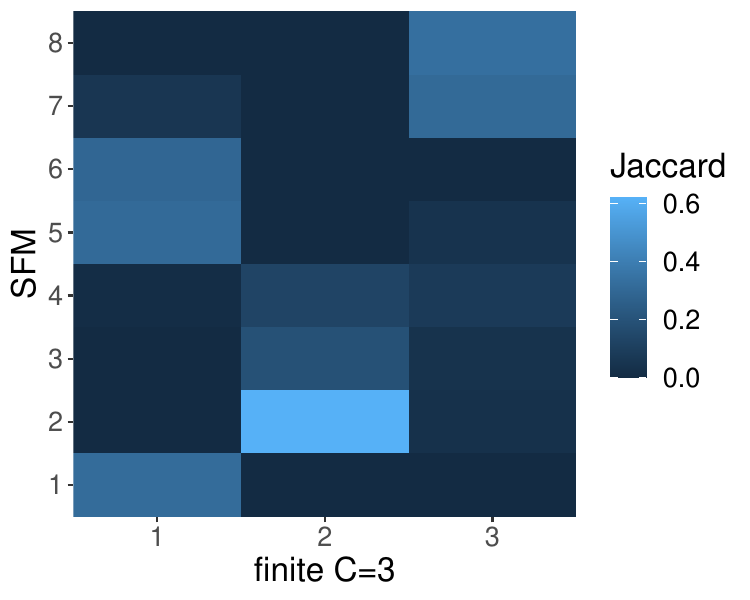}
\caption{Jaccard similarity between the set of observations in each of the 8 clusters recovered with the SFM extension of our model (y axis) and the set of observations in each of the 3 clusters from our finite mixture model.}
\label{part_jacc}
\end{minipage}
\end{figure}

\textcolor{black}{The analysis of the data assuming $C=3$ clusters gives sensible and interpretable results as discussed above. The clear allocation of individuals to clusters, and the movement patterns revealed by the cluster specific network representatives, along with the type of content characterising each cluster, indicate that three clusters is a sensible choice for the data. However, to investigate the number of clusters further, 
we implement the SFM extension of our model.} 

\textcolor{black}{For the SFM, we specify an upper bound of clusters $C_{max}=10$ and a Gamma$(1,400)$ hyperprior to impose a high degree of sparsity. We run the MCMC for 500,000 iterations with a burn-in of 150,000 iterations. In Figure \ref{cplus}, we present the traceplot for the number of clusters in each iteration of the MCMC. To summarise the results from the posterior draws obtained for the cluster membership of the networks, we use the R package \texttt{GreedyEPL} \citep{rastelli2018optimal} that gives a final, optimal partition of the observations. The final partition obtained from the posterior draws of the SFM model suggests the presence of 8 clusters in the data, as can also be seen from the traceplot in Figure \ref{cplus}. This is despite the high sparsity imposed through the hyperprior. However, there are several clusters containing only few observations (between 5 to 14 observations) and only three clusters contain more than 20 observations each, which is not appealing for certain inferences. In particular, one of the main innovations of our model is the ability to infer a cluster specific network representative; however, inferring a network representative for clusters that contain only few observations is not necessarily meaningful.}

\textcolor{black}{To compare the final partitions obtained under the finite mixture model with $C=3$ and the SFM extension, we use a set similarity metric, specifically the Jaccard similarity, which is defined as the size of the intersection divided by the size of the union of the sets, ranging from 0 (no similarity) to 1 (perfect similarity). In Figure \ref{part_jacc} we visualise the pairwise comparisons of the partitions obtained under our finite mixture model with $C=3$ clusters and the SFM model extension of our model using the Jaccard similarity metric. We notice that the network observations allocated to the three clusters of the finite model, are spread out among the clusters of the SFM extension, with only cluster 2 of the finite model being more similar to cluster 2 of the SFM model.}

\textcolor{black}{To account for the cluster configurations obtained across all the iterations of the MCMC, we visualise the proportion of times that each individual is allocated to one of the identified clusters for the SFM, similarly to the visualisation obtained for the finite mixture model in Figure \ref{memberships} (right). Under the SFM extension of our model, Figure \ref{prop_sfm} shows a greater variability in the allocation of networks to clusters, without a particularly dominant cluster of networks, compared to the finite mixture model in Figure \ref{memberships} (right).
\\
\begin{figure}[ht!]
\centering
\includegraphics[height=2.1in]{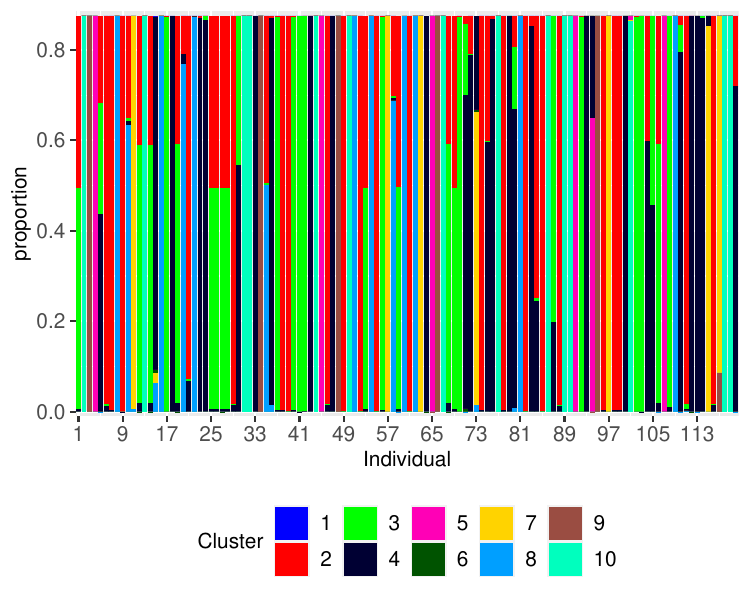}
\vspace{-3mm}
\caption{Proportion of times (y axis) an individual's network (x axis) is allocated to each cluster inferred by the SFM model, after a burn-in of 150,000 iterations.}\label{prop_sfm}
\end{figure}}

\textcolor{black}{As noted in Section \ref{sec53}, the inference for the number of clusters with the SFM model is highly influenced by the hyperprior of $e_0$ specified. This also holds for infinite Dirichlet process mixture models, where the posterior distribution of the number of clusters is highly influenced by the precision parameter $\alpha$ of the Dirichlet process, as stated in \cite{fruhwirth2019here}. Thus, different hyperprior settings could potentially lead to different conclusions about the number of clusters in the data, with some clusters being hard to interpret in the context of our application, which is a fundamental objective of our modelling framework. We adopt the view that the number of clusters should in part be informed by the applied questions of interest.}

\textcolor{black}{Lastly, we formulate our model to allow inferences for matrices of false positive $P$ and false negative $Q$ probabilities with an SBM structure as proposed in \cite{le2018estimating} and discussed in Section \ref{sec41}. The issue arising with this assumption for the Tacita data is the absence of a block structure for the network representative of one of the mixture components as the results showed in this section. Our model makes simultaneous inferences for the block membership of the nodes together with the other model parameters within the MCMC. In contrast, \cite{le2018estimating} have a two-stage algorithm to infer the block membership of the nodes under the assumption of a single true network in the population. In our setting, this leads to identifiability issues when making inferences for the block specific false positive and false negative probabilities, as suggested by the results of our MCMC, not presented herein due to length limitations. In light of this, exploring alternative MCMC formulations for making inferences for $P,Q$ matrices presents an interesting direction for future work.
}

\subsection{Connectivity patterns in the brain}\label{sec62}
As described in Section \ref{sec22}, 
this example involves a population of 300 undirected networks corresponding to 10 brain-scans taken for 30 healthy individuals via diffusion magnetic resonance imaging (dMRI). \textcolor{black}{The 300 networks are treated as independent observations, similarly to the studies of \cite{lunagomez} and \cite{arroyo2021inference}.} The nodes of the networks correspond to regions of the brain, and edges denote connections recorded among these regions. Specifically, the network data consist of 200 nodes according to the CC200 atlas \citep{craddock2012whole}. Our goal is to identify a cluster of individuals with brain connectivity patterns that differ compared to the majority of the networks in the population, and characterise these with respect to a model parameterisation. We thus implement our outlier cluster detection algorithm, as discussed in Section \ref{sec44}.

We initialise our algorithm similarly to the initialisation performed for the Tacita application in Section \ref{sec61}. Hence, we determine an initial membership of networks in two different clusters by implementing a k-means algorithm using the R package \texttt{kmed} \cite{kmed}. This is done for three distance matrices that correspond to the Jaccard, the wavelets and the $l_2$ distance metrics. We combine the results 
using majority vote, to obtain the initial cluster membership of each network. \textcolor{black}{We now consider three distance metrics versus the four distance metrics considered for the Tacita application in Section \ref{sec61}, as the prespecified number of clusters is now $C=2$, and we wish to avoid complications caused by ties.} 
We initialise the network representative by generating its edges through independent Bernoulli draws, with probabilities equal to the proportion of times the corresponding edge is observed in the network data. We initialise the block membership of the representative's nodes by SBM estimation on the initial representative using the R package \texttt{blockmodels} \cite{blockmodels}. For the rest of the parameters of the model we consider three different random initialisations, and run the MCMC 
for each initialisation for 1,000,000 iterations.

\begin{figure}[h!]
\centering
\includegraphics[width=0.49\textwidth]{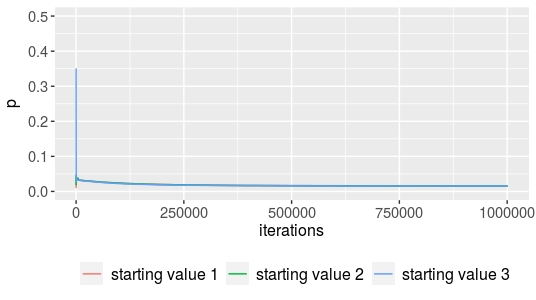}
\hfill 
\includegraphics[width=0.49\textwidth]{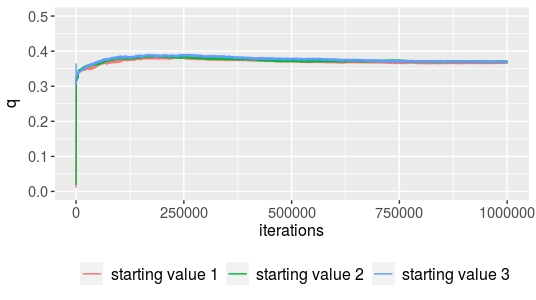}
\caption{Left: Trace plot for false positive probability $p$ for majority cluster for 1,000,000 iterations and three different initialisations. Right: Trace plot for false negative probability $q$ for majority cluster for 1,000,000 iterations and three different initialisations.} \label{trac_maj}
\end{figure}

In Figure \ref{trac_maj} we present the trace plots of the false positive and false negative probabilities for the majority cluster for all iterations under three different initialisations. In the Supplementary Material, Section 3.2, Figure 11 \citep{mantziousupp}, we also present the trace plots of the false positive and false negative probabilities for the outlier cluster. We observe that under the three different initialisations the algorithm converges very quickly to the same region. 
This is encouraging and suggests that a high posterior region has been identified.

\begin{figure}
\centering
\includegraphics[width=0.32\textwidth]{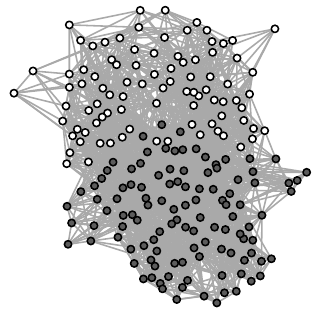}
\hfill 
\includegraphics[width=0.32\textwidth]{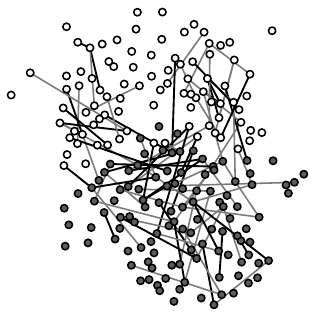}
\hfill 
\includegraphics[width=0.32\textwidth]{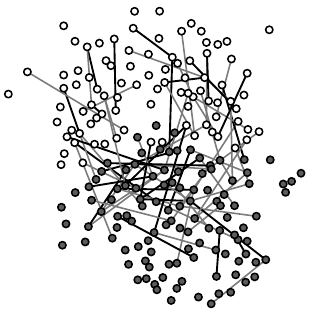}
\vspace{3mm}
\caption{Left: Posterior mode of representative network from $1^{\text{st}}$ initialisation. Middle: Network with not in common edges between posterior modes of representatives from $1^{\text{st}}$ and $2^{\text{nd}}$ initialisation. Right: Network with not in common edges between posterior modes of representatives from $1^{\text{st}}$ and $3^{\text{rd}}$ initialisation. The nodes' colours correspond to the block structure identified under each initialisation.}\label{post_modes_brain}
\end{figure}

We also compare the results from the three initialisations by obtaining the posterior mode of the representative network for the last 50,000 iterations. In Figure \ref{post_modes_brain} (left) we obtain the posterior mode for the network representative under the first initialisation (posterior mass of 0.08). The colours of the nodes correspond to the block membership. 
In Figure \ref{post_modes_brain} (middle), we present the not in common edges between the posterior modes of the first and second initialisation to facilitate comparisons. The black edges correspond to the edges present in the posterior mode of the first initialisation and not present in the posterior mode of the second initialisation, and the light gray edges correspond to the edges present in the posterior mode under the second initialisation and not present in the posterior mode of the first initialisation. In Figure \ref{post_modes_brain} (right), we similarly present the not in common edges between the posterior modes of the first and third initialisation. The posterior mode of the second initialisation has posterior mass of 0.08, while the posterior mode of the third initialisation has posterior mass of 0.16.

There are three interesting findings with respect to the representative inferred under the three different initialisations. First, there is only a small proportion of edges not in common among the posterior modes of the three different initialisations, considering the density of the graphs.  Second, our algorithm infers the same block structure for the three posterior modes of the representative networks. Third, the posterior masses for the posterior mode representatives are small, but is expected due to the high dimensional space of the networks. 
In general the results are very encouraging given the size of the networks considered.

We further observe a smaller posterior mean for the false negative probability in the outlier cluster (0.32) compared to the majority cluster (0.37), suggesting that the edges not observed in the network data of the outlier cluster are more likely to be correct compared to the majority cluster. 
This finding also suggests that the network data in the outlier cluster are sparser compared to the majority cluster. Also, the small posterior means of the false positive probability for both clusters (0.016 for the majority cluster and 0.025 for the outlier cluster) indicate that the edges observed in the network data are likely correct.

\begin{figure}[h!]
\centering
\includegraphics[height=3in,width=5in]{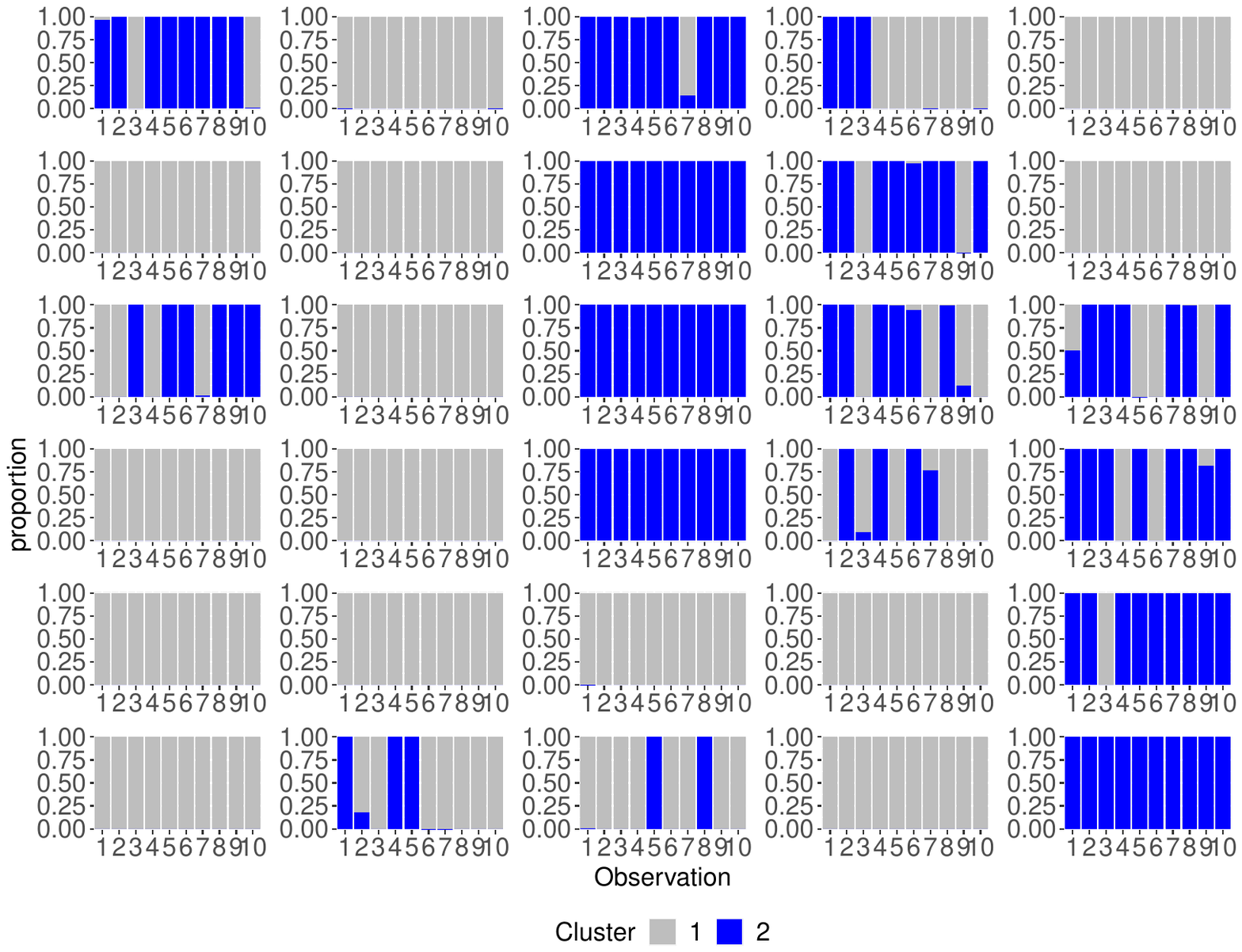}
\vspace{-3mm}
\caption{Proportion of times that each of the 10 brain scans from the 30 individuals is allocated in cluster labelled 1 and/or 2.}\label{memb_brain}
\end{figure}

In Figure \ref{memb_brain}, we present the results for the cluster membership $z$ of the network observations for the first initialisation. Specifically, we calculate the proportion of times that a network observation is allocated to the majority or the outlier cluster, labelled by 1 or 2 respectively, from the last 100,000 iterations. Each subfigure in Figure \ref{memb_brain} shows the cluster allocation of the 10 brain scans obtained for the same individual. We see that our model mostly allocates scans of the same individual to the same cluster.
Thus, our model detects similarities among the brain scans of the same individual, giving credence to our model clustering the networks sensibly. 

This is a common finding with \cite{arroyo2021inference} who performed semi-supervised classification on the brain network population. However, our approach is different to \cite{arroyo2021inference} in two ways. First, we implement an unsupervised method to infer underlying clusters of networks. We only pre-define the number of cluster in the population. Second, the interpretation of the results of our model-based clustering method compared to \cite{arroyo2021inference}
differs significantly, as it reveals a cluster of individuals whose brain connectivity patterns are different to a majority group, and are interpreted through a parametric model.

\textcolor{black}{A common characteristic of human brain networks discussed in the Neuroscience literature, is the exhibition of small-world structures \citep{bassett2006small}. Networks exhibiting small-worldness are characterised by two main network properties, the clustering coefficient, indicating the level of clique formation in the network, and the average shortest path length, indicating the average length of the shortest paths connecting the nodes. It has been identified that small-worldness is associated with individual cognitive performance \citep{liao2017small}.  For example, it has been found that higher intelligence corresponds to shorter path lengths in the brain network \citep{liao2017small}.}

\begin{figure}[h!]
\centering
\includegraphics[height=2in]{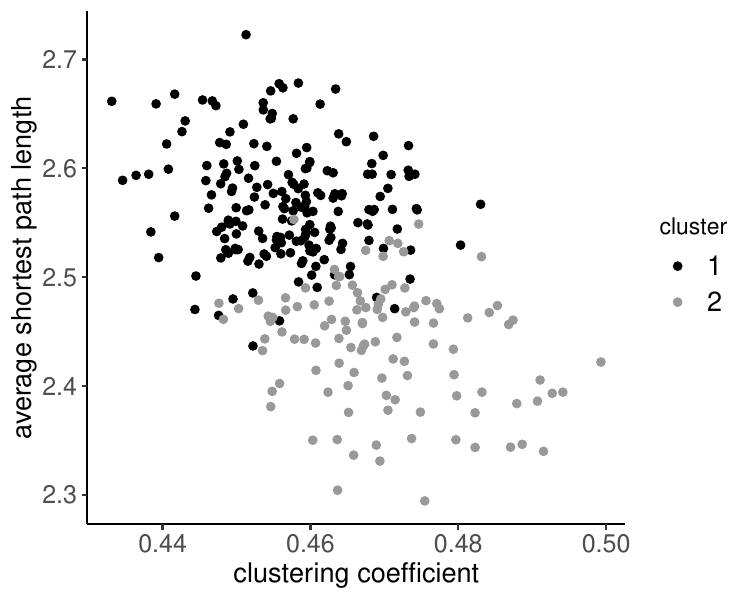}
\vspace{-3mm}
\caption{Cluster configuration of brain networks with respect to small-world properties, with x axis corresponding to clustering coefficient and y axis corresponding to average shortest path length. Colours correspond to cluster membership of the brain networks inferred by our algorithm.}\label{smallworld}
\end{figure}

\textcolor{black}{In light of this, we investigate how our model clusters the individuals with respect to these two network properties. For each individual brain scan, we calculate the clustering coefficient and the average shortest path length, and use a scatterplot to plot these against each other in Figure \ref{smallworld}. 
The shading of the dots correspond to the cluster membership of each brain scan inferred by our model. We observe a clear distinction between the individual brain scans in each cluster with respect to these two network properties, which are both of interest to the Neuroscience community.}

\textcolor{black}{We note here that assuming $C=2$ clusters, i.e. assuming a single outlier cluster, is appropriate given the objective of our analysis, which is to infer an outlier cluster under a single representative network. Thus, we do not implement the SFM extension of our model in this case. In addition, as the results in this section indicate, the assumption of $C=2$ gives meaningful and interpretable results, utilising information from the Neuroscience literature.
}

\section{Discussion}\label{sec7}

In this paper we introduced a mixture model for \textcolor{black}{populations of network} data that allows us to identify clusters of networks in a population.
To achieve this, we formulated a mixture of measurement error models, 
and developed a Bayesian framework that allows us to make inferences for all model parameters jointly.
This framework permits a diverse specification of the network model for the representative networks in the population, determined according to the type of information we want to exploit based on the data.

Through extensive simulations, we observed our method reliably inferred the model parameters and cluster membership for moderate-sized networks, even for regimes with high noise levels. This is an interesting result, as for high noise levels there is great variability in the structure of the simulated network population, making inference a challenging task. The results suggested that our model can perform well for a range of real data applications. 
Simulations also examined the model performance for large network and population sizes. This had not been explored by \cite{signorelli2020model} who also develop a model-based approach for clustering \textcolor{black}{populations of network} data. 
We observed that the absolute errors of the posterior means for the parameters were small, even for large networks and a relatively small sample size. This suggests that our model does not require a large number of observations to make accurate inferences. \textcolor{black}{We also compared our model with the two model-based methods for heterogeneous network populations proposed by \cite{durante2017} and \cite{signorelli2020model} respectively. Our model performed similarly to \cite{durante2017}, and better than \cite{signorelli2020model}. 
Importantly, the parameterisation of our model allows greater interpretability of the results compared to both aforementioned approaches.} 

\textcolor{black}{We also present an extension to our model that incorporates uncertainty in the number of clusters using the approach suggested in \cite{fruhwirth2019here}. The implementation of this approach on simulated data shows that it can accurately recover the true number of clusters as well as the true cluster membership of the simulated networks.}

The clustering performed on the Tacita application revealed three different movement patters of the users. This can be deduced by the network representatives inferred for the clusters, which are primarily characterised by their density. The cluster described by the sparser representative, reveals movements among closely located displays on campus. 
The second and third cluster identified, enclose denser networks in the population, and the representative of each cluster reveals a specific movement pattern of the individuals therein. As the majority of the networks are very sparse it was encouraging to see that our model was able to separate out the denser networks and further distinguish two different clusters among this subset of individuals. \textcolor{black}{In addition, the type of content shown on the displays visited by the users of each cluster has a meaningful interpretation considering the movement patterns indicated by the network representative in each cluster.}

Analysis of the brain network data
led to some interesting findings. First, the results suggest we identified a high posterior region for the false positive and false negative probabilities of each cluster. Second, a similar posterior mode for the network representative has been inferred under three different initialisations. These are especially encouraging given the high dimensional space spanned by 200-node network representatives. \cite{lunagomez} who also obtain a network representative in terms of a Fr\'{e}chet mean for the same brain network population resorted to divide and conquer methods to be able to make inferences, which was not required here. \textcolor{black}{Another interesting finding is that our model partitions the brain networks with respect to two network properties of particular interest to the Neuroscience community.}

Our model could potentially incorporate covariates at the node or edge level to inform the inference. In some network applications, it is common to have additional information about the nodes or the edges of the network.
For example, the Tacita mobile application 
also records the type of content shown by the display at the time visited by the user. The incorporation of this additional information 
could potentially lead to interesting additional findings. 
It would be also interesting to perform a follow up analysis on data recorded by the Tacita application after the emergence of the COVID-19 pandemic. This would allow us to investigate whether there is a change in the movement patterns of users before and after the pandemic.

\textcolor{black}{In our analysis, the brain network data have been studied as independent network observations, similar to the studies of \cite{lunagomez} and \cite{arroyo2021inference}. However, in this data set, multiple brain measurements of the same individual are included, and thus the assumption of independence might not be satisfied. The impact of the independence assumption for \textcolor{black}{populations of network} observations when dependencies between observations exist has not been studied in the network literature, due to a lack of existing methods for modelling multiple dependent network data. Thus, an interesting direction for future work would be to modify our model in order to capture dependence between network observations.}

An interesting result from both the Tacita and the brain networks application, is the high false negative probabilities inferred for the clusters, attributed to the networks' sparsity. Network sparsity is a common issue in many real world network applications. One way to deal with the sparsity would be to consider shrinkage priors 
e.g. formulating the Horseshoe priors \citep{carvalho2009handling}. Another way to account for network sparsity is to assume the networks are partially observed. In the literature, partially observed networks have been considered under two different perspectives. The coarsening approach focuses on incorporating the coarsening mechanism, that allows us to only partially observe the networks, in the model, and efficiently impute the partially observed data thereafter \citep{heitjan1991ignorability,handcock2010modeling,heitjan1990inference,kim2012imputation}. Another approach focuses on the missingness of certain edges and performs edge prediction \citep{koskinen2013bayesian,marchette2015utilizing,zhao2017link,airoldi2013estimating}. Under the first approach, one way we could incorporate the coarsening mechanism is through the assumption of a sampling design, while under the second approach we could have a two-stage method which would first involve performing link prediction, and second performing inference. All the aforementioned approaches require significant modifications of our model and present interesting avenues for future research.  

\textcolor{black}{A key challenge arising with the analysis of network data, and especially with data sets consisting of \textcolor{black}{multiple network} observations, is the development of methods with good scaling properties as the number of networks' nodes and sample size increase. In our study, we were able to explore the performance of our model for networks with up to 100 nodes and sample sizes of 315 network observations, as well as apply our model and get meaningful results for a Neuroscience application involving networks with 200 nodes and a sample size of 300 observations. The network and sample sizes considered in our study are larger than those commonly considered in the network literature, and specifically in the analysis of \cite{durante2017} and \cite{signorelli2020model}. However, there is increasing availability of massive networks for which MCMC approaches like ours are not necessarily practical to implement with respect to the computational cost. The inference of network representatives through MCMC approaches for populations of networks with very large node sets can become very challenging. 
In such cases, other methods, such as Variational Bayes, could constitute alternatives for making inferences 
}

\textcolor{black}{An additional challenge not considered here is when a population of networks contain both small clusters, together with some much larger clusters. In general, classification techniques are known to struggle when class imbalance is large (i.e. when one class dominates the sample), and so similar challenges could occur in this setting as a result. In particular, correctly inferring cluster membership of small clusters, as well as associated properties of the cluster representatives, could be compromised when the vast majority of networks belong to a single cluster. Addressing this area is an important direction for future research. However, it is encouraging to see that in the brain networks application, small networks clusters are able to be inferred, which indicates the potential of our model to deal with a modest amount of imbalance in cluster sizes.}

Our paper contributes to the growing literature on clustering complex types of data. Examples of such studies are the study of \cite{lu2014analysis} who focus on identifying clusters of juggling cycles using Functional Principal Component Analysis (FPCA) on diverse data objects, \cite{song2007clustering} who perform model-based clustering on time-dependent gene expression data using Functional Data Analysis, and  \cite{shen2013shape} who propose an agglomerative clustering approach for clustering shape data according to their structure. In summary, the flexibility of our modelling framework has been shown to address diverse applied research questions with the potential to be widely applicable to many fields.

\section*{Acknowledgment}
The authors would like to thank Mateusz Mikusz and Petteri Nurmi for sharing the Tacita data set and for the constructive discussions about the data.

\section*{Supplement}

\textbf{Supplement to "Bayesian model-based clustering for populations of network data"}

\noindent Supplement contains additional details and results for the model, the simulations and the data applications.

\noindent\textbf{Code for Bayesian model-based clustering for populations of network data}

\noindent This file contains main code for MCMC algorithms, as well as code for simulation experiments and data analysis.

\bibliography{references}

\end{document}


\maketitle
In this document we provide supplementary material to the article "Bayesian model-based clustering for multiple network data". In Section 1 we provide more details about the MCMC scheme introduced in Section 4 of the main article. In Section 2 we provide additional results from the simulation studies performed in Section 5 of the main article. In Section 3, we provide additional results for the real data applications discussed in Section 6 of the main article. In Section 4 we provide details of the MCMC algorithm implemented.

\section{Additional details for the MCMC scheme}

In this Section we provide the full conditional posteriors for the model parameters that are updated through a Gibbs sampler, as discussed in Section 4.3 of the main article.

The full conditional posterior for the probability of a network to belong to cluster $\boldsymbol{\tau}$ is given by

\begin{displaymath}
P(\boldsymbol{\tau}|\boldsymbol{A_{\mathcal{G}^{*}}},\boldsymbol{p},\boldsymbol{q},\boldsymbol{z},\boldsymbol{w},\boldsymbol{b},\boldsymbol{\theta},A_{\mathcal{G}_{1}},\ldots,A_{\mathcal{G}_{N}})\propto 
 P(\boldsymbol{z}|\boldsymbol{\tau}) \cdot P(\boldsymbol{\tau}|\boldsymbol{\psi})
\end{displaymath}
where $P(\boldsymbol{z}|\boldsymbol{\tau})=\text{Multinomial}(1;\tau_{1},\ldots,\tau_{C})$ and $ P(\boldsymbol{\tau}|\boldsymbol{\psi})=\text{Dirichlet}(\boldsymbol{\psi})$, as specified in Section 4.2 of the main article. Thus, we have

\begin{align}
\begin{gathered}
P(\boldsymbol{\tau}|\boldsymbol{A_{\mathcal{G}^{*}}},\boldsymbol{p},\boldsymbol{q},\boldsymbol{z},\boldsymbol{w},\boldsymbol{b},\boldsymbol{\theta},A_{\mathcal{G}_{1}},\ldots,A_{\mathcal{G}_{N}})\propto
\prod_{j=1}^{N} \tau_{z_{j}} \cdot \Gamma (\psi C) \cdot \Gamma (\psi)^{-C} \cdot \prod_{c=1}^{C}\tau_{c}^{\psi -1}  \\ \propto \Gamma (\psi C) \cdot \Gamma (\psi)^{-C} \cdot \tau_{z_{1}} \cdots \tau_{z_{N}} \cdot \tau_{1}^{\psi -1} \cdots \tau_{C}^{\psi -1} \\ \propto \Gamma (\psi C) \cdot \Gamma (\psi)^{-C} \cdot \tau_{1}^{\eta_{1}} \cdots \tau_{C}^{\eta_{C}} \cdot \tau_{1}^{\psi -1} \cdots \tau_{C}^{\psi -1} \propto \tau_{1}^{\eta_{1}+\psi-1}\cdots \tau_{C}^{\eta_{C}+\psi-1} 
\end{gathered}
\end{align}
where $\eta_{c}=\sum_{j=1}^{N} 1_{c} (z_{j})$, $c=1,\ldots,C$, denotes the number of network data that belong to cluster $c$. Thence we obtain,
\begin{displaymath}
P(\boldsymbol{\tau}|\boldsymbol{A_{\mathcal{G}^{*}}},\boldsymbol{p},\boldsymbol{q},\boldsymbol{z},\boldsymbol{w},\boldsymbol{b},\boldsymbol{\theta},A_{\mathcal{G}_{1}},\ldots,A_{\mathcal{G}_{N}})= \text{Dirichlet}(\psi + \eta_{1}, \ldots ,\psi + \eta_{C}).
\end{displaymath}

The derivation of the full conditional posterior for the vector of the nodes' block-membership probabilities $\boldsymbol{w_{c}}$ for cluster $c$, is similar to the derivation of the the full conditional posterior for $\boldsymbol{\tau}$, as already described above, thus we have
\begin{gather*}
P(\boldsymbol{w_{c}}|A_{\mathcal{G}^{*}_{c}},p_{c},q_{c},\boldsymbol{z},\boldsymbol{\tau},\boldsymbol{b_{c}},\boldsymbol{\theta_{c}},A_{\mathcal{G}_{1}},\ldots,A_{\mathcal{G}_{N}}) \propto P(\boldsymbol{b_{c}}|\boldsymbol{w_{c}}) \cdot P(\boldsymbol{w_{c}}|\boldsymbol{\chi}) 
\end{gather*}
where $P(\boldsymbol{b_{c}}|\boldsymbol{w_{c}})=\text{Multinomial}(\boldsymbol{w_{c}})$ and $P(\boldsymbol{w_{c}}|\boldsymbol{\chi})=\text{Dirichlet}(\boldsymbol{\chi})$, as specified in Section 4.1 of the main article. Hence we obtain

\begin{gather*}
P(\boldsymbol{w_{c}}|A_{\mathcal{G}^{*}_{c}},p_{c},q_{c},\boldsymbol{z},\boldsymbol{\tau},\boldsymbol{b_{c}},\boldsymbol{\theta_{c}},A_{\mathcal{G}_{1}},\ldots,A_{\mathcal{G}_{N}}) \propto \prod_{i=1}^{n} w_{c,b_{i}}\cdot \Gamma (\chi K) \cdot \Gamma (\chi)^{-K} \cdot  \prod_{k=1}^{K} w_{c,k}^{\chi -1} \\ \propto \Gamma (\chi K) \cdot \Gamma (\chi)^{-K} \cdot w_{c,b_{1}}\cdots w_{c,b_{n}} \cdot w_{c,1}^{\chi-1} \cdots w_{c,K}^{\chi-1} \\ \propto \Gamma (\chi K) \cdot \Gamma (\chi)^{-K} \cdot w_{c,1}^{h_{1}} \cdots w_{c,K}^{h_{K}} \cdot w_{c,1}^{\chi-1} \cdots w_{c,K}^{\chi-1} \propto w_{c,1}^{(h_{1}+\chi)-1} \cdots w_{c,K}^{(h_{K}+\chi)-1}.
\end{gather*}

\noindent where $h_{k}$ denotes the number of the nodes that belong to block k. Thus the full conditional posterior for $\boldsymbol{w_{c}}$ is
\begin{displaymath}
P(\boldsymbol{w_{c}}|A_{\mathcal{G}^{*}_{c}},p_{c},q_{c},\boldsymbol{z},\boldsymbol{\tau},\boldsymbol{b_{c}},\boldsymbol{\theta_{c}},A_{\mathcal{G}_{1}},\ldots,A_{\mathcal{G}_{N}})=\text{Dirichlet}(\chi + h_{1}, \ldots ,\chi + h_{K}).
\end{displaymath}

The full conditional posterior for the vector of the block-specific probabilities of an edge occurrence, $\boldsymbol{\theta_{c}}$, for the network representative of cluster $c$ is
\begin{gather*}
P(\boldsymbol{\theta_{c}}|A_{\mathcal{G}^{*}_{c}},p_{c},q_{c},\boldsymbol{z},\boldsymbol{\tau},\boldsymbol{b_{c}},\boldsymbol{w_{c}},A_{\mathcal{G}_{1}},\ldots,A_{\mathcal{G}_{N}}) \propto P(A_{\mathcal{G}^{*}_{c}}|\boldsymbol{w_{c}},\boldsymbol{b_{c}},\boldsymbol{\theta_{c}}) \cdot P(\boldsymbol{\theta_{c}}|\boldsymbol{\epsilon},\boldsymbol{\zeta})
\end{gather*}
\noindent where $P(A_{\mathcal{G}^{*}_{c}}|\boldsymbol{w_{c}},\boldsymbol{b_{c}},\boldsymbol{\theta_{c}})=\text{SBM}(\boldsymbol{w_{c}},\boldsymbol{b_{c}},\boldsymbol{\theta_{c}})$ and $P(\boldsymbol{\theta_{c}}|\boldsymbol{\epsilon},\boldsymbol{\zeta})=\text{Beta}(\boldsymbol{\epsilon},\boldsymbol{\zeta})$, as specified in Section 4.1 of the main article. Thus,

\begin{gather*}
P(\boldsymbol{\theta_{c}}|A_{\mathcal{G}^{*}_{c}},p_{c},q_{c},\boldsymbol{z},\boldsymbol{\tau},\boldsymbol{b_{c}},\boldsymbol{w_{c}},A_{\mathcal{G}_{1}},\ldots,A_{\mathcal{G}_{N}}) \\ \propto \prod_{(i,j):i<j} \theta_{c,b_{i}b_{j}}^{A_{\mathcal{G}^{*}_{c}}(i,j)} (1-\theta_{c,b_{c,i}b_{c,j}})^{1-A_{\mathcal{G}^{*}_{c}}(i,j)} \cdot \prod_{k=1}^{K}\prod_{l=1}^{K} \theta_{c,kl}^{\epsilon_{kl}-1}(1-\theta_{c,kl})^{\zeta_{kl}-1} \\ \propto \prod_{k=1}^{K}\prod_{l=1}^{K}\theta_{c,kl}^{A_{\mathcal{G}^{*}_{c}}[kl]} (1-\theta_{c,kl})^{n_{c,kl}-A_{\mathcal{G}^{*}_{c}}[kl]} \theta_{c,kl}^{\epsilon_{kl}-1}(1-\theta_{c,kl})^{\zeta_{kl}-1} \\ \propto \prod_{k=1}^{K}\prod_{l=1}^{K}\theta_{c,kl}^{A_{\mathcal{G}^{*}_{c}}[kl]+\epsilon_{kl}-1} (1-\theta_{c,kl})^{n_{c,kl}-A_{\mathcal{G}^{*}_{c}}[kl]+\zeta_{kl}-1},
\end{gather*}

\noindent where $A_{\mathcal{G}^{*}_{c}}[kl]=\sum_{(i,j):b_{c,i}=k,b_{c,j}=l} A_{\mathcal{G}^{*}_{c}}(i,j)$ represents the sum of the entries for the pairs of nodes of the network representative for cluster $c$ that have block membership $k,l$ respectively, and $n_{c,kl}=\sum_{(i,j):i \neq j} \mathbb{I}(b_{c,i}=k,b_{c,j}=l)$ is the number of the pairs of nodes of the representative of cluster $c$ that have membership $k,l$ accordingly. Hence we obtain
\begin{displaymath}
P(\boldsymbol{\theta_{c}}|A_{\mathcal{G}^{*}_{c}},p_{c},q_{c},\boldsymbol{z},\boldsymbol{\tau},\boldsymbol{b_{c}},\boldsymbol{w_{c}},A_{\mathcal{G}_{1}},\ldots,A_{\mathcal{G}_{N}})=\text{Beta}(A_{\mathcal{G}^{*}_{c}}[kl]+\epsilon_{kl},\zeta_{kl}+n_{c,kl}-A_{\mathcal{G}^{*}_{c}}[kl]).
\end{displaymath}

\section{Additional details for the Simulation Studies}

\subsection{Additional details for simulation study for moderate-sized networks}

In this Section we provide the results for the simulation regimes presented in Section 5.1 of the main article (Table 1 in the main article). Specifically, in Tables 1-13 we present the posterior means and credible intervals for the false positive probabilities $p_c$, false negative probabilities $q_c$, and block specific edge probabilities $\boldsymbol{\theta_c}$. In Tables 14-15 we present the posterior means for the probability of a node to belong to a block $\boldsymbol{w_c}$.

In addition, Tables 16-17 show the proportion of times that the Hamming distance between the true representatives and the posterior representatives is less than or equal to 1, 5 and 10 respectively, for each simulation regime. Table 18 shows the mean clustering entropy and mean clustering purity calculated for each simulation regime, as discussed in Section 5.1 of the main article.

\begin{table}[ht!]
\centering
\begin{tabular}{cccccc}
\textbf{}&& \multicolumn{2}{c}{\textbf{\textbf{$\text{SBM}_1$}}}&\multicolumn{2}{c}{\textbf{$\text{SBM}_2$}}\\ \hline \hline
$\boldsymbol{p_c}$         & \multicolumn{1}{c}{$\boldsymbol{q_c}$} & \multicolumn{1}{c}{\textbf{\begin{tabular}[c]{@{}c@{}}posterior\\ mean of $\boldsymbol{p_c}$ \end{tabular}}} & \multicolumn{1}{c}{\textbf{\begin{tabular}[c]{@{}c@{}}posterior\\ mean of $\boldsymbol{q_c}$ \end{tabular}}} & \multicolumn{1}{c}{\textbf{\begin{tabular}[c]{@{}c@{}}posterior\\ mean of $\boldsymbol{p_c}$ \end{tabular}}} & \multicolumn{1}{c}{\textbf{\begin{tabular}[c]{@{}c@{}}posterior\\ mean of $\boldsymbol{q_c}$ \end{tabular}}} \\ \hline \hline
\multirow{2}{*}{0.1} & 0.2& (0.09,0.10,0.11) & (0.20,0.21,0.21) & (0.10,0.10,0.10)& (0.21,0.19,0.19)\\
& 0.3& (0.10,0.10,0.10) & (0.31,0.29,0.30)& (0.11,0.10,0.10)& (0.30,0.31,0.30)\\ \hline
\multirow{2}{*}{0.2} & 0.1& (0.19,0.19,0.20)& (0.10,0.10,0.11) & (0.20,0.20,0.20)& (0.10,0.09,0.10) \\
& 0.3& (0.20,0.20,0.19)& (0.29,0.29,0.30)& (0.20,0.20,0.19) & (0.30,0.32,0.30)\\ \hline
\multirow{2}{*}{0.3} & 0.1& (0.31,0.30,0.31)& (0.10,0.10,0.10)& (0.31,0.30,0.31)& (0.10,0.10,0.10)\\
& 0.2& (0.30,0.29,0.30)& (0.20,0.19,0.21)& (0.30,0.30,0.28) & (0.20,0.19,0.20)  \\ \hline            
\end{tabular}
\caption{Posterior means for false positive probabilities $p_c$ and false negative probabilities $q_c$, for $c\in\{1,2,3\}$.}\label{post_mean_p_q_new}
\end{table}

\begin{table}[ht!]
\centering
\begin{tabular}{ccccc}
\textbf{}&& \multicolumn{3}{c}{\textbf{\textbf{$\text{SBM}_1$}}}\\ \hline \hline
$\boldsymbol{p_c}$         & \multicolumn{1}{c}{$\boldsymbol{q_c}$} & \multicolumn{1}{c}{\textbf{\begin{tabular}[c]{@{}c@{}}credible\\ interval for $\boldsymbol{p_1}$ \end{tabular}}} & \multicolumn{1}{c}{\textbf{\begin{tabular}[c]{@{}c@{}}credible\\ interval for $\boldsymbol{p_2}$ \end{tabular}}} & \multicolumn{1}{c}{\textbf{\begin{tabular}[c]{@{}c@{}}credible\\ interval for $\boldsymbol{p_3}$ \end{tabular}}} \\ \hline \hline
\multirow{2}{*}{0.1} & 0.2& (0.09,0.10) & (0.09,0.11) &  (0.10,0.11)\\
& 0.3& (0.09,0.10) & (0.10,0.11)& (0.09,0.11)\\ \hline
\multirow{2}{*}{0.2} & 0.1& (0.18,0.20)& (0.18,0.20) & (0.19,0.21) \\
& 0.3& (0.19,0.21) & (0.19,0.21) & (0.18,0.20) \\ \hline
\multirow{2}{*}{0.3} & 0.1& (0.29,0.32) & (0.29,0.31) & (0.30,0.32)\\
& 0.2& (0.29,0.32) & (0.28,0.31) & (0.29,0.31)  \\ \hline            
\end{tabular}
\caption{95 \% credible intervals for false positive probabilities $p_c$, for $c\in\{1,2,3\}$, under SBM 1.}\label{cred_p_sbm1_new}
\end{table} 

\begin{table}[ht!]
\centering
\begin{tabular}{ccccc}
\textbf{}&& \multicolumn{3}{c}{\textbf{\textbf{$\text{SBM}_2$}}}\\ \hline \hline
$\boldsymbol{p_c}$         & \multicolumn{1}{c}{$\boldsymbol{q_c}$} & \multicolumn{1}{c}{\textbf{\begin{tabular}[c]{@{}c@{}}credible\\ interval for $\boldsymbol{p_1}$ \end{tabular}}} & \multicolumn{1}{c}{\textbf{\begin{tabular}[c]{@{}c@{}}credible\\ interval for $\boldsymbol{p_2}$ \end{tabular}}} & \multicolumn{1}{c}{\textbf{\begin{tabular}[c]{@{}c@{}}credible\\ interval for $\boldsymbol{p_3}$ \end{tabular}}} \\ \hline \hline
\multirow{2}{*}{0.1} & 0.2&  (0.09,0.11) & (0.09,0.11) & (0.09,0.11)\\
& 0.3& (0.10,0.11) & (0.09,0.10) & (0.09,0.11)\\ \hline
\multirow{2}{*}{0.2} & 0.1& (0.19,0.21) & (0.19,0.21) & (0.19,0.21) \\
& 0.3& (0.19,0.21) & (0.19,0.21) & (0.18,0.20)\\ \hline
\multirow{2}{*}{0.3} & 0.1& (0.30,0.32) & (0.29,0.31) & (0.30,0.32)\\
& 0.2& (0.29,0.31) & (0.29,0.31) & (0.27,0.30) \\ \hline            
\end{tabular}
\caption{95 \% credible intervals for false positive probabilities $p_c$, for $c\in\{1,2,3\}$, under SBM 2.}\label{cred_p_sbm2_new}
\end{table} 

\begin{table}[ht!]
\centering
\begin{tabular}{ccccc}
\textbf{}&& \multicolumn{3}{c}{\textbf{\textbf{$\text{SBM}_1$}}}\\ \hline \hline
$\boldsymbol{p_c}$         & \multicolumn{1}{c}{$\boldsymbol{q_c}$} & \multicolumn{1}{c}{\textbf{\begin{tabular}[c]{@{}c@{}}credible\\ interval for $\boldsymbol{q_1}$ \end{tabular}}} & \multicolumn{1}{c}{\textbf{\begin{tabular}[c]{@{}c@{}}credible\\ interval for $\boldsymbol{q_2}$ \end{tabular}}} & \multicolumn{1}{c}{\textbf{\begin{tabular}[c]{@{}c@{}}credible\\ interval for $\boldsymbol{q_3}$ \end{tabular}}} \\ \hline \hline
\multirow{2}{*}{0.1} & 0.2& (0.19,0.21) & (0.20,0.21) & (0.19,0.22)\\
& 0.3& (0.30,0.32) & (0.28,0.30) & (0.28,0.31)\\ \hline
\multirow{2}{*}{0.2} & 0.1& (0.09,0.10) & (0.09,0.11) & (0.10,0.12) \\
& 0.3& (0.28,0.30) & (0.28,0.31) & (0.28,0.31) \\ \hline
\multirow{2}{*}{0.3} & 0.1& (0.09,0.11) & (0.09,0.11) & (0.10,0.11) )\\
& 0.2& (0.19,0.21) & (0.18,0.20) & (0.20,0.22) \\ \hline            
\end{tabular}
\caption{95 \% credible intervals for false negative probabilities $q_c$, for $c\in\{1,2,3\}$, under SBM 1.}\label{cred_q_sbm1_new}
\end{table} 
\clearpage
\newpage
\begin{table}[ht!]
\centering
\begin{tabular}{ccccc}
\textbf{}&& \multicolumn{3}{c}{\textbf{\textbf{$\text{SBM}_2$}}}\\ \hline \hline
$\boldsymbol{p_c}$         & \multicolumn{1}{c}{$\boldsymbol{q_c}$} & \multicolumn{1}{c}{\textbf{\begin{tabular}[c]{@{}c@{}}credible\\ interval for $\boldsymbol{q_1}$ \end{tabular}}} & \multicolumn{1}{c}{\textbf{\begin{tabular}[c]{@{}c@{}}credible\\ interval for $\boldsymbol{q_2}$ \end{tabular}}} & \multicolumn{1}{c}{\textbf{\begin{tabular}[c]{@{}c@{}}credible\\ interval for $\boldsymbol{q_3}$ \end{tabular}}} \\ \hline \hline
\multirow{2}{*}{0.1} & 0.2&  (0.20,0.22) & (0.18,0.20) & (0.18,0.20)\\
& 0.3& (0.29,0.31) & (0.30,0.32) & (0.29,0.32)\\ \hline
\multirow{2}{*}{0.2} & 0.1& (0.09,0.11) & (0.09,0.10) & (0.10,0.11) \\
& 0.3&  (0.29,0.31) & (0.30,0.33) & (0.29,0.31)\\ \hline
\multirow{2}{*}{0.3} & 0.1& (0.09,0.11) & (0.09,0.11) & (0.09,0.11)\\
& 0.2&  (0.19,0.21) & (0.18,0.20) & (0.19,0.21) \\ \hline            
\end{tabular}
\caption{95 \% credible intervals for false negative probabilities $q_c$, for $c\in\{1,2,3\}$, under SBM 2.}\label{cred_q_sbm2_new}
\end{table} 

\begin{table}[ht!]
\centering
\begin{tabular}{ccccc}
\textbf{}&& \multicolumn{3}{c}{\textbf{\textbf{$\text{SBM}_1$}}}\\ \hline \hline
$\boldsymbol{p_c}$         & \multicolumn{1}{c}{$\boldsymbol{q_c}$} & \multicolumn{1}{c}{\textbf{\begin{tabular}[c]{@{}c@{}}posterior means of \\ $\boldsymbol{(\theta_{11}^{(1)},\theta_{12}^{(1)},\theta_{22}^{(1)})}$ \end{tabular}}} & \multicolumn{1}{c}{\textbf{\begin{tabular}[c]{@{}c@{}}posterior means of \\ $\boldsymbol{(\theta_{11}^{(2)},\theta_{12}^{(2)},\theta_{22}^{(2)})}$ \end{tabular}}} & \multicolumn{1}{c}{\textbf{\begin{tabular}[c]{@{}c@{}}posterior means of \\ $\boldsymbol{(\theta_{11}^{(3)},\theta_{12}^{(3)},\theta_{22}^{(3)})}$ \end{tabular}}} \\ \hline \hline
\multirow{2}{*}{0.1} & 0.2& (0.84,0.22,0.90) & (0.83,0.21,0.82) & (0.79,0.18,0.74) \\
& 0.3& (0.84,0.22,0.90) & (0.83,0.21,0.82) & (0.79,0.18,0.74)\\ \hline
\multirow{2}{*}{0.2} & 0.1& (0.84,0.22,0.90)  & (0.83,0.21,0.82) & (0.79,0.18,0.74)  \\
& 0.3& (0.84,0.22,0.90)  & (0.83,0.21,0.82)  & (0.74,0.18,0.79) \\ \hline
\multirow{2}{*}{0.3} & 0.1& (0.84,0.22,0.90)  & (0.82,0.21,0.83) & (0.74,0.18,0.79) \\
& 0.2& (0.90,0.22,0.84)  & (0.82,0.21,0.83)  & (0.79,0.18,0.74)  \\ \hline            
\end{tabular}
\caption{Posterior means for block specific edge probabilities $\boldsymbol{\theta_c}$, for $c\in\{1,2,3\}$, under SBM 1.}\label{post_mean_theta_sbm1_new}
\end{table} 

\begin{table}[ht!]
\centering
\begin{tabular}{ccccc}
\textbf{}&& \multicolumn{3}{c}{\textbf{\textbf{$\text{SBM}_2$}}}\\ \hline \hline
$\boldsymbol{p_c}$         & \multicolumn{1}{c}{$\boldsymbol{q_c}$} & \multicolumn{1}{c}{\textbf{\begin{tabular}[c]{@{}c@{}}posterior means of \\ $\boldsymbol{(\theta_{11}^{(1)},\theta_{12}^{(1)},\theta_{22}^{(1)})}$ \end{tabular}}} & \multicolumn{1}{c}{\textbf{\begin{tabular}[c]{@{}c@{}}posterior means of \\ $\boldsymbol{(\theta_{11}^{(2)},\theta_{12}^{(2)},\theta_{22}^{(2)})}$ \end{tabular}}} & \multicolumn{1}{c}{\textbf{\begin{tabular}[c]{@{}c@{}}posterior means of \\ $\boldsymbol{(\theta_{11}^{(3)},\theta_{12}^{(3)},\theta_{22}^{(3)})}$ \end{tabular}}} \\ \hline \hline
\multirow{2}{*}{0.1} & 0.2& (0.68,0.02,0.78) & (0.60,0.05,0.71) & (0.78,0.03,0.83)  \\
& 0.3& (0.68,0.02,0.78) & (0.60,0.05,0.71)  & (0.78,0.03,0.83)\\ \hline
\multirow{2}{*}{0.2} & 0.1& (0.68,0.02,0.78)  & (0.60,0.05,0.71)  & (0.78,0.03,0.83) \\
& 0.3& (0.68,0.02,0.78) &  (0.71,0.05,0.60) & (0.83,0.03,0.78) \\ \hline
\multirow{2}{*}{0.3} & 0.1& (0.68,0.02,0.78)  & (0.60,0.05,0.71)  & (0.78,0.03,0.83) \\
& 0.2& (0.68,0.02,0.78)  & (0.71,0.05,0.60)  & (0.83,0.03,0.78)  \\ \hline            
\end{tabular}
\caption{Posterior means for block specific edge probabilities $\boldsymbol{\theta_c}$, for $c\in\{1,2,3\}$, under SBM 2.}\label{post_mean_theta_sbm2_new}
\end{table} 

\begin{table}[ht!]
\centering
\begin{tabular}{ccccc}
\textbf{}&& \multicolumn{3}{c}{\textbf{\textbf{$\text{SBM}_1$}}}\\ \hline \hline
$\boldsymbol{p_c}$         & \multicolumn{1}{c}{$\boldsymbol{q_c}$} & \multicolumn{1}{c}{\textbf{\begin{tabular}[c]{@{}c@{}}credible interval\\ for $\boldsymbol{\theta_{11}^{(1)}}$ \end{tabular}}} & \multicolumn{1}{c}{\textbf{\begin{tabular}[c]{@{}c@{}}credible interval\\ for $\boldsymbol{\theta_{12}^{(1)}}$ \end{tabular}}} & \multicolumn{1}{c}{\textbf{\begin{tabular}[c]{@{}c@{}}credible interval\\ for $\boldsymbol{\theta_{22}^{(1)}}$ \end{tabular}}} \\ \hline \hline
\multirow{2}{*}{0.1} & 0.2& (0.73,0.94) & (0.15,0.30) & (0.82,0.97) \\
& 0.3& (0.73,0.94) & (0.15,0.30) & (0.82,0.97) \\ \hline
\multirow{2}{*}{0.2} & 0.1& (0.73,0.94) & (0.15,0.30) & (0.82,0.97)  \\
& 0.3& (0.73,0.94) & (0.15,0.30) & (0.82,0.97)  \\ \hline
\multirow{2}{*}{0.3} & 0.1& (0.73,0.94) & (0.15,0.30) & (0.82,0.97) \\
& 0.2& (0.82,0.97) & (0.15,0.30) & (0.73,0.94)\\ \hline            
\end{tabular}
\caption{95 \% credible intervals for block specific edge probabilities $\boldsymbol{\theta_1}$ of cluster 1, under SBM 1.}\label{cred_theta1_sbm1_new}
\end{table}

\begin{table}[ht!]
\centering
\begin{tabular}{ccccc}
\textbf{}&& \multicolumn{3}{c}{\textbf{\textbf{$\text{SBM}_2$}}}\\ \hline \hline
$\boldsymbol{p_c}$         & \multicolumn{1}{c}{$\boldsymbol{q_c}$} & \multicolumn{1}{c}{\textbf{\begin{tabular}[c]{@{}c@{}}credible interval\\ for $\boldsymbol{\theta_{11}^{(1)}}$ \end{tabular}}} & \multicolumn{1}{c}{\textbf{\begin{tabular}[c]{@{}c@{}}credible interval\\ for $\boldsymbol{\theta_{12}^{(1)}}$ \end{tabular}}} & \multicolumn{1}{c}{\textbf{\begin{tabular}[c]{@{}c@{}}credible interval\\ for $\boldsymbol{\theta_{22}^{(1)}}$ \end{tabular}}} \\ \hline \hline
\multirow{2}{*}{0.1} & 0.2& (0.59,0.77) & (0.00,0.04) & (0.58,0.96) \\
& 0.3& (0.60,0.77) & (0.00,0.04) & (0.59,0.96)  \\ \hline
\multirow{2}{*}{0.2} & 0.1& (0.60,0.77) & (0.00,0.04) & (0.59,0.96) \\
& 0.3& (0.60,0.77) & (0.00,0.04) & (0.59,0.96)  \\ \hline
\multirow{2}{*}{0.3} & 0.1& (0.60,0.77) & (0.00,0.04) & (0.59,0.96) \\
& 0.2& (0.59,0.77) & (0.00,0.04) & (0.58,0.96)\\ \hline            
\end{tabular}
\caption{95 \% credible intervals for block specific edge probabilities $\boldsymbol{\theta_1}$ of cluster 1, under SBM 2.}\label{cred_theta1_sbm2_new}
\end{table} 

\begin{table}[ht!]
\centering
\begin{tabular}{ccccc}
\textbf{}&& \multicolumn{3}{c}{\textbf{\textbf{$\text{SBM}_1$}}}\\ \hline \hline
$\boldsymbol{p_c}$         & \multicolumn{1}{c}{$\boldsymbol{q_c}$} & \multicolumn{1}{c}{\textbf{\begin{tabular}[c]{@{}c@{}}credible interval\\ for $\boldsymbol{\theta_{11}^{(2)}}$ \end{tabular}}} & \multicolumn{1}{c}{\textbf{\begin{tabular}[c]{@{}c@{}}credible interval\\ for $\boldsymbol{\theta_{12}^{(2)}}$ \end{tabular}}} & \multicolumn{1}{c}{\textbf{\begin{tabular}[c]{@{}c@{}}credible interval\\ for $\boldsymbol{\theta_{22}^{(2)}}$ \end{tabular}}} \\ \hline \hline
\multirow{2}{*}{0.1} & 0.2& (0.73,0.92) & (0.14,0.29) & (0.70,0.92) \\
& 0.3& (0.73,0.92) & (0.14,0.29) & (0.70,0.92) \\ \hline
\multirow{2}{*}{0.2} & 0.1&  (0.73,0.92) & (0.14,0.29) & (0.70,0.92) \\
& 0.3& (0.73,0.92) & (0.14,0.29) & (0.70,0.92)   \\ \hline
\multirow{2}{*}{0.3} & 0.1& (0.70,0.92) & (0.14,0.29) & (0.73,0.92)  \\
& 0.2& (0.70,0.92) & (0.14,0.29) & (0.73,0.92) \\ \hline            
\end{tabular}
\caption{95 \% credible intervals for block specific edge probabilities $\boldsymbol{\theta_2}$ of cluster 2, under SBM 1.}\label{cred_theta2_sbm1_new}
\end{table} 

\begin{table}[ht!]
\centering
\begin{tabular}{ccccc}
\textbf{}&& \multicolumn{3}{c}{\textbf{\textbf{$\text{SBM}_2$}}}\\ \hline \hline
$\boldsymbol{p_c}$         & \multicolumn{1}{c}{$\boldsymbol{q_c}$} & \multicolumn{1}{c}{\textbf{\begin{tabular}[c]{@{}c@{}}credible interval\\ for $\boldsymbol{\theta_{11}^{(2)}}$ \end{tabular}}} & \multicolumn{1}{c}{\textbf{\begin{tabular}[c]{@{}c@{}}credible interval\\ for $\boldsymbol{\theta_{12}^{(2)}}$ \end{tabular}}} & \multicolumn{1}{c}{\textbf{\begin{tabular}[c]{@{}c@{}}credible interval\\ for $\boldsymbol{\theta_{22}^{(2)}}$ \end{tabular}}} \\ \hline \hline
\multirow{2}{*}{0.1} & 0.2& (0.47,0.72) & (0.01,0.09) & (0.58,0.83)  \\
& 0.3& (0.47,0.72) & (0.01,0.09) & (0.58,0.83)\\ \hline
\multirow{2}{*}{0.2} & 0.1&  (0.47,0.72) & (0.01,0.09) & (0.58,0.83)\\
& 0.3&(0.47,0.72) & (0.01,0.09) &  (0.58,0.83) \\ \hline
\multirow{2}{*}{0.3} & 0.1&  (0.47,0.73) & (0.01,0.09) &(0.57,0.83) \\
& 0.2&(0.47,0.72)  & (0.01,0.09) &  (0.58,0.83)\\ \hline            
\end{tabular}
\caption{95 \% credible intervals for block specific edge probabilities $\boldsymbol{\theta_2}$ of cluster 2, under SBM 2.}\label{cred_theta2_sbm2_new}
\end{table} 
\clearpage
\newpage
\begin{table}[ht!]
\centering
\begin{tabular}{ccccc}
\textbf{}&& \multicolumn{3}{c}{\textbf{\textbf{$\text{SBM}_1$}}}\\ \hline \hline
$\boldsymbol{p_c}$         & \multicolumn{1}{c}{$\boldsymbol{q_c}$} & \multicolumn{1}{c}{\textbf{\begin{tabular}[c]{@{}c@{}}credible interval\\ for $\boldsymbol{\theta_{11}^{(3)}}$ \end{tabular}}} & \multicolumn{1}{c}{\textbf{\begin{tabular}[c]{@{}c@{}}credible interval\\ for $\boldsymbol{\theta_{12}^{(3)}}$ \end{tabular}}} & \multicolumn{1}{c}{\textbf{\begin{tabular}[c]{@{}c@{}}credible interval\\ for $\boldsymbol{\theta_{22}^{(3)}}$ \end{tabular}}} \\ \hline \hline
\multirow{2}{*}{0.1} & 0.2& (0.68,0.90) & (0.11,0.26) & (0.63,0.85)\\
& 0.3& (0.68,0.90) & (0.11,0.26) & (0.63,0.85) \\ \hline
\multirow{2}{*}{0.2} & 0.1& (0.68,0.90) & (0.12,0.26) & (0.63,0.85) \\
& 0.3&  (0.68,0.90)  & (0.12,0.26) & (0.63,0.85) \\ \hline
\multirow{2}{*}{0.3} & 0.1& (0.68,0.90) & (0.12,0.26) & (0.63,0.85) \\
& 0.2& (0.68,0.90) & (0.11,0.26) & (0.63,0.85)  \\ \hline            
\end{tabular}
\caption{95 \% credible intervals for block specific edge probabilities $\boldsymbol{\theta_3}$ of cluster 3, under SBM 1.}\label{cred_theta3_sbm1_new}
\end{table} 

\begin{table}[ht!]
\centering
\begin{tabular}{ccccc}
\textbf{}&& \multicolumn{3}{c}{\textbf{\textbf{$\text{SBM}_2$}}}\\ \hline \hline
$\boldsymbol{p_c}$         & \multicolumn{1}{c}{$\boldsymbol{q_c}$} & \multicolumn{1}{c}{\textbf{\begin{tabular}[c]{@{}c@{}}credible interval\\ for $\boldsymbol{\theta_{11}^{(3)}}$ \end{tabular}}} & \multicolumn{1}{c}{\textbf{\begin{tabular}[c]{@{}c@{}}credible interval\\ for $\boldsymbol{\theta_{12}^{(3)}}$ \end{tabular}}} & \multicolumn{1}{c}{\textbf{\begin{tabular}[c]{@{}c@{}}credible interval\\ for $\boldsymbol{\theta_{22}^{(3)}}$ \end{tabular}}} \\ \hline \hline
\multirow{2}{*}{0.1} & 0.2&(0.59,0.96) & (0.00,0.06) & (0.75,0.89)\\
& 0.3& (0.58,0.96) & (0.00,0.06) & (0.75,0.89) \\ \hline
\multirow{2}{*}{0.2} & 0.1& (0.59,0.96) & (0.00,0.06) & (0.75,0.89) \\
& 0.3&  (0.59,0.96) & (0.00,0.06) &(0.75,0.89)   \\ \hline
\multirow{2}{*}{0.3} & 0.1&(0.59,0.96) & (0.00,0.06) & (0.75,0.90) \\
& 0.2& (0.59,0.96) & (0.00,0.06) &  (0.75,0.89) \\ \hline            
\end{tabular}
\caption{95 \% credible intervals for block specific edge probabilities $\boldsymbol{\theta_3}$ of cluster 3, under SBM 2.}\label{cred_theta3_sbm2_new}
\end{table}

\begin{table}[ht!]
\centering
\begin{tabular}{ccccc}
\textbf{}&& \multicolumn{3}{c}{\textbf{\textbf{$\text{SBM}_1$}}}\\ \hline \hline
$\boldsymbol{p_c}$         & \multicolumn{1}{c}{$\boldsymbol{q_c}$} & \multicolumn{1}{c}{\textbf{\begin{tabular}[c]{@{}c@{}}posterior means\\ of $(\boldsymbol{w_{1}^{(1)}},\boldsymbol{w_{2}^{(1)}})$ \end{tabular}}} & \multicolumn{1}{c}{\textbf{\begin{tabular}[c]{@{}c@{}}posterior means\\ of $(\boldsymbol{w_{1}^{(2)}},\boldsymbol{w_{2}^{(2)}})$ \end{tabular}}} & \multicolumn{1}{c}{\textbf{\begin{tabular}[c]{@{}c@{}}posterior means\\ of $(\boldsymbol{w_{1}^{(3)}},\boldsymbol{w_{2}^{(3)}})$\end{tabular}}} \\ \hline \hline
\multirow{2}{*}{0.1} & 0.2&(0.48,0.52) & (0.52,0.48) & (0.48,0.52)\\
& 0.3& (0.48,0.52)  & (0.52,0.48)& (0.48,0.52) \\ \hline
\multirow{2}{*}{0.2} & 0.1& (0.48,0.52) & (0.52,0.48) & (0.48,0.52) \\
& 0.3&  (0.48,0.52) & (0.52,0.48) &(0.52,0.48)   \\ \hline
\multirow{2}{*}{0.3} & 0.1&(0.48,0.52) & (0.48,0.52) & (0.52,0.48) \\
& 0.2& (0.52,0.48) & (0.48,0.52) &  (0.48,0.52) \\ \hline            
\end{tabular}
\caption{Posterior means for the probability of a node to belong to a block $\boldsymbol{w_c}$ for $c\in\{1,2,3\}$, under SBM 1.}\label{post_mean_w_sbm1_new}
\end{table}

\begin{table}[ht!]
\centering
\begin{tabular}{ccccc}
\textbf{}&& \multicolumn{3}{c}{\textbf{\textbf{$\text{SBM}_2$}}}\\ \hline \hline
$\boldsymbol{p_c}$         & \multicolumn{1}{c}{$\boldsymbol{q_c}$} & \multicolumn{1}{c}{\textbf{\begin{tabular}[c]{@{}c@{}}posterior means\\ of $(\boldsymbol{w_{1}^{(1)}},\boldsymbol{w_{2}^{(1)}})$ \end{tabular}}} & \multicolumn{1}{c}{\textbf{\begin{tabular}[c]{@{}c@{}}posterior means\\ of $(\boldsymbol{w_{1}^{(2)}},\boldsymbol{w_{2}^{(2)}})$ \end{tabular}}} & \multicolumn{1}{c}{\textbf{\begin{tabular}[c]{@{}c@{}}posterior means\\ of $(\boldsymbol{w_{1}^{(3)}},\boldsymbol{w_{2}^{(3)}})$\end{tabular}}} \\ \hline \hline
\multirow{2}{*}{0.1} & 0.2&(0.70,0.30) & (0.52,0.48) & (0.30,0.70)\\
& 0.3& (0.70,0.30)  & (0.52,0.48)& (0.30,0.70) \\ \hline
\multirow{2}{*}{0.2} & 0.1& (0.70,0.30) & (0.52,0.48) & (0.30,0.70) \\
& 0.3&  (0.70,0.30) & (0.48,0.52) &(0.30,0.70)   \\ \hline
\multirow{2}{*}{0.3} & 0.1&(0.70,0.30) & (0.52,0.48) & (0.30,0.70) \\
& 0.2& (0.70,0.30) & (0.48,0.52) &  (0.30,0.70) \\ \hline            
\end{tabular}
\caption{Posterior means for the probability of a node to belong to a block $\boldsymbol{w_c}$ for $c\in\{1,2,3\}$, under SBM 2.}\label{post_mean_w_sbm2_new}
\end{table}

\begin{table}[ht!]
\centering
\begin{tabular}{ccccc}
\textbf{}&& \multicolumn{3}{c}{\textbf{\textbf{$\text{SBM}_1$}}}\\ \hline \hline
$\boldsymbol{p_c}$         & \multicolumn{1}{c}{$\boldsymbol{q_c}$} & \multicolumn{1}{c}{$\boldsymbol{d_{H}<=1}$} & \multicolumn{1}{c}{$\boldsymbol{d_{H}<=5}$} & \multicolumn{1}{c}{$\boldsymbol{d_{H}<=10}$} \\ \hline \hline
\multirow{2}{*}{0.1} & 0.2&(1.00,1.00,1.00)  & (1.00,1.00,1.00)  & (1.00,1.00,1.00) \\
& 0.3& (1.00,1.00,1.00)   & (1.00,1.00,1.00) & (1.00,1.00,1.00)  \\ \hline
\multirow{2}{*}{0.2} & 0.1& (1.00,1.00,1.00)  & (1.00,1.00,1.00)  & (1.00,1.00,1.00)  \\
& 0.3&  (1.00,1.00,1.00)  & (1.00,1.00,1.00)  &(1.00,1.00,1.00)    \\ \hline
\multirow{2}{*}{0.3} & 0.1&(1.00,1.00,1.00)  & (1.00,1.00,1.00)  & (1.00,1.00,1.00)  \\
& 0.2& (1.00,1.00,1.00)  & (1.00,1.00,1.00)  &  (1.00,1.00,1.00) \\ \hline            
\end{tabular}
\caption{Proportion of times that the Hamming distance between the posterior representatives and the true representatives is less or equal than 1, 5 and 10 respectively, under SBM 1.}\label{dh_sbm1_new}
\end{table}

\begin{table}[ht!]
\centering
\begin{tabular}{ccccc}
\textbf{}&& \multicolumn{3}{c}{\textbf{\textbf{$\text{SBM}_2$}}}\\ \hline \hline
$\boldsymbol{p_c}$         & \multicolumn{1}{c}{$\boldsymbol{q_c}$} & \multicolumn{1}{c}{$\boldsymbol{d_{H}<=1}$} & \multicolumn{1}{c}{$\boldsymbol{d_{H}<=5}$} & \multicolumn{1}{c}{$\boldsymbol{d_{H}<=10}$} \\ \hline \hline
\multirow{2}{*}{0.1} & 0.2&(1.00,1.00,1.00)  & (1.00,1.00,1.00)  & (1.00,1.00,1.00) \\
& 0.3& (1.00,1.00,1.00)   & (1.00,1.00,1.00) & (1.00,1.00,1.00)  \\ \hline
\multirow{2}{*}{0.2} & 0.1& (1.00,1.00,1.00)  & (1.00,1.00,1.00)  & (1.00,1.00,1.00)  \\
& 0.3&  (1.00,1.00,1.00)  & (1.00,1.00,1.00)  &(1.00,1.00,1.00)    \\ \hline
\multirow{2}{*}{0.3} & 0.1&(1.00,1.00,1.00)  & (1.00,1.00,1.00)  & (1.00,1.00,1.00)  \\
& 0.2& (1.00,1.00,1.00)  & (1.00,1.00,1.00)  &  (1.00,1.00,1.00) \\ \hline            
\end{tabular}
\caption{Proportion of times that the Hamming distance between the posterior representatives and the true representatives is less or equal than 1, 5 and 10 respectively, under SBM 2.}\label{dh_sbm2_new}
\end{table}

\begin{table}[ht!]
\centering
\begin{tabular}{cccccc}
\textbf{}&& \multicolumn{2}{c}{\textbf{\textbf{$\text{SBM}_1$}}}&\multicolumn{2}{c}{\textbf{$\text{SBM}_2$}}\\ \hline \hline
$\boldsymbol{p_c}$         & \multicolumn{1}{c}{$\boldsymbol{q_c}$} & \multicolumn{1}{c}{\textbf{Mean Entropy}} & \multicolumn{1}{c}{\textbf{Mean Purity}} & \multicolumn{1}{c}{\textbf{Mean Entropy}} & \multicolumn{1}{c}{\textbf{Mean Purity}} \\ \hline \hline
\multirow{2}{*}{0.1} & 0.2& 0& 1& 0& 1\\
& 0.3& 0& 1& 0& 1\\ \hline
\multirow{2}{*}{0.2} & 0.1& 0& 1& 0& 1\\
& 0.3& 0& 1& 0& 1\\ \hline
\multirow{2}{*}{0.3} & 0.1& 0& 1& 0& 1\\
& 0.2& 0& 1& 0& 1\\ \hline            
\end{tabular}
\caption{Mean clustering entropy and clustering purity}\label{entr_pur_new}
\end{table} 

\newpage

\subsection{Additional details for simulations involving varying network sizes and population sizes}

In Figures 1-4, we illustrate the representatives of clusters 2 and 3, with 25, 50, 75 and 100 nodes respectively, generated for the simulation study presented in Section 5.2 of the main article.

\begin{figure}[h!]
\includegraphics[height=1.7in,width=0.49\textwidth]{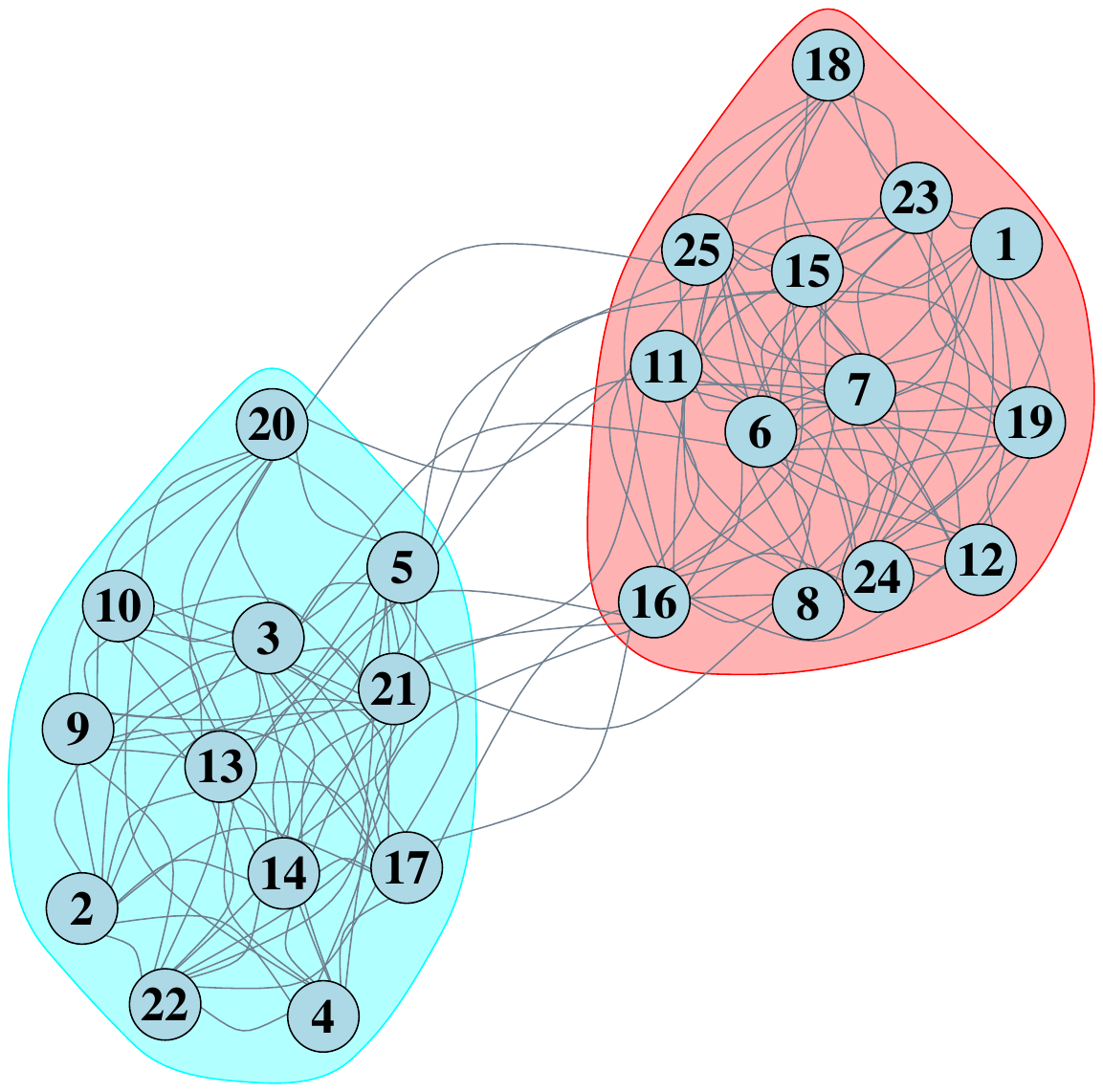}
\hfill 
\includegraphics[height=1.7in,width=0.49\textwidth]{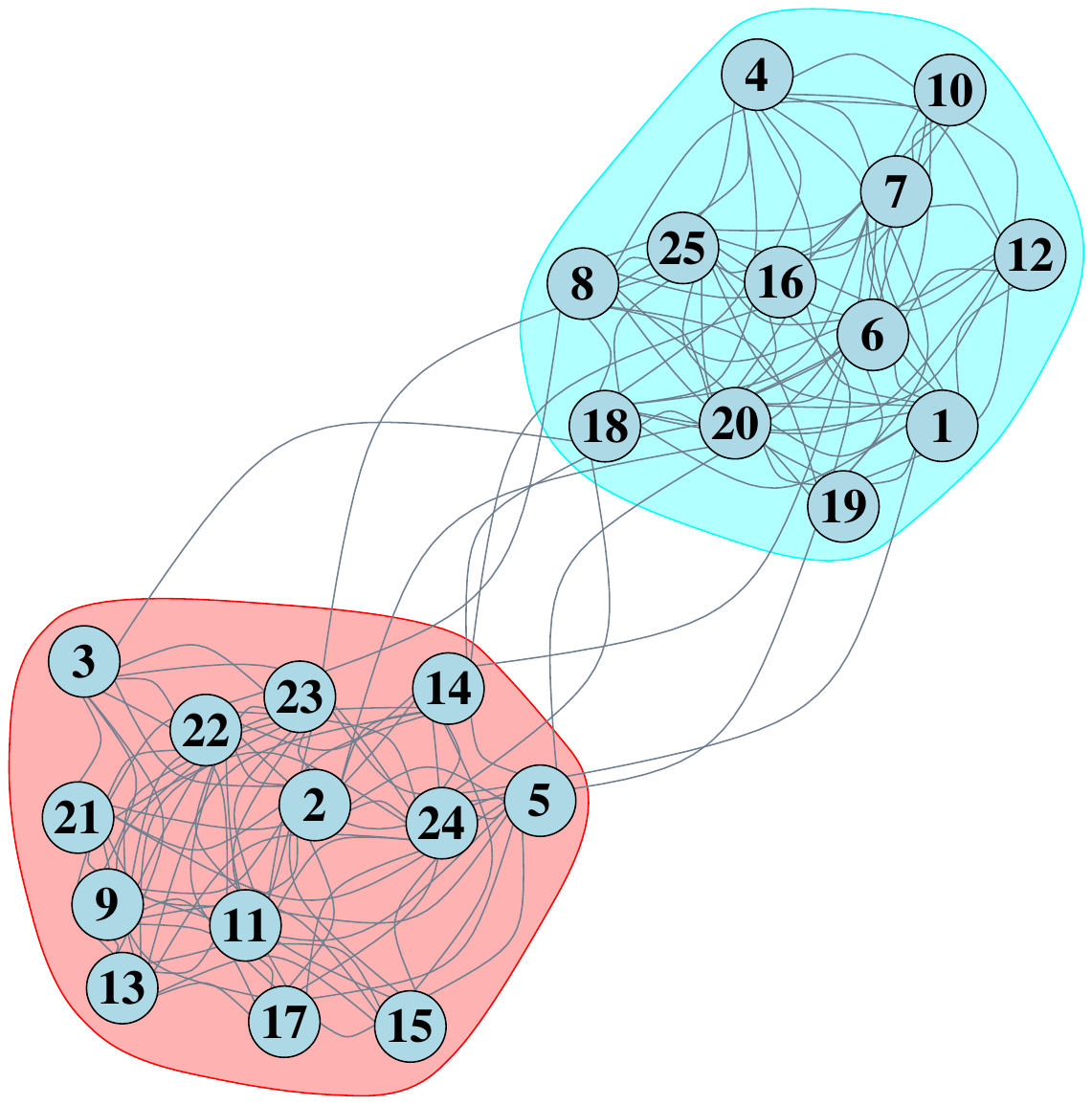}
\caption{25-node representatives of cluster 2 (left) and cluster 3 (right).}
\end{figure}

\begin{figure}[h!]
\includegraphics[height=1.7in,width=0.49\textwidth]{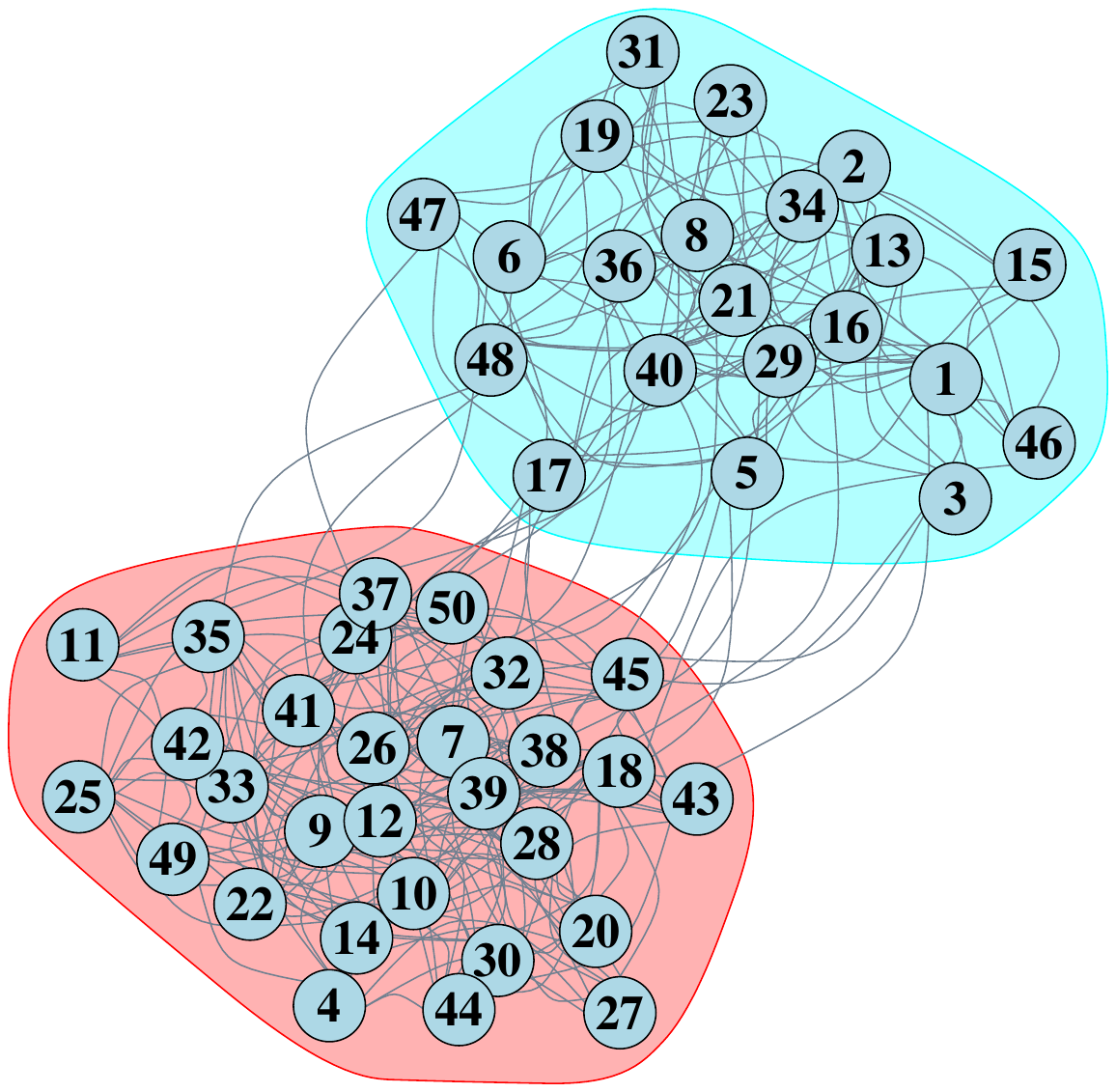}
\hfill 
\includegraphics[height=1.7in,width=0.49\textwidth]{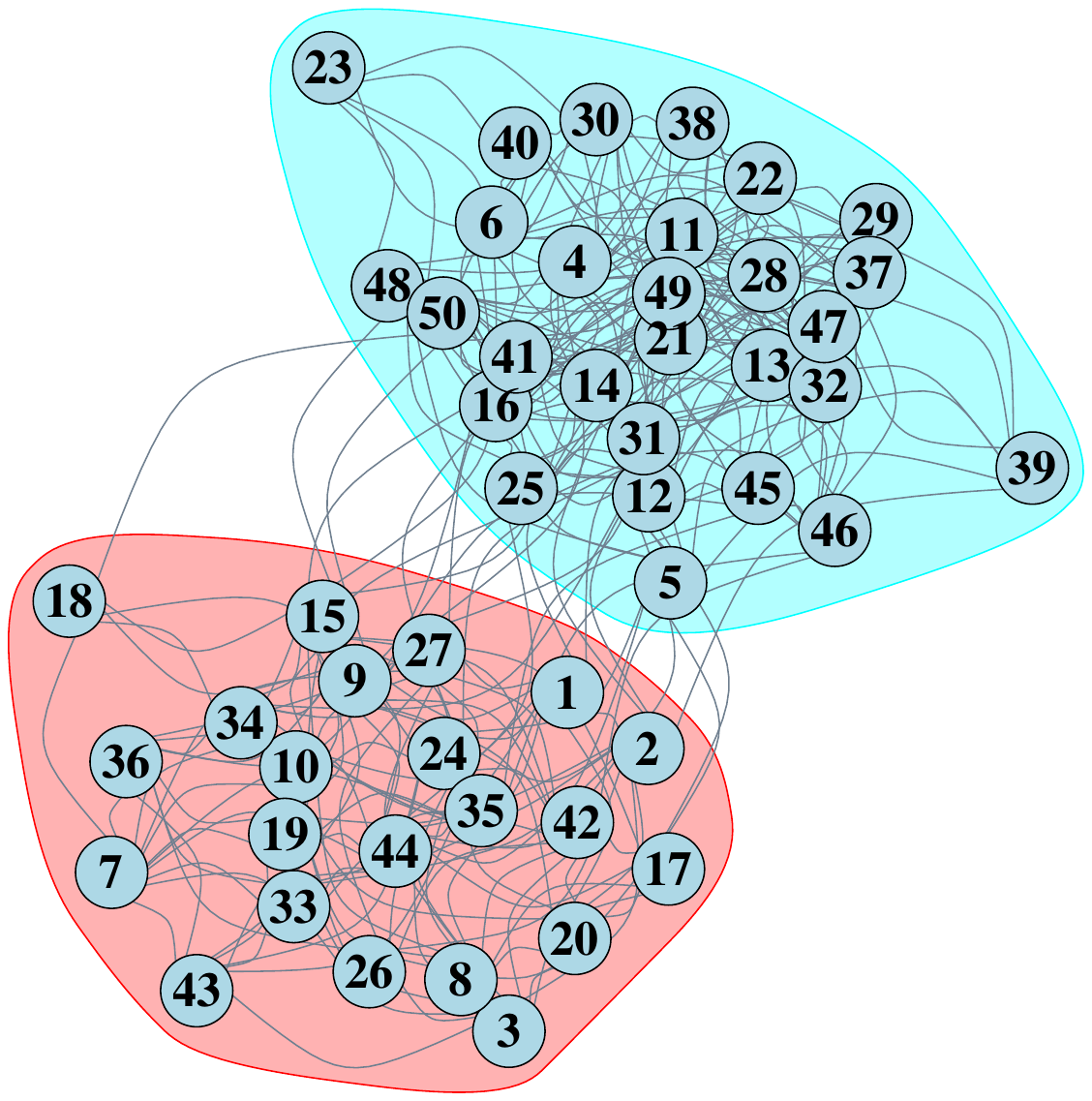}
\caption{50-node representatives of cluster 2 (left) and cluster 3 (right).}
\end{figure}

\begin{figure}[h!]
\includegraphics[height=1.7in,width=0.49\textwidth]{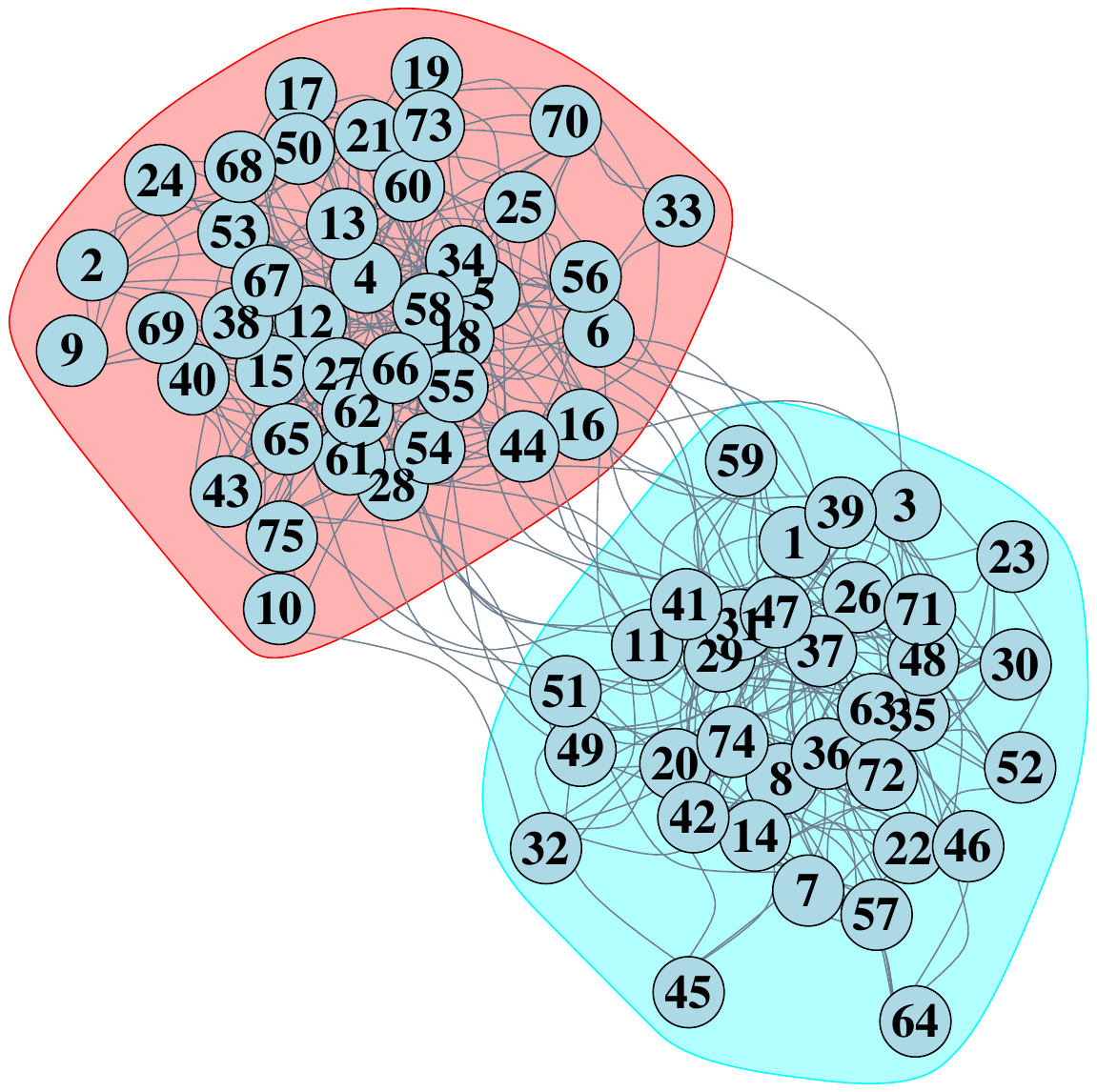}
\hfill 
\includegraphics[height=1.7in,width=0.49\textwidth]{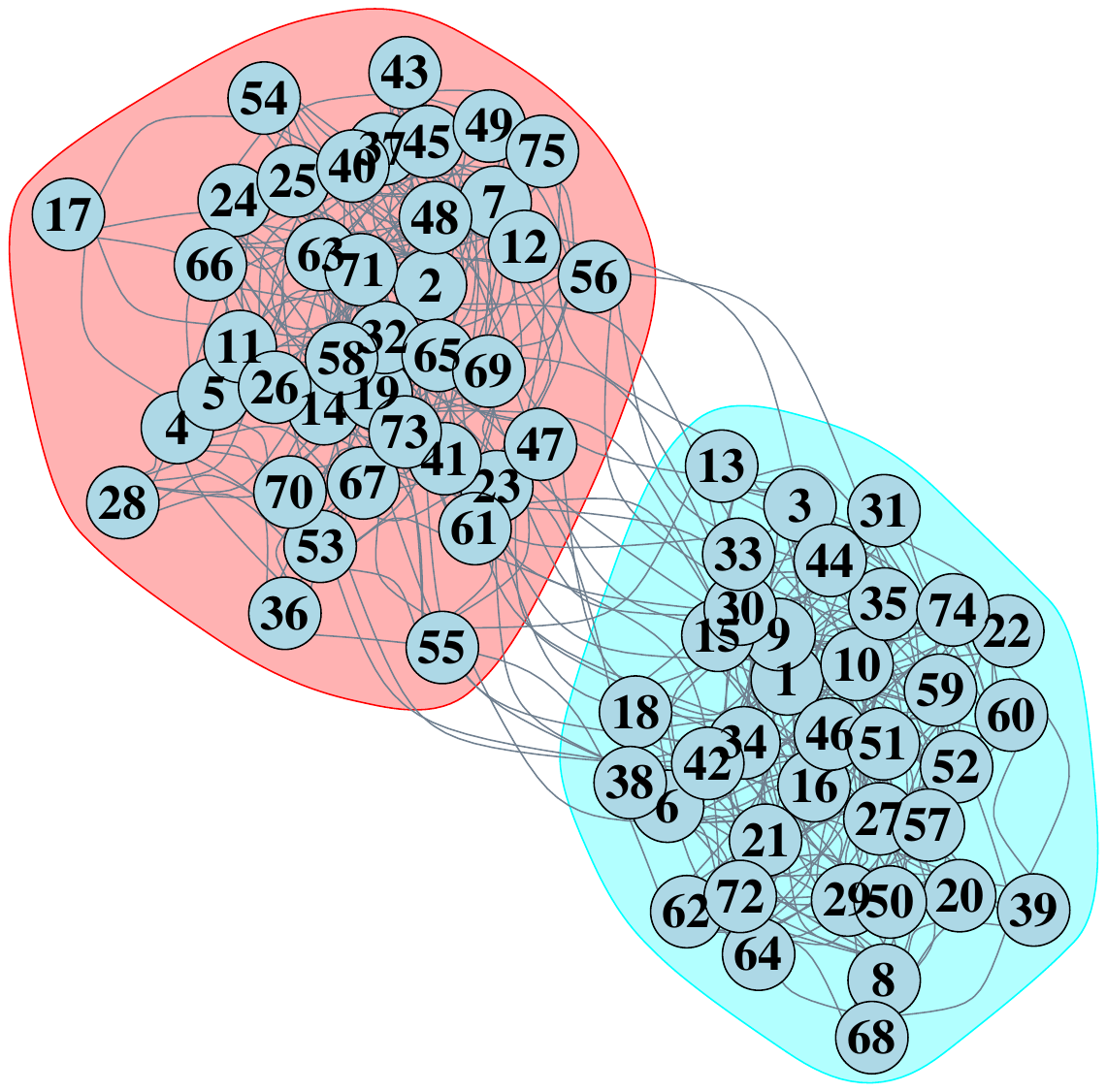}
\caption{75-node representatives of cluster 2 (left) and cluster 3 (right).}
\end{figure}

\begin{figure}[h!]
\includegraphics[height=1.7in,width=0.49\textwidth]{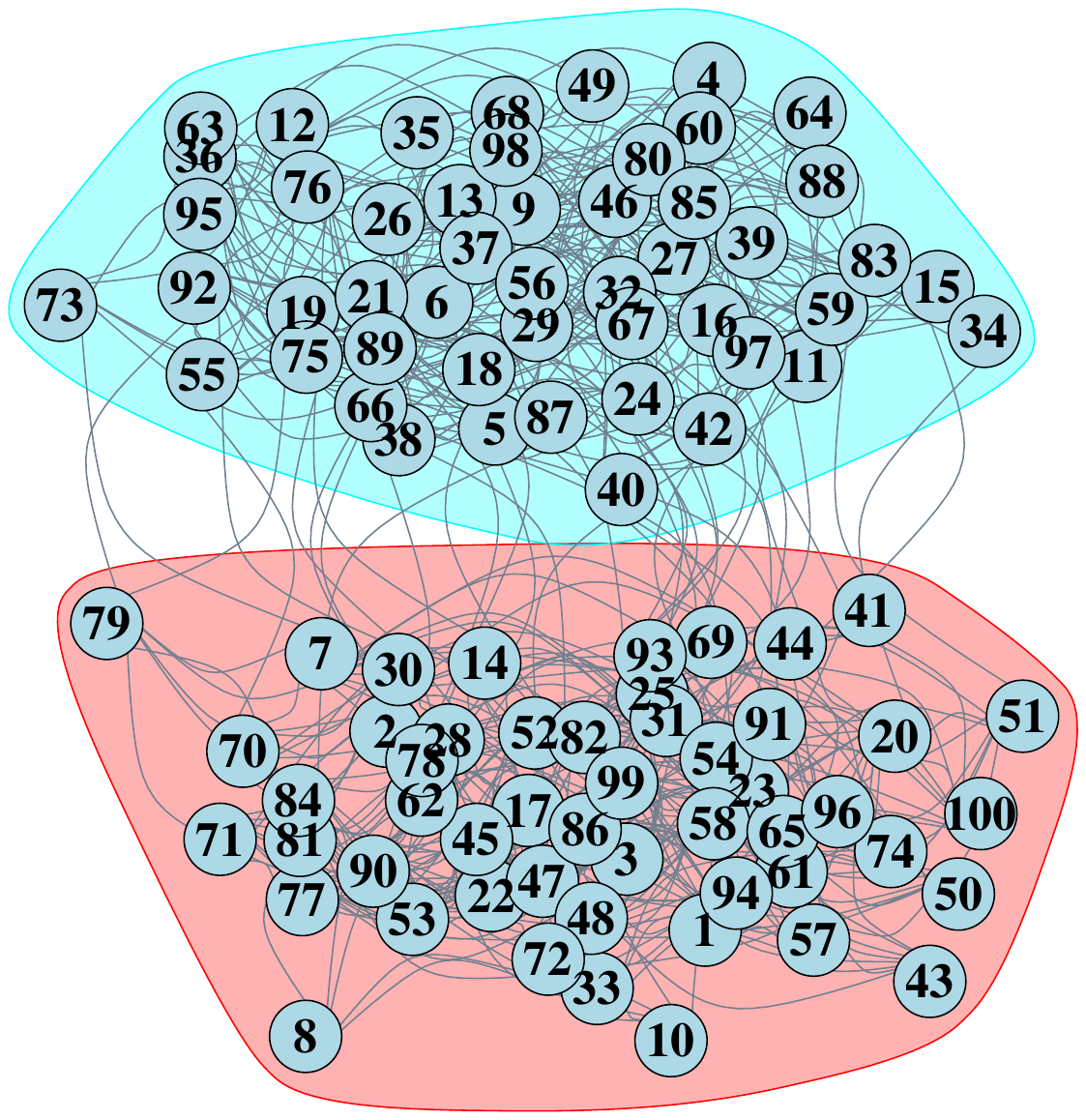}
\hfill 
\includegraphics[height=1.7in,width=0.49\textwidth]{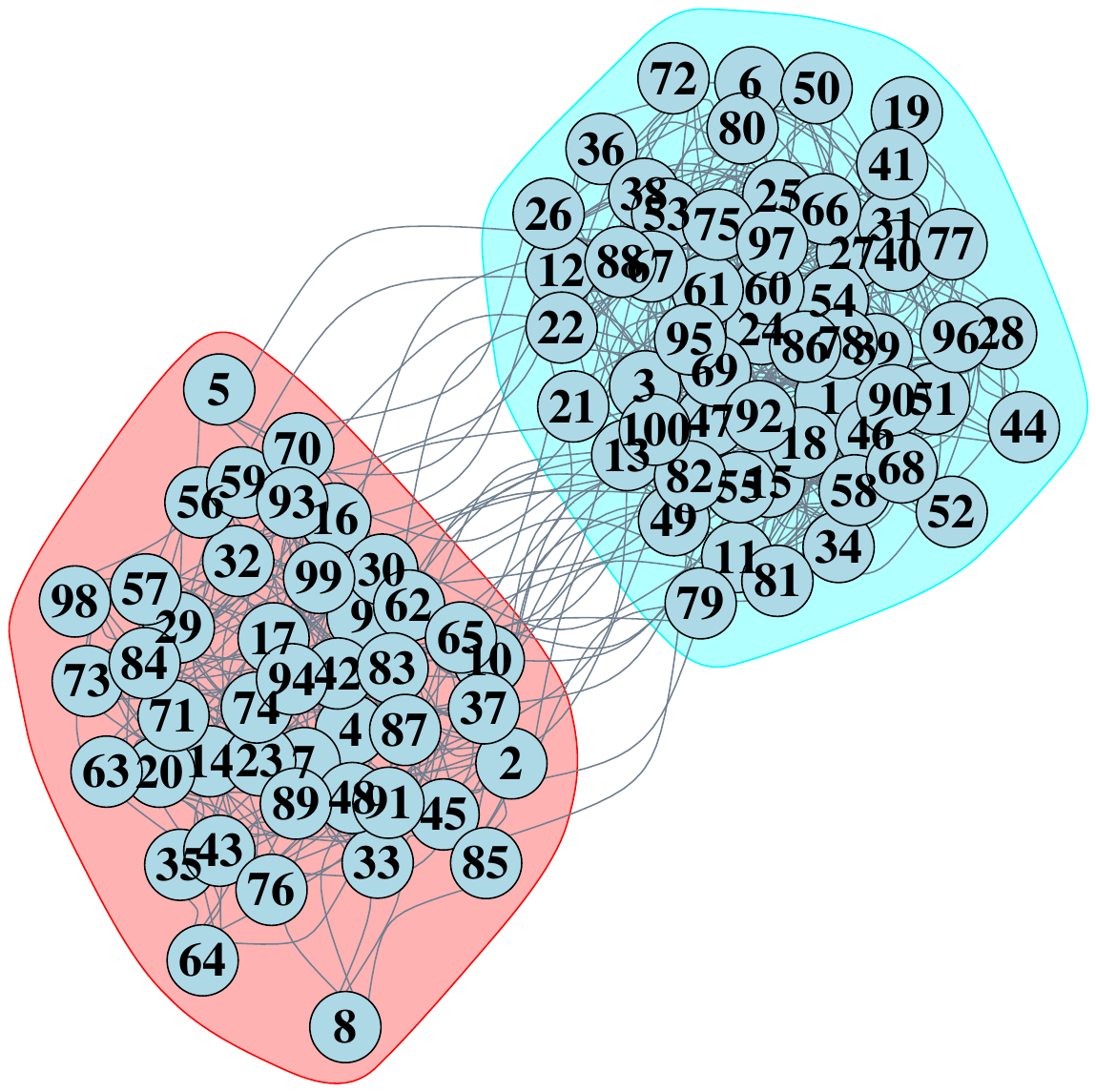}
\caption{100-node representatives of cluster 2 (left) and cluster 3 (right).}
\end{figure}

Similarly to the posterior summaries obtained with moderate network sizes, we obtain the proportion of times that the Hamming distance between posterior draws of the representative and the true representative is less than or equal to 1, 5 and 10, for each cluster. We consider the final 350,000 posterior draws after a burn-in of 150,000 iterations. The results are summarised in Figure 5. The multiple subfigures correspond to the 25, 50, 75 and 100 node representatives.

\begin{figure}[htb!]
\includegraphics[height=1.8in,width=0.32\textwidth]{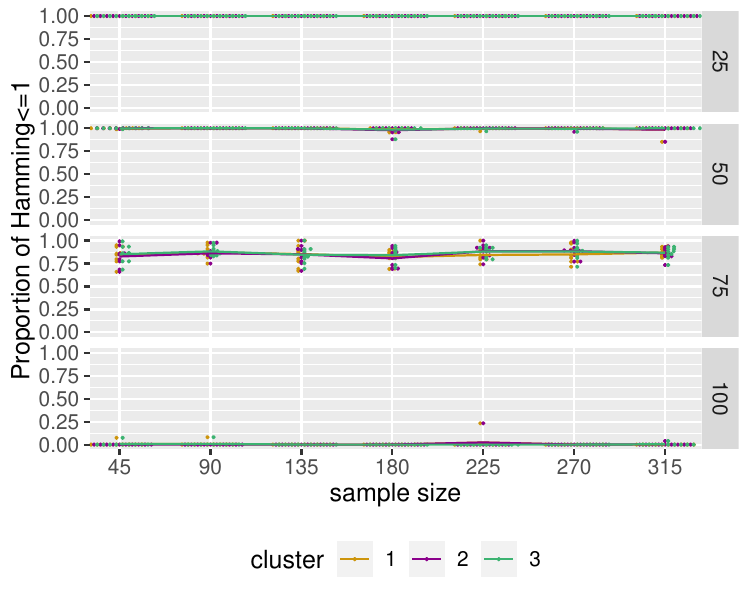}
\hfill 
\includegraphics[height=1.8in,width=0.32\textwidth]{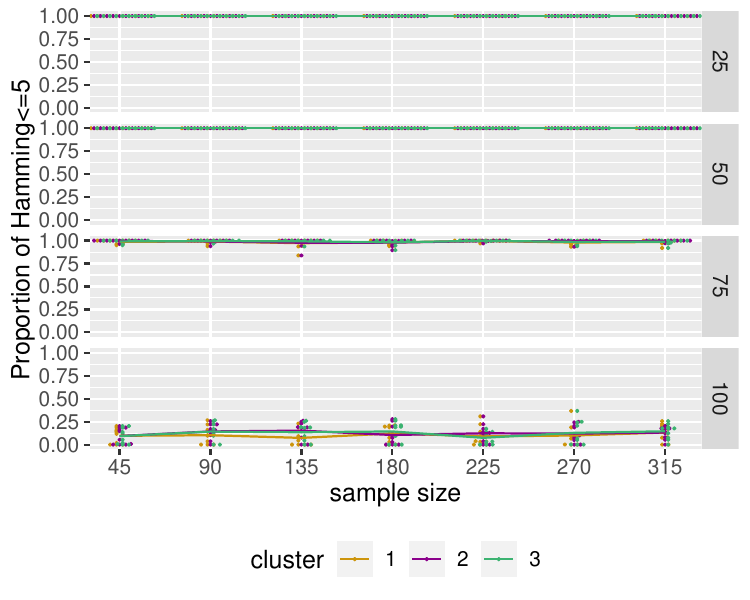}
\hfill 
\includegraphics[height=1.8in,width=0.32\textwidth]{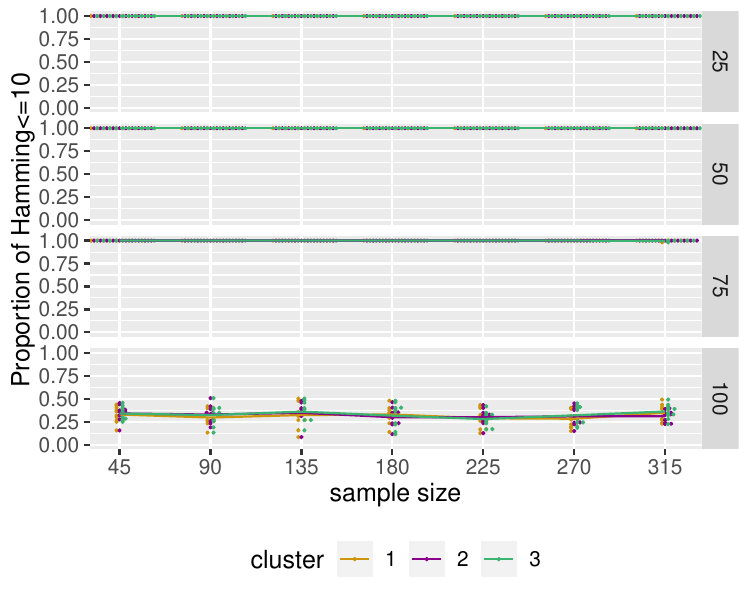}
\vspace{3mm}
\caption{Left: Proportion of times the Hamming distance is less or equal to 1 (y axis) for 25-node, 50-node, 75-node and 100-node network representatives and varying population sizes (x axis). Middle: Proportion of times the Hamming distance is less or equal to 5 (y axis) for 25-node, 50-node, 75-node and 100-node network representatives and varying population sizes (x axis). Right: Proportion of times the Hamming distance is less or equal to 10 (y axis) for 25-node, 50-node, 75-node and 100-node network representatives and varying population sizes (x axis).}\label{post_modes_brain}
\end{figure}
\newpage
\section{Additional details for real data application}

\subsection{Additional details for the analysis in Section 6.1}
In this section we present details of the analysis performed on the data collected by the Tacita mobile application monitoring the movement of individuals across a University campus. We present the results of the exploratory data analysis (EDA) conducted,
as well as additional results from fitting our mixture model to the data. Table 19 presents the labels of the nodes corresponding to each display located on Lancaster University campus.

\begin{table}[h]
\centering
\begin{tabular}{lc}
  \hline
 \textbf{Display name} & \textbf{Node label} \\ 
  \hline
 SCC (C-floor) & 1 \\ 
  Infolab Foyer & 2 \\ 
 Faraday Left & 3\\ 
Engineering Foyer (far) & 4 \\ 
  LZ1 & 5 \\ 
  LZ3 & 6\\ 
 Furness 1 & 7 \\ 
   Furness 2 & 8 \\ 
 Furness College & 9 \\ 
  Engineering Foyer (near) & 10 \\ 
   LEC 1 & 11 \\ 
 LEC 2 & 12 \\ 
   County College & 13 \\ 
   Lonsdale College & 14 \\ 
   Grizedale College & 15 \\ 
   The Base & 16 \\ 
   Faraday B & 17 \\ 
   Faraday C & 18 \\ 
   Bowland JCR & 19 \\ 
   \hline
\end{tabular}
\quad
\begin{tabular}{lc}
  \hline
 \textbf{Display name} & \textbf{Node label}\\ 
  \hline
 ISS & 20 \\ 
 Pendle College & 21 \\ 
  New Engineering & 22 \\ 
  Fylde College & 23 \\ 
  Graduate College & 24 \\ 
  Library A & 25 \\ 
  Library B & 26 \\ 
  Library C & 27 \\ 
  Bowland Main B & 28 \\ 
  Bowland North B & 29\\ 
  Hotel Conference & 30\\ 
  Chemistry A & 31 \\ 
  Psychology & 32 \\ 
  Physics & 33 \\ 
  Law 2 & 34 \\ 
  Law 1 & 35 \\ 
  Welcome Screen 1 & 36 \\ 
  Welcome Screen 2 & 37 \\ 
   \hline
\end{tabular}
\caption{Node label assigned to each display on Lancaster University campus.}
\end{table}

\begin{figure}[h!]
\centering
\includegraphics[height=2in,width=0.49\textwidth]{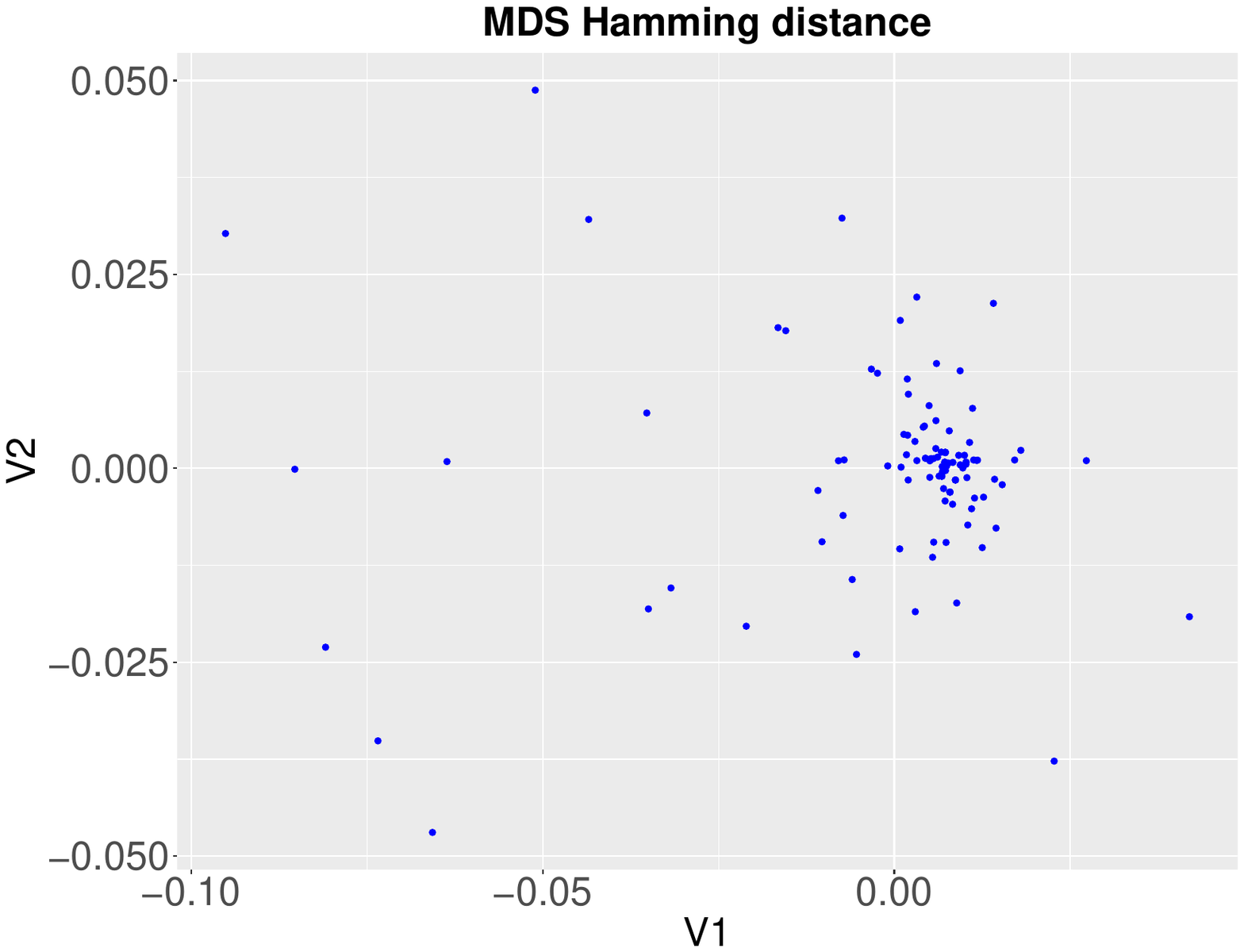}
\includegraphics[height=2in,width=0.49\textwidth]{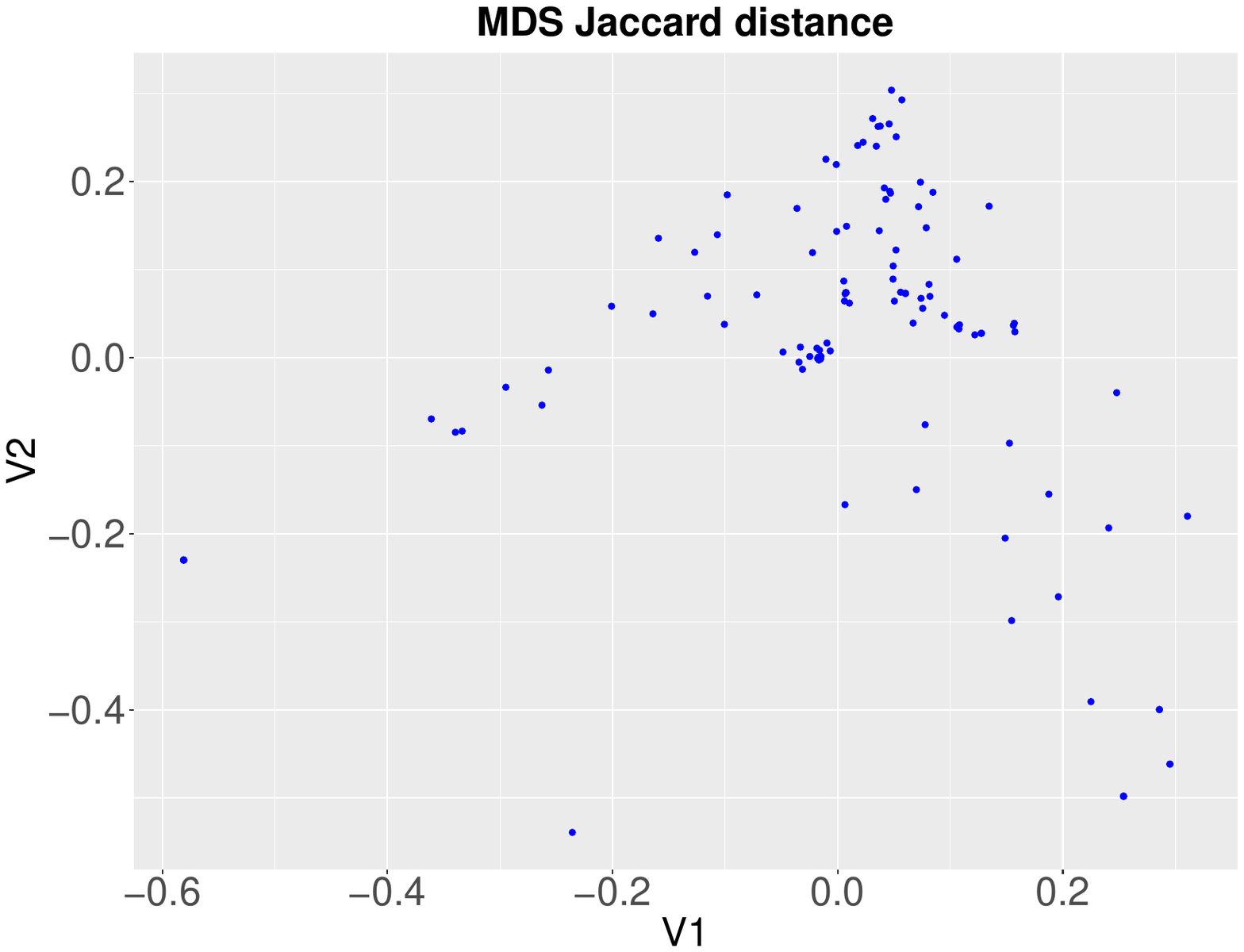}
\caption{Left: MDS for Hamming distance matrix for the Tacita data. Right: MDS for Jaccard distance matrix for the Tacita data.}\label{MDS1}
\vfill
\centering
\includegraphics[height=2in,width=0.49\textwidth]{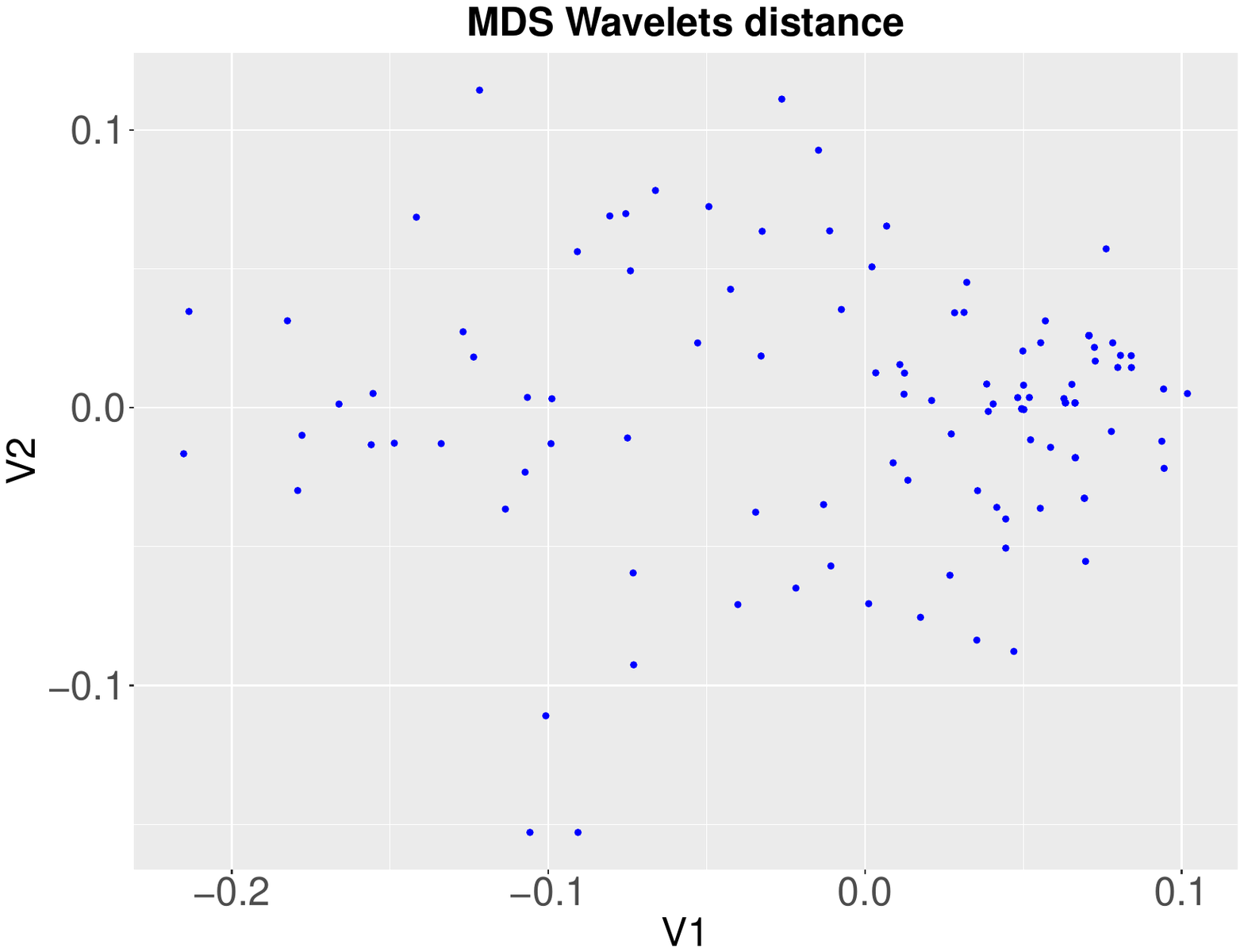}
\includegraphics[height=2in,width=0.49\textwidth]{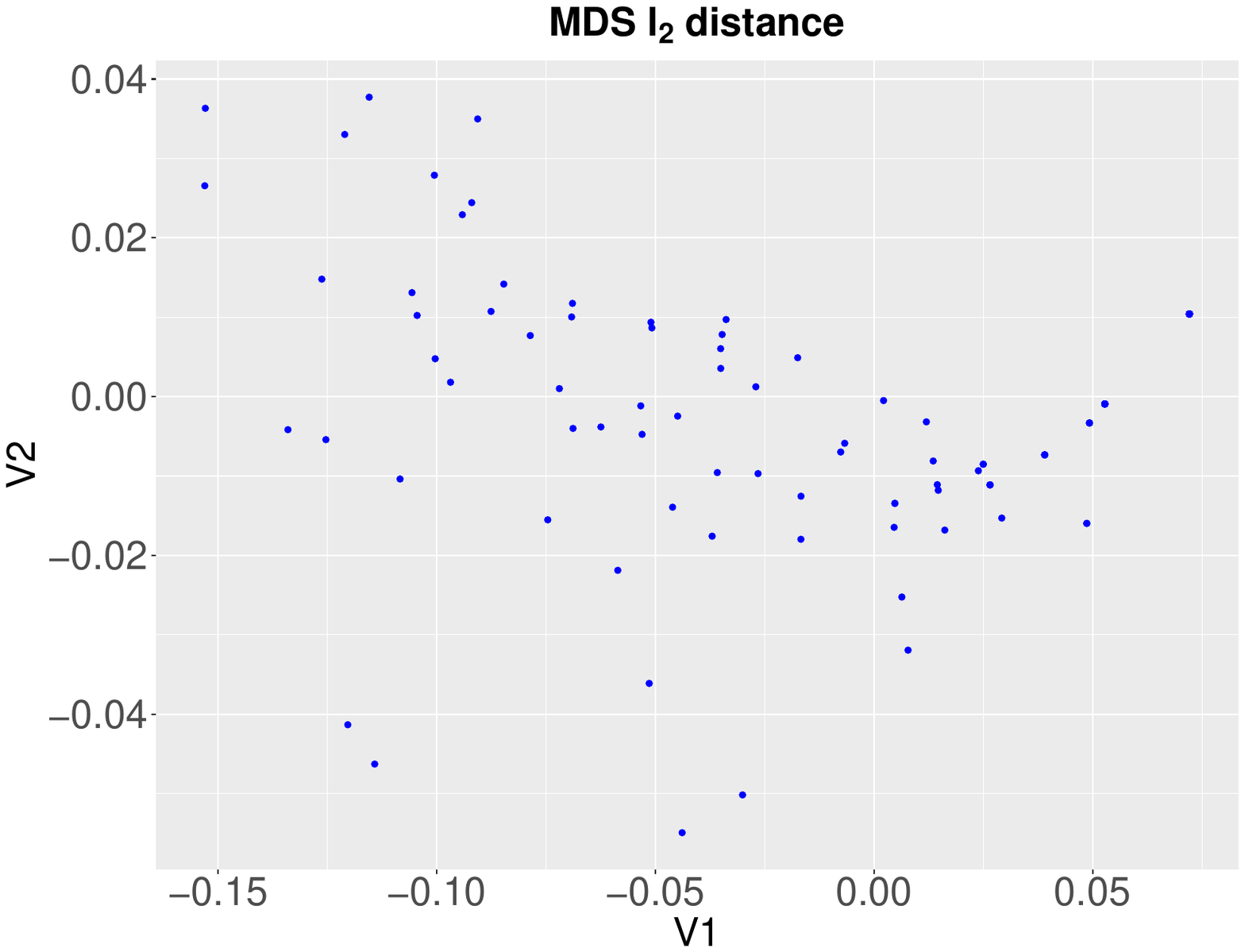}
\caption{Left: MDS for wavelets distance matrix for the Tacita data.
Right: MDS for $l_{2}$ distance matrix for the Tacita data.}\label{MDS2}
\end{figure}

As discussed in the main article, we performed EDA on the Tacita multiple network data through the use of network distance metrics. Specifically, for each distance metric, we derive a distance matrix that encloses the pairwise distances of the networks in the population. Then, the $(i,j)$ element of a distance matrix corresponds to the distance between graphs $\mathcal{G}_{i}$ and $\mathcal{G}_{j}$, for the specified distance metric. We consider various distance metrics, as different metrics can give us different information with respect to the presence of clusters in the network population. We consider the Hamming, the Jaccard, the $l_2$ distance and the distance based on wavelets. To graphically represent the distance matrices for each distance metric, we use a Multidimensional Scaling (MDS) plot. The MDS algorithm maps objects in a 2-d space, respecting their pairwise distances. In Figures 6 and 7, we plot the MDS representation under the distance matrices obtained under the Hamming, the Jaccard, the $l_{2}$, and the wavelets distance metrics. 

We also present some additional results after fitting our mixture model to the Tacita data. Figures 8 and 9 show the trace plots of 400,000 posterior draws for the false positive probabilities, $p_c$ ,and false negative probabilities, $q_c$, with $c\in\{1,2,3\}$, after a burn-in of 100,000 iterations. In addition, Figure 10 shows the proportion of times that each of the 37 nodes of the representative of cluster $c=1$ is allocated to either block 1 or 2. 

\begin{figure}[h!]
\centering
\includegraphics[height=2in,width=0.29\textwidth]{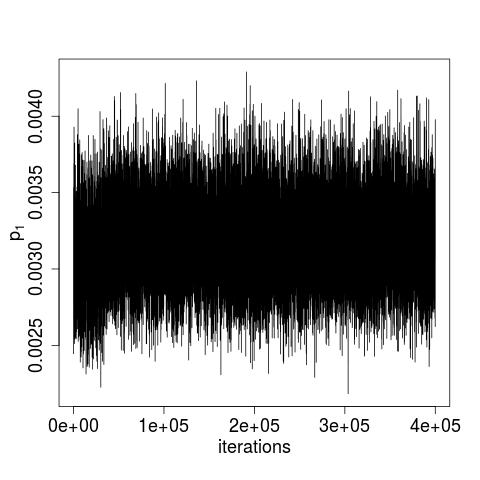}
\includegraphics[height=2in,width=0.29\textwidth]{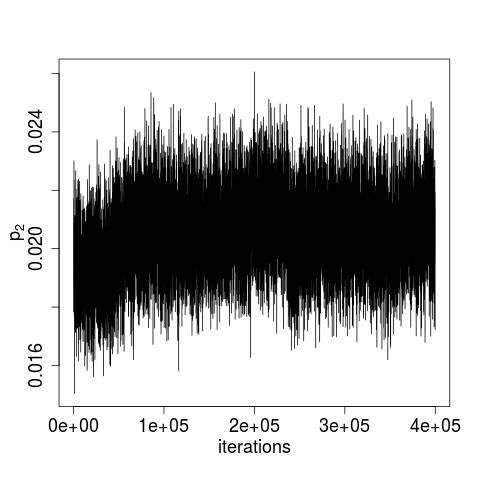}
\includegraphics[height=2in,width=0.29\textwidth]{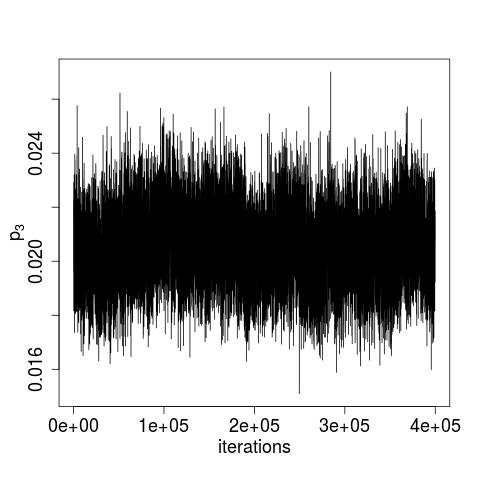}
\caption{Trace plots of false positive probabilities $p_c$ for $c=\{1,2,3\}$, for 400,000 iterations of the MCMC after a burn-in of 100,000 iterations.}\label{fp}
\vfill
\includegraphics[height=2in,width=0.29\textwidth]{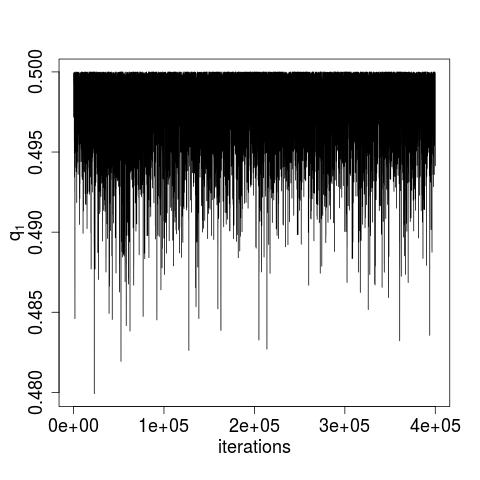}
\includegraphics[height=2in,width=0.29\textwidth]{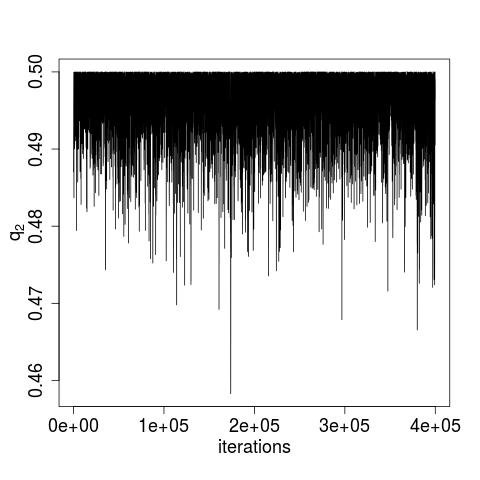}
\includegraphics[height=2in,width=0.29\textwidth]{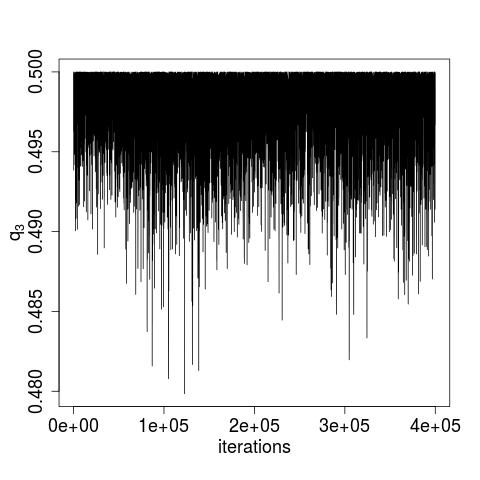}
\caption{Trace plots of false negative probabilities $q_c$ for $c=\{1,2,3\}$, for 400,000 iterations of the MCMC  after a burn-in of 100,000 iterations.}\label{fq}
\end{figure}
\clearpage
\newpage
\begin{figure}[hbt!]
\centering
\includegraphics[height=2in,width=3in]{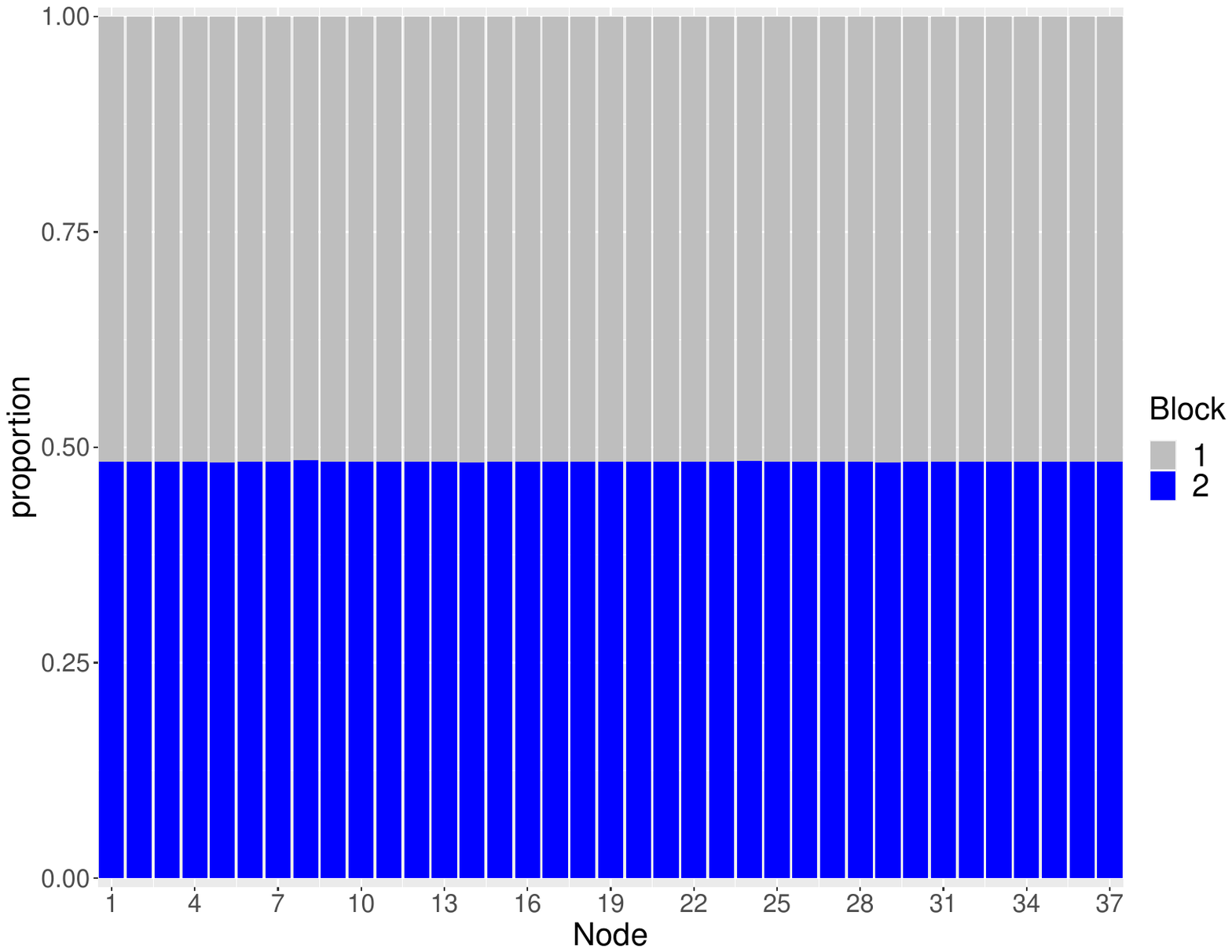}
\vspace{-3mm}
\caption{Proportion of times that each node (x axis) of representative of cluster 1, is allocated to each of the two blocks, after a burn-in of 100,000 iterations.}\label{block}
\end{figure}

\subsection{Additional details for the analysis in Section 6.2}

In Figure 11 we show the trace plots for the false positive, $p_{out}$, and false negative, $q_{out}$, probabilities, for the outlier cluster of networks detected, under three different initialisations, and 1,000,000 iterations of the MCMC.

\begin{figure}[h!]
\centering
\includegraphics[width=4in,height=2in]{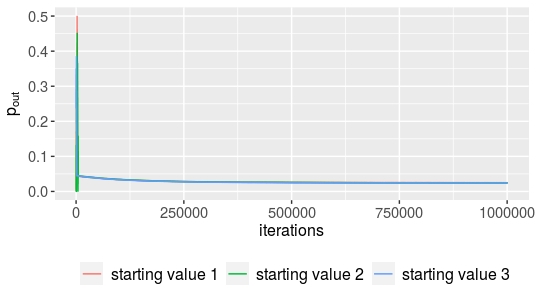}
\vfill 
\includegraphics[width=4in,height=2in]{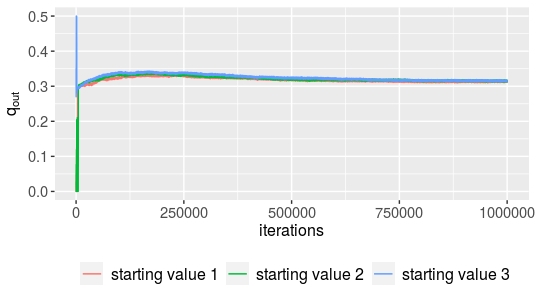}
\caption{Top: Trace plot of false positive probability for outlier cluster $p_{out}$ for 1,000,000 iterations and three different initialisations. Bottom: Trace plot of false negative probability for outlier cluster $q_{out}$ for 1,000,000 iterations and three different initialisations.}\label{trc_out}
\end{figure}

\clearpage
\newpage
\IncMargin{1em}

\section{MCMC algorithm}

Below, we present details of how the MCMC algorithm is implemented to make posterior inferences from the proposed model. In Section 4.5 of the main article, we further discuss the Sparse Finite Mixture extension to the algorithm presented herein.

\begin{algorithm}[H]
\KwInput{$\mathcal{G}_1,\mathcal{G}_2,\ldots,\mathcal{G}_N;C,K,M, w_{0}, \theta_{0}, \alpha_{0}, \beta_{0}, \gamma_{0}, \delta_{0}, \epsilon_{0}, \zeta_{0}, \psi,\chi$}
\KwOutput{Posterior distributions of $A_{\mathcal{G}^{*}_{1}},\dots,A_{\mathcal{G}^{*}_{C}}$, $p_{1},\ldots,p_{C}$, $q_{1},\ldots,q_{C}$, $\tau_{1},\ldots,\tau_{C}$, $z_{1},\ldots,z_{N}$, $\boldsymbol{\theta_{1}},\ldots,\boldsymbol{\theta_{C}}$, $\boldsymbol{w_{1}},\ldots,\boldsymbol{w_{C}}$, $\boldsymbol{b_{1}},\ldots,\boldsymbol{b_{C}}$}
\textbf{Initialisation:} starting values
$A_{\mathcal{G}^{*}_{1}}^{(0)},\dots,A_{\mathcal{G}^{*}_{C}}^{ (0)}$, $p_{1}^{(0)},\ldots,p_{C}^{(0)}$, $q_{1}^{(0)},\ldots,q_{C}^{(0)}$, $\tau_{1}^{(0)},\ldots,\tau_{C}^{(0)}$, $z_{1}^{(0)},\ldots,z_{N}^{(0)}$, $\boldsymbol{\theta_{1}}^{(0)},\ldots,\boldsymbol{\theta_{C}}^{(0)}$, $\boldsymbol{w_{1}}^{(0)},\ldots,\boldsymbol{w_{C}}^{(0)}$, $\boldsymbol{b_{1}}^{(0)},\ldots,\boldsymbol{b_{C}}^{(0)}$
\BlankLine
\For{$i\leftarrow 1$ \KwTo M}{
\textbf{Gibbs step:} Update $\tau_{1},\ldots,\tau_{C}$
\newline
compute: $\eta_{c}=\sum_{j=1}^{N} \mathbb{1}_{c}(z_{j}^{(i-1)})$ for $c=1,\ldots,C$
\newline
sample: $\tau_{1}^{(i)},\ldots,\tau_{C}^{(i)}\sim Dir(\psi+\eta_{1},\ldots,\psi+\eta_{C})$
\newline
\For{$c\leftarrow 1$ \KwTo C}{
\textbf{MH step with a mixture of kernels:} Update $A_{\mathcal{G}^{*}_{c}}$ or $p_{c}$ or $q_{c}$
\newline 
sample: $v\sim \text{Multinomial}(\xi_{1},\ldots,\xi_{L})$
\newline
Depending on the value of $v$, update either $A_{\mathcal{G}^{*}_{c}}$ or $p_{c}$ or $q_{c}$ as per the Measurement Error model with SBM structure, where the sum in likelihood is over the networks $\{j: z_{j}^{(i-1)}=c\}$
\BlankLine
\textbf{Gibbs step:} Update $\boldsymbol{w_{c}}$
\newline
compute: $h_{k}=\sum_{j=1}^{n} \mathbb{1}_{k}(b_{j}^{(i-1)})$ 
\newline
sample: $\boldsymbol{w_{c}}^{(i)}\sim Dir(\chi+h_{1},\ldots,\chi+h_{K})$
\BlankLine
\textbf{Gibbs step:} Update $\boldsymbol{\theta_{c}}$
\newline
compute: $A[st]=\sum_{(u,v):b_{u}=s,b_{v}=t} A_{\mathcal{G}_{c}^{*}}^{(i)}(u,v)$ and $n_{st}=\sum_{(u,v):u \neq v} \mathbb{I}(b_{u}=s,b_{v}=t) \text{ for } s,t \in \{1,\ldots,K\}$
\vspace{1mm}
\newline
sample: $\theta_{c,st}^{(i)} \sim$ Beta($A[st]+\epsilon_{0},\zeta_{0}+n_{st}-A[st]$)
\BlankLine
\textbf{Gibbs step:} Update $\boldsymbol{b_{c}}$
\newline
\For{$j\leftarrow 1$ \KwTo n}{
compute: $p_{kj}=w_{c,k}^{(i)} \cdot \prod_{m=1}^{n}\theta_{kb_{c,m}^{(i-1)}}^{(i)A(j,m)}(1-\theta_{kb_{c,m}^{(i-1)}}^{(i)})^{1-A(j,m)}$ for $k=1,\ldots,K$
\newline
sample: $b_{c,j}^{(i)} \sim \text{Multin}(p_{lj},\ldots,p_{Kj})$
}
}
\BlankLine
\textbf{Gibbs step:} Update $z_{1},\ldots,z_{N}$
\newline
\For{$j\leftarrow 1$ \KwTo N}{
compute: $p_{cj}=\tau_{c}^{(i)}\cdot \prod_{(u,v):u<v} \Big( (1-q_{c}^{(i)})^{A_{\mathcal{G}_{j}}(u,v)}q_{c}^{(i)(1-A_{\mathcal{G}_{j}}(u,v))} \Big)^{A_{\mathcal{G}_{c}^{*}}^{(i)}(u,v)} \cdot \Big  (p_{c}^{(i)A_{\mathcal{G}_{j}}(u,v)}(1-p_{c}^{(i)})^{1-A_{\mathcal{G}_{j}}(u,v)} \Big)^{1-A_{\mathcal{G}_{c}^{*}}^{(i)}(u,v)}$ for $c=1,\ldots,C$
\newline
sample: $z_{j}^{(i)}\sim \text{Multin}(p_{1j},\ldots,p_{Cj})$
}
}
\caption{MCMC Algorithm for Clustering Network Populations}
\end{algorithm}